\documentclass[useAMS,usenatbib]{mn2e}
\usepackage{txfonts}
\usepackage{graphicx} 
\usepackage{rotating}
\usepackage{caption}
\usepackage{subfig}

\title[Non-Gaussianity in Pulsar Timing Analysis]{Bayesian estimation of non-Gaussianity in pulsar timing analysis}
\author[L. Lentati et al.]{\parbox{\textwidth}{L. Lentati\thanks{E-mail:
ltl21@cam.ac.uk}, M. P.  Hobson, P. Alexander}\vspace{0.4cm}\\ %
Astrophysics Group, Cavendish Laboratory, JJ Thomson Avenue,  Cambridge, CB3 0HE, UK}

\begin{document}

\maketitle

\label{firstpage}

\begin{abstract}
We introduce a method for performing a robust Bayesian analysis of non-Gaussianity present in pulsar timing data, simultaneously with the pulsar timing model, and additional stochastic parameters such as those describing red spin noise and dispersion measure variations.  The parameters used to define the presence of non-Gaussianity are zero for Gaussian processes, giving a simple method of defining the strength of non-Gaussian behaviour.  We use simulations to show that assuming Gaussian statistics when the noise in the data is drawn from a non-Gaussian distribution can significantly increase the uncertainties associated with the pulsar timing model parameters.  We then apply the method to the publicly available 15 year Parkes Pulsar Timing Array data release 1 dataset for the binary pulsar J0437$-$4715.  In this analysis we present a significant detection of non-Gaussianity in the uncorrelated non-thermal noise, but we find that it does not yet impact the timing model or stochastic parameter estimates significantly compared to analysis performed assuming Gaussian statistics.  The methods presented are, however, shown to be of immediate practical use for current European Pulsar Timing Array (EPTA) and International Pulsar Timing Array (IPTA) datasets. 
\newline
\newline
\end{abstract}

\begin{keywords}
methods: data analysis, pulsars: general, pulsars:individual
\end{keywords}

\section{Introduction}

Millisecond pulsars (MSPs) have for some time been known to exhibit exceptional rotational stability, with decade long observations providing timing measurements with accuracies similar to atomic clocks (e.g. \cite{1994ApJ...428..713K,1997A&A...326..924M}).  Such stability lends itself well to the pursuit of a wide range of scientific goals, e.g. observations of the pulsar PSR B1913+16 showed a loss of energy at a rate consistent with that predicted for gravitational waves \citep{1989ApJ...345..434T}, whilst the double pulsar system PSR J0737-3039A/B has provided precise measurements of several `post Keplerian' parameters allowing for additional stringent tests of general relativity \citep{2006Sci...314...97K}.  

For a detailed review of pulsar timing refer to e.g. \cite{2004hpa..book.....L}.  In brief, the arrival times of pulses (TOAs) for a particular pulsar will be recorded by an observatory in a series of discrete observations over a period of time.  These arrival times must all be transformed into a common frame of reference, the solar system barycenter, in order to correct for the motion of the Earth.

A model for the pulsar can then be fitted to the TOAs; this characterises the properties of the pulsar's orbital motion, as well as its timing properties such as its orbital frequency and spin down.  This is most commonly carried out using the {\sc TEMPO2} pulsar-timing packages  \citep{2006MNRAS.369..655H, 2006MNRAS.372.1549E, 2009MNRAS.394.1945H}, or more recently, the Bayesian pulsar timing package TempoNest \citep{2014MNRAS.437.3004L}.

When performing this fitting process, both TEMPO2 and TempoNest assume purely Gaussian statistics in the properties of the uncorrelated noise.  In realistic datasets, however, this assumption is not necessarily correct.  If the underlying probability density function (PDF) for the noise is not Gaussian, for example, if there is an excess of outliers relative to a Gaussian distribution, modifiers to the TOA error bars that scale their size are used to find the best approximation to a Gaussian distribution.  This can be performed using a single modifier for a given receiving system determined across an entire dataset, or as in the `fixData' plugin for {\sc TEMPO2} \citep{2011MNRAS.418..561C}, where the modifier is determined separately for a series of short time lags.  While the latter of these two approaches can better account for a non-Gaussian distribution in the uncorrelated noise, it does so at the expense of a potentially large number of additional free parameters, and ultimately does not address the core issue, that the underlying distribution is not Gaussian. 

Both approaches then have the direct consequence of decreasing the precision with which one can estimate the timing parameters, and any other signals of interest, such as intrinsic red spin noise due to rotational irregularities in the neutron star \citep{2010ApJ...725.1607S} or correlated noise due to a stochastic gravitational wave background (GWB) generated by, for example, coalescing black holes (e.g. \citealt{2003ApJ...583..616J,2001astro.ph..8028P}).  Indeed, currently all published limits on the signals induced by a GWB have been obtained under the assumption that the statistics of the TOA errors are Gaussian (see e.g. \citealt{2013ApJ...762...94D,2011MNRAS.414.3117V}).

In this paper we introduce a method of performing a robust Bayesian analysis of non-Gaussianity present in pulsar timing data, simultaneously with the pulsar timing model, and additional stochastic parameters such as those describing the red noise, and dispersion measure variations.  The parameters used to define the presence of non-Gaussianity are zero for Gaussian processes, giving a simple method of defining the strength of non-Gaussian behaviour.
In Section \ref{Section:Bayes} we will describe the basic principles of our Bayesian approach to data analysis, giving a brief overview of how it may be used to perform model selection, and introduce {\sc MultiNest}.  In Sections \ref{Section:NonGaussLike} and \ref{Section:toy} we introduce the non-Gaussian likelihood we will use in our pulsar timing analysis, and apply it to a simple toy problem. In Section \ref{Section:PulsarNonGaussian} we then extend this likelihood to the subject of pulsar timing, and apply it to both simulated and real data in Sections \ref{Section:Pulsarsims} and \ref{Section:RealData}  respectively, before finally offering some concluding remarks in Section \ref{Section:Conclusion}.

This research is the result of the common effort to directly detect gravitational waves using pulsar timing, known as the European Pulsar Timing Array (EPTA) \citep{2008AIPC..983..633J} \footnote{www.epta.eu.org/}.

\section{Bayesian Inference}
\label{Section:Bayes}

Given a set of data $D$, Bayesian inference provides a consistent approach to the estimation of a set of parameters $\Theta$ in a model or hypothesis $H$.  In particular,   Bayes' theorem states that:

\begin{equation}
\mathrm{Pr}(\Theta \mid D, H) = \frac{\mathrm{Pr}(D\mid \Theta, H)\mathrm{Pr}(\Theta \mid H)}{\mathrm{Pr}(D \mid H)},
\end{equation}
where $\mathrm{Pr}(\Theta \mid D, H) \equiv \mathrm{Pr}(\Theta)$ is the posterior probability distribution of the parameters,  $\mathrm{Pr}(D\mid \Theta, H) \equiv L(\Theta)$ is the likelihood, $\mathrm{Pr}(\Theta \mid H) \equiv \pi(\Theta)$ is the prior probability distribution, and $\mathrm{Pr}(D \mid H) \equiv Z$ is the Bayesian Evidence.

Since the evidence is independent of the parameters $\Theta$ it is typically ignored when one is only interested in performing parameter estimation.  In this case inferences are obtained by taking samples from the (unnormalised) posterior using, for example, standard Markov chain Monte Carlo (MCMC) sampling methods.  

For model selection, however, the evidence is key, and is defined simply as the factor required to normalise the posterior over $\Theta$:

\begin{equation}
Z = \int L(\Theta)\pi(\Theta) \mathrm{d}^n\Theta,
\label{eq:Evidence}
\end{equation}
where $n$ is the dimensionality of the parameter space.  

As the evidence is just the average of the likelihood over the prior, it will be larger for a simpler model with a compact parameter space if more of that parameter space is likely. More complex models where large areas of parameter space have low likelihood values will have a smaller evidence even if the likelihood function is very highly peaked, unless they are significantly better at explaining the data.  Thus, the evidence automatically implements Occam's razor.

The question of model selection between two models $H_0$ and $H_1$ can be answered via the model selection ratio $R$, commonly referred to as the `Bayes Factor':

\begin{equation}
R= \frac{\mathrm{Pr}(H_1\mid D)}{\mathrm{Pr}(H_0\mid D)} = \frac{\mathrm{Pr}(D \mid H_1)\mathrm{Pr}(H_1)}{\mathrm{Pr}(D\mid H_0)\mathrm{Pr}(H_0)} = \frac{Z_1}{Z_0}\frac{\mathrm{Pr}(H_1)}{\mathrm{Pr}(H_0)},
\label{Eq:Rval}
\end{equation}
where $\mathrm{Pr}(H_1)/\mathrm{Pr}(H_0)$ is the a priori probability ratio for the two models, which in this work we will set to unity but occasionally requires further consideration.

The Bayes factor then allows us to obtain the probability of one model compared the other simply as:

\begin{equation}
P = \frac{R}{1+R}.
\end{equation}

In practice when performing Bayesian analysis we do not work with the likelihood, but the log likelihood.  In this case the quantity of interest is the log Bayes Factor, which is simply the difference in the log evidence for the two models.  For example, a difference in the log evidence of 3 for two competing models gives a Bayes factor of $\sim 20$, which in turn gives a probability of $\sim 95\%$.  We use the difference in the log evidence in Sections \ref{Section:Pulsarsims} and \ref{Section:RealData} to perform model selection between our Gaussian and non-Gaussian models.

\subsection{Nested sampling and evidence evaluation}

While many techniques exist for calculating the evidence, such as thermodynamic integration \citep{Thermo}, it remains a challenging task both numerically and computationally, with evidence evaluation at least an order-of-magnitude more costly than parameter estimation.

Nested sampling \citep{2004AIPC..735..395S} is an approach designed to make the calculation of the evidence more efficient, and also produces posterior inferences as a by-product.  The {\sc MultiNest} algorithm (\citealt{2009MNRAS.398.1601F, 2008MNRAS.384..449F}) builds upon this nested sampling framework, and provides an efficient means of sampling from posteriors that may contain multiple modes and/or large (curving) degeneracies, and also calculates the evidence.  Since its release {\sc MultiNest} has been used successfully in a wide range of astrophysical problems, including inferring the properties of a potential stochastic gravitational wave background in pulsar timing array data \citep{2013PhRvD..87j4021L}, and is also used in the Bayesian pulsar timing package {\sc TempoNest}. This technique has greatly reduced the computational cost of Bayesian parameter estimation and model selection, and is employed in this paper.

\section{A non-Gaussian likelihood}
\label{Section:NonGaussLike}

In this section we will outline the method adopted for including non-Gaussian behaviour in our analysis.  We use the approach developed in \cite{2001PhRvD..64f3512R}, which is based on the energy eigenmode wavefunctions of a simple harmonic oscillator. We will describe this in brief below in order to aid future discussion.

We begin by considering our data, the vector $\mathbf{d}$ of length $N_d$, as the sum of some signal $\mathbf{s}$ and noise $\mathbf{n}$ such that:
\begin{equation}
\mathbf{d} = \mathbf{s} + \mathbf{n}.
\end{equation}
We can then construct the likelihood that the residuals after subtracting our model signal from the data follows an uncorrelated Gaussian distribution of width $\sigma$ as:

\begin{equation}
\label{Eq:Gausslike}
\mathrm{Pr}(\mathbf{d} | \sigma) = \frac{1}{\sqrt{(2\pi)^{N_d} \mathrm{det}(\mathbf{N})}}\exp\left[-\frac{1}{2}(\mathbf{d} -\mathbf{s})^T\mathbf{N}^{-1}(\mathbf{d} -\mathbf{s})\right],
\end{equation}
with $\mathbf{N}$ the diagonal noise covariance matrix for the residuals, such that $N_{ii} = \sigma^2$, and $\mathrm{det}(\mathbf{N})$ the determinant of $\mathbf{N}$.

We now extend this to the general case in order to allow for non-Gaussian distributions by modelling our PDF as the sum of a set of Gaussians, modified by Hermite polynomials $H_n(x)$ (see e.g. \cite{1989AnnStat.17} for previous uses of Hermite polynomials in describing departures from Gaussianity), defined as:
\begin{equation}
H_n(x) = (-1)^n \exp(x^2) \frac{\mathrm{d}^n}{\mathrm{d}x^n} \exp(-x^2).
\end{equation}

Therefore, for a general random variable $x$ the PDF for fluctuations in $x$ can be written:
\begin{equation}
\label{Eq:PrX}
\mathrm{Pr}(x | \sigma, \bmath{\alpha}) = \exp\left[-\frac{x^2}{2\sigma^2}\right]\left|\sum_{n=0}^{\infty}\alpha_nC_nH_n\left(\frac{x}{\sqrt{2}\sigma}\right)\right|^2
\end{equation}
with $\alpha_n$ free parameters that describe the relative contributions of each term to the sum, and
\begin{equation}
C_{n} = \frac{1}{(2^n n! \sqrt{2\pi}\sigma)^{1/2}},
\end{equation}
is a normalization factor.  Equation \ref{Eq:PrX} forms a complete set of PDFs, normalised such that:

\begin{equation}
\int_{-\infty}^\infty \; \mathrm{d}x \; \exp\left[-\frac{x^2}{\sigma^2}\right]C_nH_n\left(\frac{x}{\sqrt{2}\sigma}\right)C_mH_m\left(\frac{x}{\sqrt{2}\sigma}\right) = \delta_{mn},
\end{equation}
with $\delta_{mn}$ the Kronecker delta, where the ground state, $H_0$, reproduces a standard Gaussian PDF, and any non-Gaussianity in the distribution of $x$ will be reflected in non-zero values for the coefficients $\alpha_n$ associated with higher order states.

The only constraint we must place on the values of the amplitudes $\bmath{\alpha}$ is:
\begin{equation}
\sum_{n=0}^{n_{\mathrm{max}}} \left|\alpha_n\right|^2 = 1
\end{equation}
with $n_{\mathrm{max}}$ the maximum number of coefficients to be included in the model for the PDF.  This is performed most simply by setting:
 \begin{equation}
\alpha_0 = \sqrt{1 - \sum_{n=1}^{n_{\mathrm{max}}} \left|\alpha_n\right|^2}.
\end{equation}

We can therefore rewrite Eq. \ref{Eq:Gausslike} in this more general form as:
\begin{eqnarray}
\label{Eq:NonGausslike}
\mathrm{Pr}(\mathbf{d} | \sigma, \bmath{\alpha}) &=& \exp\left[-\frac{1}{2}(\mathbf{d} -\mathbf{s})^T\mathbf{N}^{-1}(\mathbf{d}  -\mathbf{s})\right]\nonumber \\
&\times&\prod_{i=1}^{N_d}\left|\sum_{n=0}^{n_{\mathrm{max}}}\alpha_nC_nH_n\left(\frac{d_i - s_i}{\sqrt{2}\sigma}\right)\right|^2.
\end{eqnarray}
The advantage of this method is that one may use a finite set of non-zero $\alpha_n$ to model the non-Gaussianity, without mathematical inconsistency.  Any truncation of the series still yields a proper distribution, in contrast to the more commonly used Edgeworth expansion (e.g. \citealt{2000ApJ...534...25C}).

\section{Application to a toy problem}
\label{Section:toy}

Before applying the formalism described in Section \ref{Section:NonGaussLike} to the practice of pulsar timing, we first demonstrate its use in a toy problem.  Here our data vector $\mathbf{d}$ contains 10000 points drawn from a non-Gaussian distribution obtained using Eq. \ref{Eq:PrX}, with parameters listed in Table \ref{Table:toyparams}.

\begin{table}
\centering
\caption{Parameters used to generate non-Gaussian noise in a simple toy problem.} 
\centering 
\begin{tabular}{c c c} 
\hline\hline 
Parameter &  Value & Parameter Estimate\\[0.5ex] 
\hline 
$\sigma$  &1  & 0.997  $\pm$  0.005\\
$\alpha_1$ &0.1 & 0.105  $\pm$  0.006\\
$\alpha_2$ & 0.2 & 0.198  $\pm$  0.006\\
$\alpha_3$ &0.4 & 0.402  $\pm$  0.005\\
\hline
\end{tabular}
\label{Table:toyparams} 
\end{table}

\begin{figure*}
\begin{center}$
\begin{array}{cc}
\hspace{-1.5cm}
\includegraphics[width=100mm]{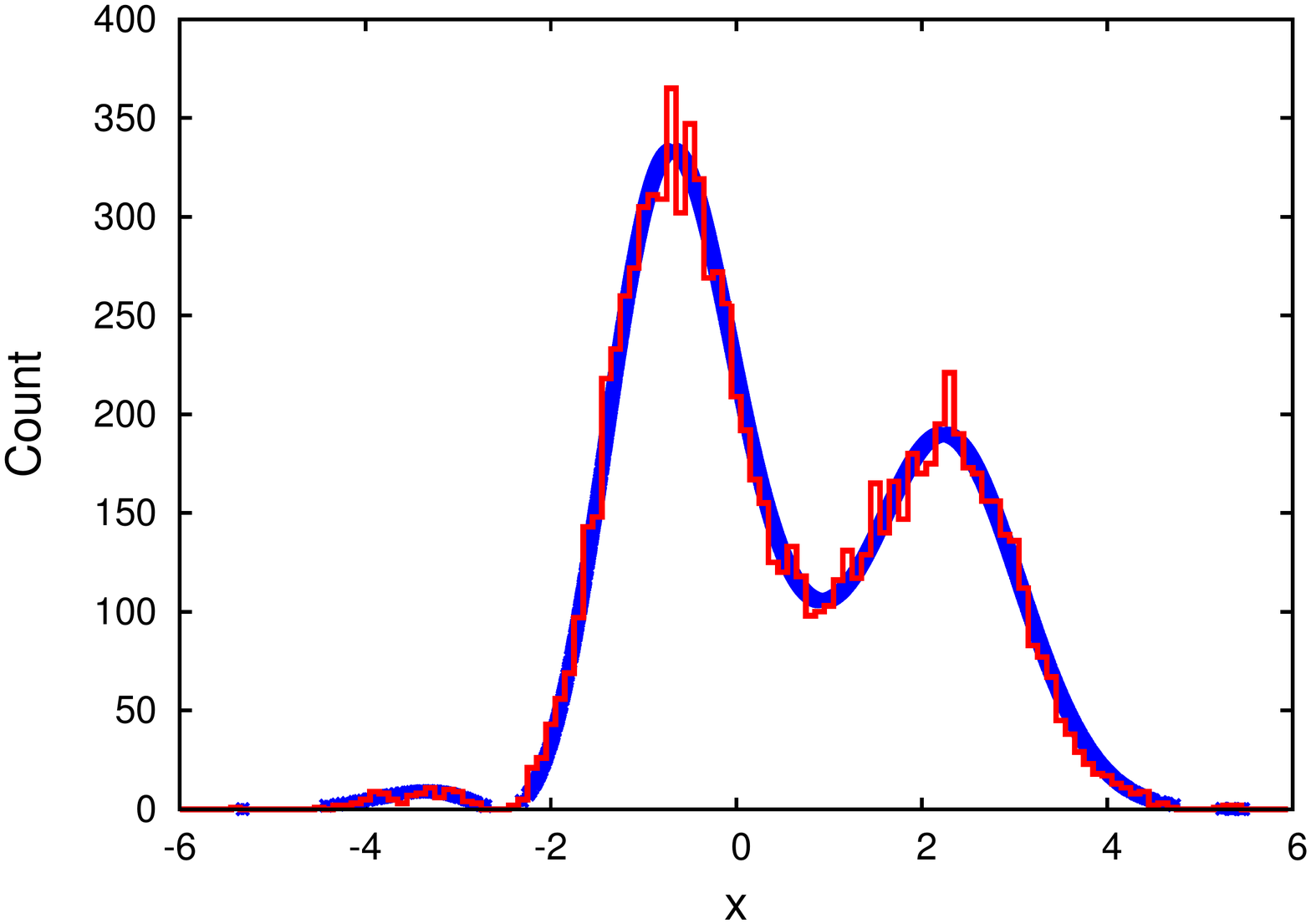} &
\includegraphics[width=100mm]{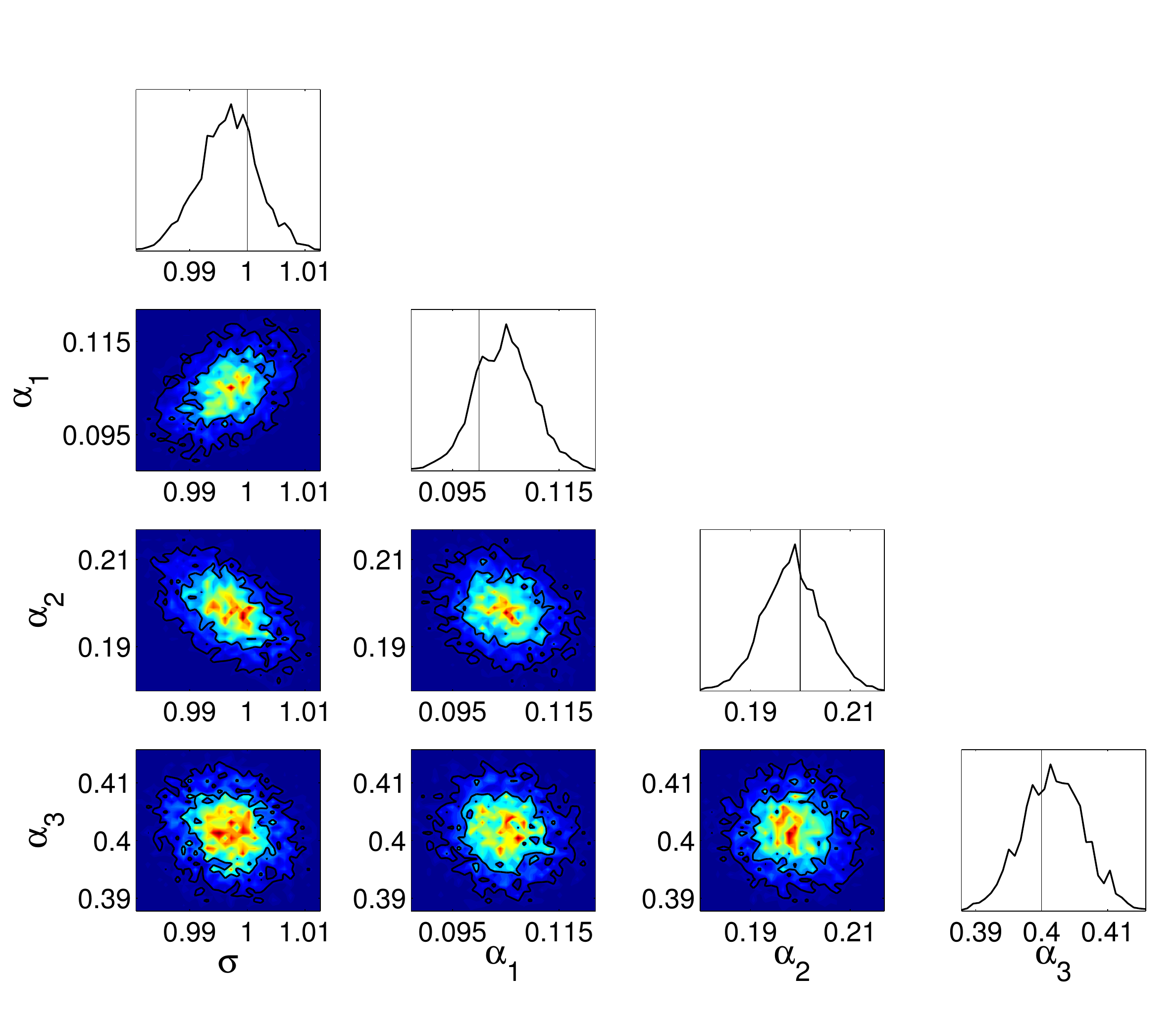} \\
\end{array}$
\end{center}
\caption{(left) The distribution of points in the vector $\mathbf{d}$ (red histogram), and the mean model solution obtained for the distribution in our analysis (smooth blue line). (right) The one and two-dimensional marginalised posterior distributions for the parameters sampled in the toy problem; vertical lines represent the values used in the simulations.  We find that the analysis successfully reproduces the target distribution.}
\label{Fig:toyprobdist}
\end{figure*}

We then sample over the 4 dimensional problem using MULTINEST using Eq. \ref{Eq:NonGausslike} as our likelihood, with $\mathbf{s} = 0$. We note here that in practice when dealing with real data the correct number of coefficients to use in the analysis will not be a known quantity.  In this case one could take the approach of performing the analysis multiple times including an increasing number of non-Gaussian terms in order to find the value that maximises the evidence, an approach we take in Section \ref{Section:RealData}.  

Fig. \ref{Fig:toyprobdist} (left) shows the distribution of points in the vector $\mathbf{d}$ (red histogram), along with the mean model solution obtained for the distribution in our analysis (smooth blue line) which we also give in Table \ref{Table:toyparams} along with the value for one standard deviation in each parameter. Fig. \ref{Fig:toyprobdist} (right) shows the one and two-dimensional marginalised posterior distributions for the parameters sampled in the toy problem, with vertical lines indicating the values used in the simulation.  We find that the analysis successfully reproduces the PDF for the values in $\mathbf{d}$.

\section{Non-Gaussian pulsar timing likelihood}
\label{Section:PulsarNonGaussian}
For any pulsar we can, as in Section \ref{Section:NonGaussLike}, write the TOAs for the pulses as a sum of both a deterministic and a stochastic component:

\begin{equation}
\mathbf{d} = \mathbf{s} + \mathbf{n},
\end{equation}
where $\mathbf{d}$ represents the $N_d$ TOAs for a single pulsar, with $\mathbf{s}$ and $\mathbf{n}$ the deterministic and stochastic contributions to the total respectively.

Writing the deterministic signal due to the timing model parameters $\bmath{\epsilon}$ as $\mathbf{\tau}(\bmath{\epsilon})$, and the uncertainty associated with a particular TOA $i$ as $\sigma_i$ we can rewrite our non-Gaussian likelihood from Eq. \ref{Eq:NonGausslike} simply as:

\begin{eqnarray}
\label{Eq:Pulsarlike}
\mathrm{Pr}(\mathbf{d} | \bmath{\epsilon}, \beta, \bmath{\alpha}) &=& \exp\left[-\frac{1}{2}(\mathbf{d} - \mathbf{\tau}(\bmath{\epsilon}))^T\mathbf{N}^{-1}(\mathbf{d}  - \mathbf{\tau}(\bmath{\epsilon}))\right]\nonumber \\
&\times&\prod_{i=1}^{N_d}\left|\sum_{n=0}^{n_{\mathrm{max}}}\alpha_nC_{i,n}H_n\left(\frac{d_i - \mathbf{\tau}(\bmath{\epsilon})_i}{\sqrt{2}\beta\sigma_i}\right)\right|^2,
\end{eqnarray}
where we have redfined the constant $C$ as:
\begin{equation}
C_{i,n} = \frac{1}{(2^n n! \sqrt{2\pi}\beta\sigma_i)^{1/2}},
\end{equation}
to allow for the more general case where each TOA $d_i$ has a different uncertainty $\sigma_i$, and we include the scaling factor $\beta$, commonly referred to as `EFAC', as a free parameter to modify these values to account for possible mis-calibration of this radiometer noise.

\subsection{Additional white noise}
\label{Section:White1}

In typical pulsar timing analysis the white noise is considered the sum of multiple terms, the radiometer noise associated with a given TOA discussed in the previous section, and additional terms that add in quadrature and represent sources of time independent noise.  These can include, for example, contributions from the high frequency tail of the pulsar's red spin noise power spectrum, or, jitter noise that results from the time averaging of a finite number of single pulses to form each TOA (see e.g. \cite{2014MNRAS.443.1463S}).  For simplicity we will refer to these quadrature terms simply as `EQUAD'.

For any given TOA the total white noise can therefore be considered to be the sum of these separate processes.  In incorporating the different white noise processes into our likelihood we first note that, for a random variable $c$, defined as $c=a+b$ with $a$ and $b$ themselves random variables, we can write the PDF for $c$ as:

\begin{equation}
\mathrm{Pr}(c) =  \mathrm{Pr}(a) \ast \mathrm{Pr}(b),
\end{equation}
where $\mathrm{Pr}(a)$  and $\mathrm{Pr}(b)$ are the PDFs for variables $a$ and $b$ respectively, and $\ast$ represents a convolution.  Using this result we can then take one of two approaches.  Firstly, we can use Eq. \ref{Eq:Pulsarlike} and simply fit for a single non-Gaussian probability density function for the combined radiometer and quadrature terms by taking the total uncertainty to be:

\begin{equation}
\label{Eq:BasicWhite}
\hat{\sigma_i}^2 =  \beta^2\sigma_i^2 + \gamma^2
\end{equation} 
where $\gamma$ represents the quadrature component of the total error bar in which case we simply replace $\beta\sigma_i$ with $\hat{\sigma}_i$ in Eq. \ref{Eq:Pulsarlike}.

Alternatively we can allow the PDFs for the radiometer and quadrature terms to be different, for example, by assuming the radiometer term is Gaussian, and that any non-Gaussianity comes from all, or a subset of, the terms added in quadrature.  In this case we will have:

\begin{eqnarray}
\label{Eq:Pulsarwhitelike}
\mathrm{Pr}(\mathbf{d} | \bmath{\epsilon}, \beta, \gamma, \bmath{\alpha}) &=& \mathrm{Pr}(\mathbf{d} | \bmath{\epsilon}, \beta) \ast \mathrm{Pr}(\mathbf{d} | \bmath{\epsilon}, \gamma, \bmath{\alpha})
\end{eqnarray}
where
\begin{equation}
\mathrm{Pr}(\mathbf{d} | \bmath{\epsilon}, \beta) = \frac{1}{\sqrt{(2\pi)^{N_d} |\mathbf{N_{\beta}}|}}\exp\left[-\frac{1}{2}(\mathbf{d} - \mathbf{\tau}(\bmath{\epsilon}))^T\mathbf{N_{\beta}}^{-1}(\mathbf{d}  - \mathbf{\tau}(\bmath{\epsilon}))\right],
\end{equation}
with $\mathbf{N_{\beta}}_{i,i} = \beta^2\sigma_i^2$, and
\begin{eqnarray}
\mathrm{Pr}(\mathbf{d} | \bmath{\epsilon}, \gamma, \bmath{\alpha}) &=& \exp\left[-\frac{1}{2}(\mathbf{d} - \mathbf{\tau}(\bmath{\epsilon}))^T\mathbf{J}^{-1}(\mathbf{d}  - \mathbf{\tau}(\bmath{\epsilon}))\right]\nonumber \\ 
&\times&\left.\prod_{i=1}^{N_d}\left|\sum_{n=0}^{n_{\mathrm{max}}}\alpha_nC_{i,n,\gamma}H_n\left(\frac{d_i - \mathbf{\tau}(\bmath{\epsilon})_i}{\sqrt{2}\gamma}\right)\right|^2\right]. \nonumber
\end{eqnarray}
with $\mathbf{J}_{i,i} = \gamma^2$.

\subsection{Sampling from the un-marginalised EQUAD}
\label{Section:EQUADDeMargin}

A second approach to including both EFAC and EQUAD parameters in the likelihood is to parameterise the EQUAD term using $N_d$ free parameters $j_i$ that each represent a shift in a given TOA $i$, and then having a prior on those parameters that describes their underlying distribution.

If we first assume a Gaussian distribution on the parameters $\mathbf{j}$ we can write our likelihood as:

\begin{eqnarray}
\label{Eq:Pulsarjitterlike}
\mathrm{Pr}(\mathbf{d} | \bmath{\epsilon}, \beta, \mathbf{j},  \mathbf{J}) &=& \mathrm{Pr}(\mathbf{d} | \bmath{\epsilon}, \beta, \mathbf{j}) \times \mathrm{Pr}(\mathbf{j} | \mathbf{J})
\end{eqnarray}
with,

\begin{eqnarray}
\mathrm{Pr}(\mathbf{d} | \bmath{\epsilon}, \beta, \mathbf{j}) &=& \frac{1}{\sqrt{(2\pi)^{N_d} \mathrm{det}~\mathbf{N_{\beta}}}} \\
&\times& \exp\left[-\frac{1}{2}(\mathbf{d} - \mathbf{\tau}(\bmath{\epsilon}) -  \mathbf{j})^T\mathbf{N_{\beta}}^{-1}(\mathbf{d}  - \mathbf{\tau}(\bmath{\epsilon})- \mathbf{j})\right] \nonumber ,
\end{eqnarray}
and

\begin{equation}
\mathrm{Pr}(\mathbf{j} | \mathbf{J}) = \frac{1}{\sqrt{(2\pi)^{N_d}\mathrm{det}~\mathbf{J}}}\exp\left[-\frac{1}{2}\mathbf{j}^T\mathbf{J^{-1}}\mathbf{j}\right],
\end{equation}
with $\mathbf{J}$ a diagonal matrix, where each element is equal to $\gamma^2$ as defined previously.

This formalism is infact completely equivalent to the convolution described in Section \ref{Section:White1}, which can be seen quite transparently by simply considering the definition of the convolution for two functions P and Q: 

\begin{equation}
\int_{-\infty}^{\infty} \mathrm{P}(\mathbf{d} - \mathbf{j})\mathrm{Q}(\mathbf{j}) \mathrm{d}\mathbf{j} =  \mathrm{P}(\mathbf{d}) \ast \mathrm{Q}(\mathbf{d}).
\end{equation}
By equating $\mathrm{Q}(\mathbf{j})$ to our prior on the parameters $\mathbf{j}$, and $\mathrm{P}(\mathbf{d} - \mathbf{j})$ to our likelihood, the integration over all $\mathbf{j}$'s is seen to be equivalent to a convolution.

For completeness, we will now show that by re-marginalising over the parameters $\mathbf{j}$ we recover the definition of the white noise given in Eqn \ref{Eq:BasicWhite} in the Gaussian case.  In order to perform this marginalisation, we first write the log of the likelihood in Eq \ref{Eq:Pulsarjitterlike}, which denoting  $(\mathbf{N_{\beta}}^{-1} + \mathbf{J}^{-1})$ as $\mathbf{\Sigma}$ and $\mathbf{N_{\beta}}^{-1}\mathbf{d}$ as $\mathbf{\bar{d}}$ is given by:
\begin{equation}
\label{Eq:LogL}
\log \mathrm{L} = -\frac{1}{2} \mathbf{d}^T\mathbf{\bar{d}} - \frac{1}{2}\mathbf{j}^T\mathbf{J^{-1}}\mathbf{j} + \mathbf{\bar{d}}^T\mathbf{j}.
\end{equation}

Taking the derivative of $\log \mathrm{L}$ with respect to $\mathbf{j}$ gives us:
\begin{equation}
\label{Eq:Gradj}
\frac{\partial \log \mathrm{L}}{\partial \mathbf{j}} =  -\mathbf{\Sigma}\mathbf{j} + \mathbf{\bar{d}},
\end{equation}
which can be solved to give us the maximum likelihood vector of coefficients $\hat{\mathbf{j}}$:
\begin{equation}
\label{Eq:jmax}
\hat{\mathbf{j}} = \mathbf{\Sigma}^{-1}\mathbf{\bar{d}}.
\end{equation}
Re-expressing Eq. \ref{Eq:LogL} in terms of $\hat{\mathbf{j}}$:

\begin{eqnarray}
\log \mathrm{L} &=& -\frac{1}{2} \mathbf{d}^T\mathbf{\bar{d}} + \frac{1}{2}\hat{\mathbf{j}}^T\mathbf{\Sigma}\hat{\mathbf{j}} 
 -  \frac{1}{2}(\mathbf{j} - \hat{\mathbf{j}})^T\mathbf{\Sigma}(\mathbf{j} - \hat{\mathbf{j}}),
\end{eqnarray}
the 3rd term in this expression can then be integrated with respect to the $N_d$ elements in $\mathbf{j}$ to give:
\begin{eqnarray}
I &=& \int_{-\infty}^{+\infty}\mathrm{d}\mathbf{j}\exp\left[-\frac{1}{2}(\mathbf{j} - \hat{\mathbf{j}})^T\mathbf{\Sigma}(\mathbf{j} - \hat{\mathbf{j}})\right] \nonumber \\
&=& (2\pi)^{N_d}~\mathrm{det} ~ \mathbf{\Sigma}^{-\frac{1}{2}}.
\end{eqnarray}
Our marginalised likelihood for EQUAD is then given as:

\begin{eqnarray}
\label{Eq:Margin}
\log \mathrm{L} &=& -\frac{1}{2}|\mathbf{\Sigma}| - \frac{1}{2}|\mathbf{J}| - \frac{1}{2}|\mathbf{N_{\beta}}| 
- \frac{1}{2}\left(\mathbf{d}^T\mathbf{\bar{d}} - \mathbf{\bar{d}}^T\mathbf{\Sigma}^{-1}\mathbf{\bar{d}}\right). 
\end{eqnarray}
Given the definitions of $\mathbf{\bar{d}}$, $J$, and $\mathbf{\Sigma}$ we can write for a single TOA $i$:

\begin{eqnarray}
\label{Eq:Margin}
\left(\mathbf{d}^T\mathbf{\bar{d}} - \mathbf{\bar{d}}^T\mathbf{\Sigma}^{-1}\mathbf{\bar{d}}\right)_i &=& d_i\left( \frac{1}{\beta^2\sigma_i^2} - \frac{1}{\beta^2\sigma_i^2}\left( \frac{1}{\beta^2\sigma_i^2}+\frac{1}{\gamma^2}\right)^{-1}\frac{1}{\beta^2\sigma_i^2}\right)d_i, \nonumber \\
&=& \frac{d_i^2}{\beta^2\sigma_i^2+\gamma^2}
\end{eqnarray}
and similarly for the determinants.

We can use this un-marginalised distribution to trivially include non-Gaussianity in the EQUAD term only by altering the prior term:

\begin{equation}
\mathrm{Pr}(\mathbf{j} |  \mathbf{J}, \mathbf{\alpha}) = \exp\left[-\frac{1}{2}\mathbf{j}^T\mathbf{J^{-1}}\mathbf{j}\right]\prod_{i=1}^{N_d}\left|\sum_{n=0}^{n_{\mathrm{max}}}\alpha_nC_{i,n,\gamma}H_n\left(\frac{j_i}{\sqrt{2}\gamma}\right)\right|^2.
\end{equation}
It is still possible to then analytically marginalise over the $j$ parameters, which will result in the convolution of a Gaussian and non-Gaussian probability density function.

\subsection{Analytic marginalisation over the EFAC parameters}
\label{Section:AnalyticEFAC}

An interesting possibility that arises when parameterising the EQUAD parameter as in Section \ref{Section:EQUADDeMargin} is that it allows us to marginalise analytically over the EFAC white noise parameters.  If we assume a uniform prior on the amplitude of the EFAC parameter $\beta$ and define:

\begin{equation}
r^2 = \sum_{i=1}^{N_d} (d_i - \mathbf{\tau}(\bmath{\epsilon})_i -  j_i)^2/\sigma_i^2
\end{equation}
we can write our probability distribution as:

\begin{equation}
\label{Eq:NumInt}
\mathrm{Pr}(\mathbf{d} | \bmath{\epsilon},  \mathbf{j}, \beta) \propto \frac{1}{\beta^{N_d}}\exp\left[-\frac{r^2}{2\beta^2}\right]
\end{equation}
which can then be integrated over $\beta$ between some finite lower limit $a$ and a finite upper limit $b$.  While this does have an analytic solution given by:

\begin{eqnarray}
\label{Eq:MarginEFAC}
&&\int_{a}^{b} \mathrm{d}\beta \frac{1}{\beta^{N_d}}\exp\left[-\frac{r^2}{2\beta^2}\right] \propto \\  &&\left(\frac{1}{r^2}\right)^{\frac{N_d-1}{2}}\left(\Gamma\left[\frac{N_d-1}{2}, \frac{r^2}{2b^2}\right] - \Gamma\left[\frac{N_d-1}{2},  \frac{r^2}{2a^2}\right]\right)
\end{eqnarray}
with $\Gamma$ the upper incomplete Gamma function, if the lower bound for the integral $a$ is non zero, the precision required to carry out the difference can be substantial for large $N_d$.  In this case it may be preferable simply to integrate using a standard numerical integration package, where the integral is rephrased to make it computable at double precision as:

\begin{equation}
\mathrm{Pr}(\mathbf{d} | \bmath{\epsilon},  \mathbf{j}, \beta) \propto \frac{\exp(M)}{\beta^{N_d}}\exp\left[-\frac{r^2}{2\beta^2}\right],
\end{equation}
with $M$ the maximum value taken by the log of Eq. \ref{Eq:NumInt}.

If one takes the case where the lower bound on the integral is 0, the integral in Eq \ref{Eq:MarginEFAC} simplifies to:

\begin{eqnarray}
\label{Eq:MarginEFAC2}
\int_{0}^{b} \mathrm{d}\beta \frac{1}{\beta^{N_d}}\exp\left[-\frac{r^2}{2\beta^2}\right] \propto \left(\frac{1}{r^2}\right)^{\frac{N_d-1}{2}}\Gamma\left[\frac{N_d-1}{2}, \frac{r^2}{2b^2}\right],
\end{eqnarray}
however, if the upper limit is taken to be arbitrarily large, but still finite, the log of the Gamma function term is effectively constant, and the result is a trivial, scale invariant log likelihood:

\begin{eqnarray}
\log L = -\frac{N_d-1}{2}\log{r^2}.
\end{eqnarray}

We note that in principle if a Gaussian prior is assumed on the EQUAD parameters $\mathbf{j}$ one could also marginalise analytically over the hyper parameter $\gamma$.  In this case we would redefine:

\begin{equation}
r^2 = \sum_{i=1}^{N_d} (j_i)^2
\end{equation}
giving us a probability distribution:

\begin{equation}
\mathrm{Pr}(\mathbf{j} | \mathbf{J}) \propto \frac{1}{\gamma^{N_d}}\exp\left[-\frac{r^2}{2\gamma^2}\right]
\end{equation}
from where we can proceed as before.  In practice however we find this approach less efficient to sample from in comparison to the convolved likelihood described in Section \ref{Section:White1}, and so we take the latter approach in Sections \ref{Section:Pulsarsims} and \ref{Section:RealData}.

\subsection{Additional red spin noise}
\label{Section:Red}

In order to include additional red noise processes we begin by taking the same approach as that given in \cite{2014MNRAS.437.3004L}, which we will describe in brief below to aid further discussion.  Writing the red noise component of the stochastic signal, which we will denote $\mathbf{d}_{\mathrm{red}}$, in terms of its Fourier coefficients $\mathbf{a_{\mathrm{red}}}$ so that $\mathbf{d}_{\mathrm{red}} = \mathbf{F_{\mathrm{red}}}\mathbf{a_{\mathrm{red}}}$ where $\mathbf{F_{\mathrm{red}}}$ denotes the Fourier transform such that for frequency $\nu$ and time $t$ we will have both:

\begin{equation}
\label{Eq:FMatrix}
F_{\mathrm{red}}(\nu,t) = \frac{1}{T}\sin\left(2\pi\nu t\right),
\end{equation}
and an equivalent cosine term.  Here $T$ represents the total observing span for the pulsar, and $\nu$ the frequency of the signal to be sampled.  Defining the number of coefficients to be sampled by $n_{\mathrm{red}}$, we can then include the set of frequencies with values $n/T$, where $n$ extends from 1 to $n_{\mathrm{red}}$.
For typical PTA data \cite{2012MNRAS.423.2642L} show that a low frequency cut off of $1/T$ is sufficient to accurately describe the expected long term variations present in the data. If necessary though it is also possible to specify arbitrary sets of frequencies such that terms with $\nu \ll 1/T$ can be included in the model, or to allow noise terms where the frequency itself is a free parameter. 

For a single pulsar the covariance matrix $\bmath{\varphi}_{\mathrm{red}}$ of the Fourier coefficients $\mathbf{a}_{\mathrm{red}}$ will be diagonal, with components

\begin{equation}
\label{Eq:BPrior}
\varphi_{\mathrm{red}, ij}= \left< a_{\mathrm{red},i}a_{\mathrm{red},j}^*\right> = \varphi_{\mathrm{red},i}\delta_{ij},
\end{equation}
where there is no sum over $i$, and the brackets $\left<..\right>$ denotes the expectation value such that the set of coefficients $\{\varphi_{i}\}$ represent the theoretical power spectrum of the red noise signal present in the timing data.

Whilst Eq. \ref{Eq:BPrior} states that the Fourier modes are orthogonal to one another, this does not mean that we assume they are orthogonal in the time domain where they are sampled, and it can be shown that this non-orthogonality is accounted for within the likelihood.  Instead, in Bayesian terms, Eq. \ref{Eq:BPrior} represents our prior knowledge of the power spectrum coefficients within the data.  We are therefore stating that, whilst we do not know the form the power spectrum will take, we know that the underlying Fourier modes are still orthogonal by definition, regardless of how they are sampled in the time domain.  It is here then that, should one wish to fit a specific model to the power spectrum coefficients at the point of sampling, such as a broken, or single power law, the set of coefficients $\{\varphi_{i}\}$ should be given by some function $f(\Theta)$, where we sample from the parameters $\Theta$ from which the power spectrum coefficients $\{\varphi_{i}\}$ can then be derived.

We can then use the signal realisation of the red noise process given by $\mathbf{Fa}$ to alter the model TOAs given by the timing model,  $\tau(\bmath{\epsilon})$ such that:

\begin{equation}
\bmath{\hat{\tau}}(\bmath{\epsilon}, \bmath{a}_{\mathrm{red}}) =  \bmath{\tau}(\bmath{\epsilon}) - \mathbf{F}_{\mathrm{red}}\mathbf{a}_{\mathrm{red}},
\end{equation}
enabling us to write the joint probability density Pr$(\bmath{\epsilon}, \beta, \bmath{\alpha}, \bmath{\varphi}, \mathbf{a}_{\mathrm{red}} \;|\; \mathbf{d})$, as:

\begin{eqnarray}
\label{Eq:Prob}
\mathrm{Pr}(\bmath{\epsilon}, \beta, \bmath{\alpha}, \bmath{\varphi}, \mathbf{a}_{\mathrm{red}} \;|\; \mathbf{d}) \; &\propto& \; \mathrm{Pr}(\mathbf{d} |  \bmath{\epsilon}, \beta, \bmath{\alpha}, \mathbf{a}_{\mathrm{red}}) \; \\\ \nonumber
&\times & \mathrm{Pr}(\mathbf{a}_{\mathrm{red}} | \bmath{\varphi}_{\mathrm{red}}) \; \mathrm{Pr}(\bmath{\varphi}_{\mathrm{red}}). \nonumber
\end{eqnarray}
For our choice of $\mathrm{Pr}(\bmath{\varphi}_{\mathrm{red}})$ we use an uninformative prior that is uniform in $\log_{10}$ space, and draw our samples from the parameter $\rho_{\mathrm{red},i} = \log_{10}(\varphi_{\mathrm{red},i})$ instead of $\varphi_{\mathrm{red},i}$.  Given this choice of prior the conditional distributions that make up Eq. \ref{Eq:Prob} can be written:

\begin{eqnarray}
\label{Eq:ProbTime}
\mathrm{Pr}(\mathbf{d} | \bmath{\epsilon}, \beta, \bmath{\alpha},\mathbf{a_{\mathrm{red}}} ) &=& \exp\left[-\frac{1}{2}(\mathbf{d} - \mathbf{\hat{\tau}}(\bmath{\epsilon}))^T\mathbf{N}^{-1}(\mathbf{d}  - \mathbf{\hat{\tau}}(\bmath{\epsilon}))\right]\nonumber \\
&\times&\prod_{i=1}^{N_d}\left|\sum_{n=0}^{n_{\mathrm{max}}}\alpha_nC_{i,n}H_n\left(\frac{d_i - \mathbf{\hat{\tau}}(\bmath{\epsilon})_i}{\sqrt{2}\beta\sigma_i}\right)\right|^2,
\end{eqnarray}
and:

\begin{equation}
\label{Eq:ProbFreq}
\mathrm{Pr}(\mathbf{a}_{\mathrm{red}}\; | \;\bmath{\rho}_{\mathrm{red}}) \; \propto \; \frac{1}{\sqrt{\mathrm{det}\bmath{\varphi}_{\mathrm{red}}}} \exp\left[-\frac{1}{2}\mathbf{a}_{\mathrm{red}}^{T}\bmath{\varphi}_{\mathrm{red}}^{-1}\mathbf{a}_{\mathrm{red}}\right].
\end{equation}

As in section \ref{Section:EQUADDeMargin} we note that we need not use a Gaussian prior on the Fourier coefficients.  If we assume that the red noise follows a power law, such that the matrix $\bmath{\varphi}$ is a function of an amplitude $A_{\mathrm{red}}$ and spectral index $\kappa_{\mathrm{red}}$:

\begin{equation}
\varphi_{ii} = \frac{A_{\mathrm{red}}^2}{12\pi^2T_s} \left(\frac{1}{1\mathrm{yr}}\right)^{-3}\left(1\mathrm{yr}~\nu_i\right)^{-\kappa_{\mathrm{red}}},
\end{equation}
with $T_s$ the total observing span of the dataset in seconds, and $\nu_i$ the frequency of the power spectrum coefficient $\varphi_{ii}$, then we can parameterise the non-Gaussianity in the coefficients $\mathbf{a}$ as before:

\begin{eqnarray}
\label{Eq:NGRed}
\mathrm{Pr}(\mathbf{a}_{\mathrm{red}}\; | \;\bmath{\rho}_{\mathrm{red}},  \mathbf{\alpha}) &=& \exp\left[-\frac{1}{2}\mathbf{a}_{\mathrm{red}}^{T}\bmath{\varphi}_{\mathrm{red}}^{-1}\mathbf{a}_{\mathrm{red}}\right]\\
&\times&\prod_{i=1}^{n_{\mathrm{red}}}\left|\sum_{n=0}^{n_{\mathrm{max}}}\alpha_nC_{i,n,\varphi_{\mathrm{red},i}}H_n\left(\frac{a_i}{\sqrt{2\varphi_{\mathrm{red}, i}}}\right)\right|^2. \nonumber
\end{eqnarray}
We note here that, if a Gaussian prior is assumed on the Fourier coefficients $\mathbf{a}$, and uniform priors assumed on the red noise amplitude hyper parameters, those amplitude parameters can be marginalised over analytically in exactly the same manner as described in Section \ref{Section:AnalyticEFAC}. 

\subsection{Including dispersion measure variations}
\label{Section:DM}

The plasma located in the interstellar medium (ISM), as well as in solar winds and the ionosphere can result in delays in the propagation of the pulse signal between the pulsar and the observatory, an effect that appears as a red noise signal in the timing residuals.  

The severity of the observed dispersion measure variations, however, is dependent upon the observing frequency, and as such we can use this additional information to isolate this component of the red noise from effects that do not have this dependence, such as red spin noise.

In particular,the group delay $t_g(\nu)$ for an observing frequency $\nu$ is given by the relation:

\begin{equation}
t_g(\nu) = K~DM/(\nu^2),
\end{equation}
where the dispersion constant $K$ is given by:

\begin{equation}
K \equiv 4.15 \times 10^{15}~\mathrm{Hz^2~cm^3~pc^{-1}~s}
\end{equation}
and the dispersion measure is defined as the integral of the electron density $n_e$ from the Earth to the pulsar:

\begin{equation}
\mathrm{DM} = \int_0^L n_e \mathrm{d}l.
\end{equation}

Dispersion measure corrections can be included in the analysis as an additional set of stochastic parameters in almost the same was as the frequency independent spin noise.  We begin by first defining a vector $\bmath{D}$ of length equal to the number of pulse profiles for a given pulsar as:

\begin{equation}
D_i = K/(\nu^2_i)
\end{equation}
for observation $i$ with observing frequency $\nu_i$.

We then write the basis vectors that describe the dispersion measure Fourier modes as:

\begin{equation}
F_{\mathrm{DM}}(\nu,t_i) = \frac{1}{T}\sin\left(2\pi\nu t_i\right)D_i
\end{equation}
and an equivalent cosine term, where $T$ is the length of the observing timespan, and $\nu$ denotes the frequency of the signal to be parameterised as before, where the set of frequencies to be included is defined in the same way as for the red spin noise.  Unlike when modelling the red spin noise, we no longer have the quadratic in the timing model to act as a proxy to the low frequency ($\nu < 1/T$) DM variations in our data.  As such these terms must be accounted for either by explicitly including these low frequencies in the model, or by including a quadratic in DM to act as a proxy, as with the red noise, defined as: 

\begin{equation}
 Q_{\mathrm{DM}}(t_i)= \delta_0 t_iD_i + \delta_1 t_i^2D_i,
\end{equation}
with $\delta_{0,1}$ free parameters to be fit for, and $t_i$ the barycentric arrival time for TOA $i$. This is most simply done by including these terms in the set of timing model parameters $\bmath{\epsilon}$.  If these terms are not included in the model then power from frequencies lower than $1/T$ will be absorbed by the Fourier coefficients included in $\mathbf{F}_{\mathrm{DM}}$, biasing the estimated power spectrum.  

As with the red noise we can then include the DM signal realisation in our model TOAs:

\begin{equation}
\bmath{\hat{\tau}}(\bmath{\epsilon}, \bmath{a}_{\mathrm{red}}, \bmath{a}_{\mathrm{DM}}) =  \bmath{\tau}(\bmath{\epsilon}) - \mathbf{F}_{\mathrm{red}}\mathbf{a}_{\mathrm{red}} - \mathbf{F}_{\mathrm{DM}}\mathbf{a}_{\mathrm{DM}},
\end{equation}
where we have factored the quadratic $Q_{\mathrm{DM}}$ into the timing model $\bmath{\tau}(\bmath{\epsilon})$. Finally we then define the matrix of DM power spectrum coefficients $\bmath{\varphi}_{\mathrm{DM}}$ such that:

\begin{equation}
\label{Eq:ProbDMFreq}
\mathrm{Pr}(\mathbf{a}_{\mathrm{DM}}\; | \;\bmath{\rho}_{\mathrm{DM}}) \; \propto \; \frac{1}{\sqrt{\mathrm{det}\bmath{\varphi}_{\mathrm{DM}}}} \exp\left[-\frac{1}{2}\mathbf{a}_{\mathrm{DM}}^{*T}\bmath{\varphi}_{\mathrm{DM}}^{-1}\mathbf{a}_{\mathrm{DM}}\right].
\end{equation}
We note here that, as with the red noise, if desired additional terms can be added into the DM Fourier matrix $\mathbf{F}_{\mathrm{DM}}$ to model, for example, additional annual variations in the data, and that as before the DM spectrum coefficients can be parameterised with a non-Gaussian prior.

\subsection{A non-Gaussian stochastic gravitational wave background}

The final application to pulsar timing of the non-Gaussian formalism developed thus far that we will consider is to a stochastic gravitational wave background (GWB) for which we follow the approach given in \citep{2013PhRvD..87j4021L}.

As with the intrinsic red noise model described in section \ref{Section:Red} we parameterise the signal in each pulsar using the Fourier basis given in equation \ref{Eq:FMatrix}.  When dealing with a signal from a GWB, however, it is crucial to include the cross correlated signal between the pulsars on the sky.  We do this by using the Hellings-Downs relation \cite{1983ApJ...265L..39H}:

\begin{eqnarray}
\Theta_{mn} &= & \frac{3}{2}\frac{1-\cos(\theta_{mn})}{2}\ln\left(\frac{1-\cos(\theta_{mn})}{2}\right) \nonumber\\ 
			&-&   \frac{1}{4}\frac{1-\cos(\theta_{mn})}{2}+\frac{1}{2} + \frac{1}{2}\delta_{mn},
\end{eqnarray}
where $\theta_{mn}$ is the angle between the pulsars $m$ and $n$ on the sky and $\Theta_{mn}$ represents the expected correlation between the TOAs given an isotropic background.  With this addition our covariance matrix for the Fourier coefficients becomes
 
\begin{equation}
\label{Eq:FreqMatrix}
\varphi_{mi,nj} = \left< a_{mi}a_{nj}^*\right> = \Theta_{mn}\varphi_{i}\delta_{ij},
\end{equation}
where there is no sum over $i$, which results in a band diagonal matrix for which calculating the inverse is extremely computationally efficient.

Our prior term $\mathrm{Pr}(\mathbf{a} | \bmath{\varphi})$ is then similar to that given in Eq. \ref{Eq:ProbFreq}, with the only difference that $\mathbf{a}$ is now the concatenated vector of Fourier coefficients for all pulsars, and $ \bmath{\varphi}$ includes the correlations between pulsars.  We can then parameterise the non-Gaussianity at each GWB frequency as in Eq. \ref{Eq:NGRed}, however with multiple pulsars we need not assume a power law prior on the coefficients if desired.

\section{Application to simulations}
\label{Section:Pulsarsims}

\begin{figure*}
\begin{center}$
\begin{array}{cc}
\hspace{-1.5cm}
\includegraphics[width=100mm]{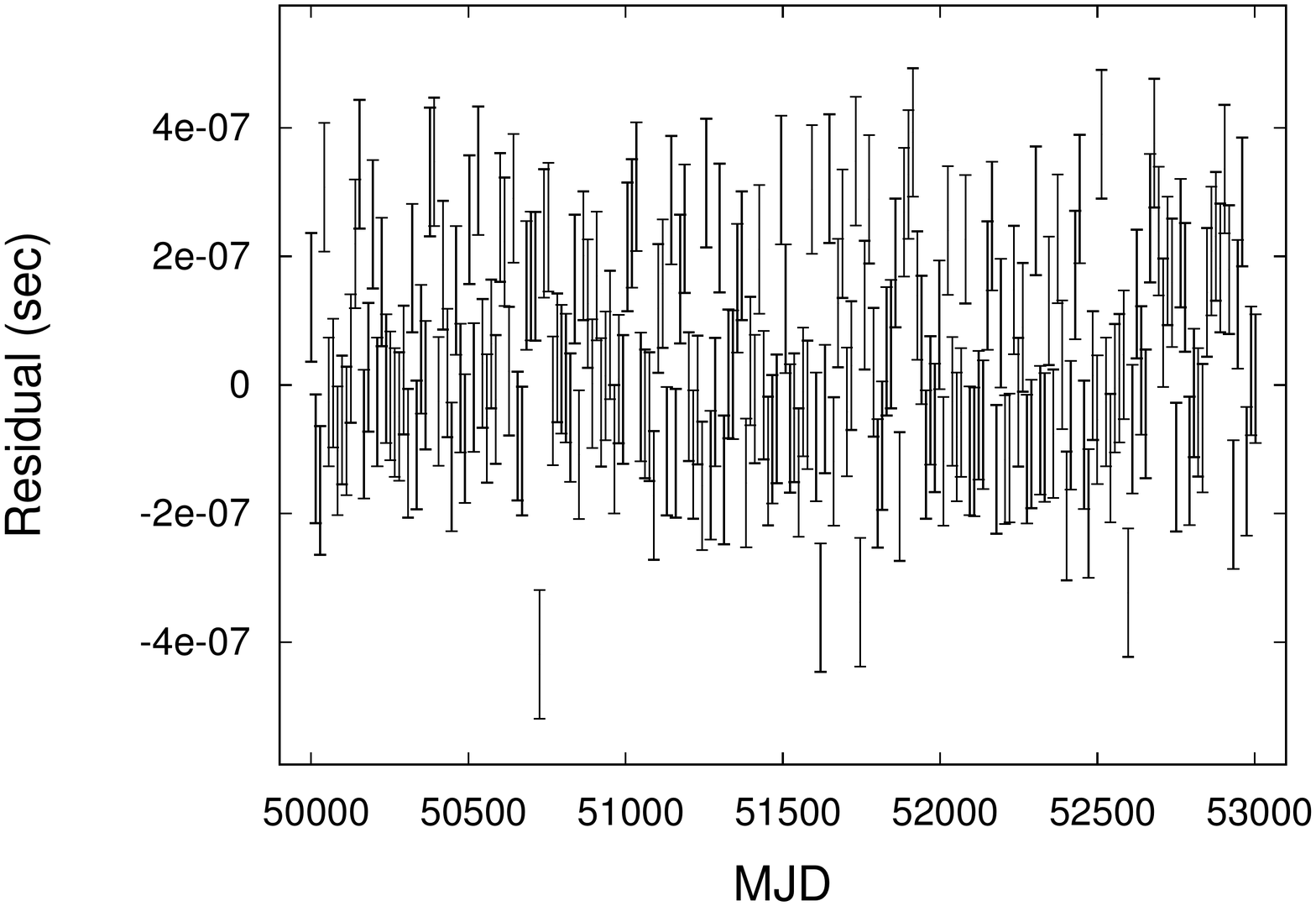} &
\includegraphics[width=100mm]{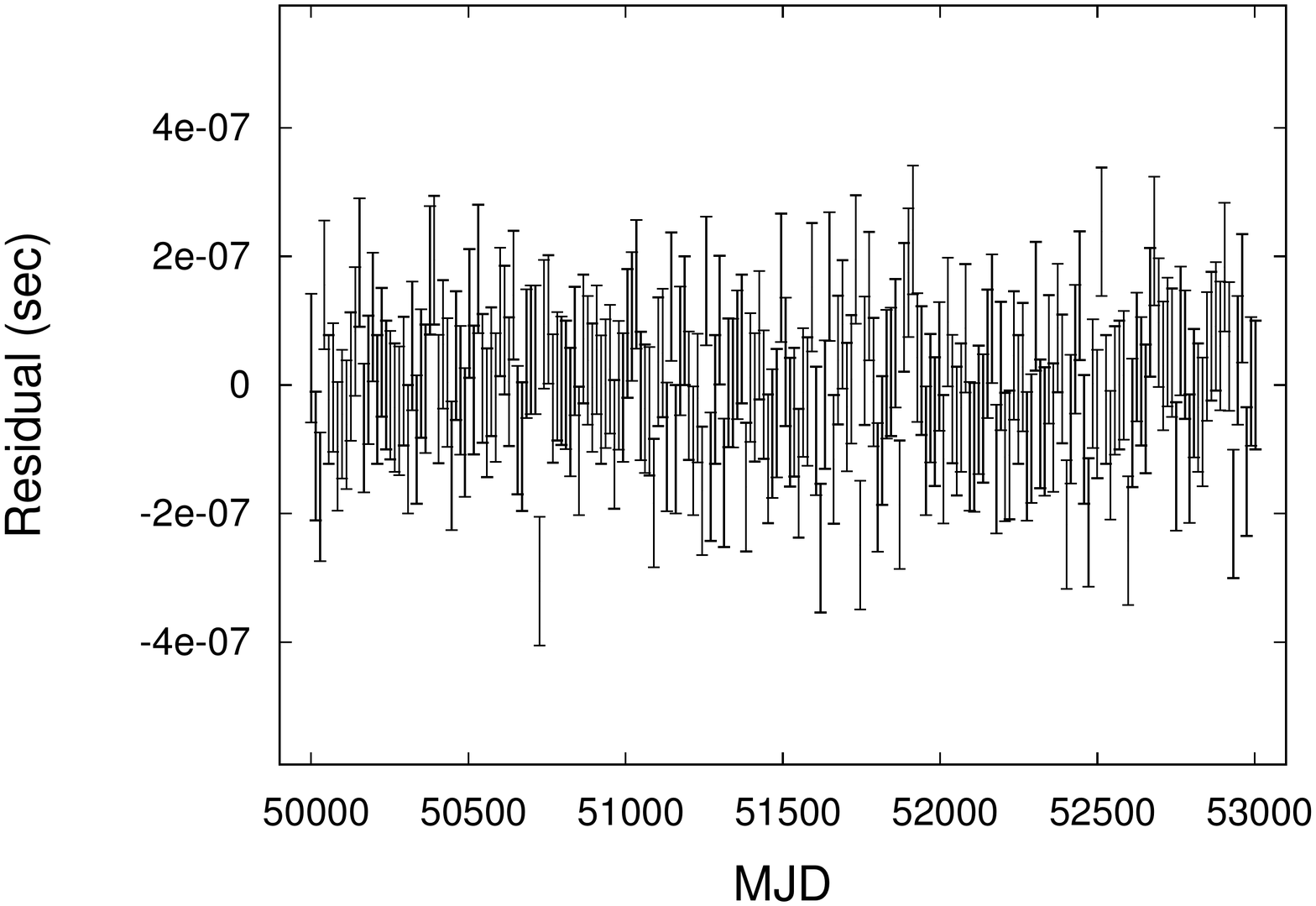}\\ 
\end{array}$
\end{center}
\vspace{-0.8cm}
\caption{Non-Gaussian (left) and Gaussian (right) timing residuals from simulations one and two respectively after subtracting the simulated timing model.\label{figure:SimRes}}
\end{figure*}

We now apply the likelihood developed in section \ref{Section:PulsarNonGaussian} to two simulations.  Simulation 1 includes non-Gaussian noise with the same distribution as in the toy model in section \ref{Section:toy}, and Simulation 2 includes only Gaussian noise.  In both cases we set the width of the ground state Gaussian distribution to be $10^{-6}$ seconds.  We then simulate an $\sim$ 8 year dataset for the isolated pulsar J0030+0451 with observations spaced $\sim$ 2 weeks apart resulting in a total of 216 TOAs onto which we add a noise realisation drawn from the two distributions using the same seed both times.  The timing model parameters used in the simulations are listed in Table \ref{Table:Sim1}, and are based on values provided as part of the first IPTA data challenge. The timing residuals that remain from the two simulations after subtracting the injected timing model are shown in Fig. \ref{figure:SimRes}.

In both simulations we will compare the parameter estimates obtained using two models.  Model 1 will include the parameters $\alpha_{1..3}$ to model any deviations from Gaussianity that the timing residuals might be subject to, and in Model 2 will set these parameters equal to zero, therefore constraining the probability density to be purely Gaussian. 

While in principle the $\bmath{\alpha}$ coefficients can be real or complex valued, as they only appear in the likelihood in the term $\left|\sum_{n}^{n_{\mathrm{max}}}\alpha_nC_nH_n\right|^2$ we restrict our search to only real values in order to eliminate potential degeneracies that would otherwise arise.  In both simulations we take the prior range on these coefficients to be $[-1,1]$.

\subsection{Simulation 1}

\begin{figure*}
\begin{center}$
\begin{array}{ccc}
\hspace{-0.5cm}
\includegraphics[width=60mm]{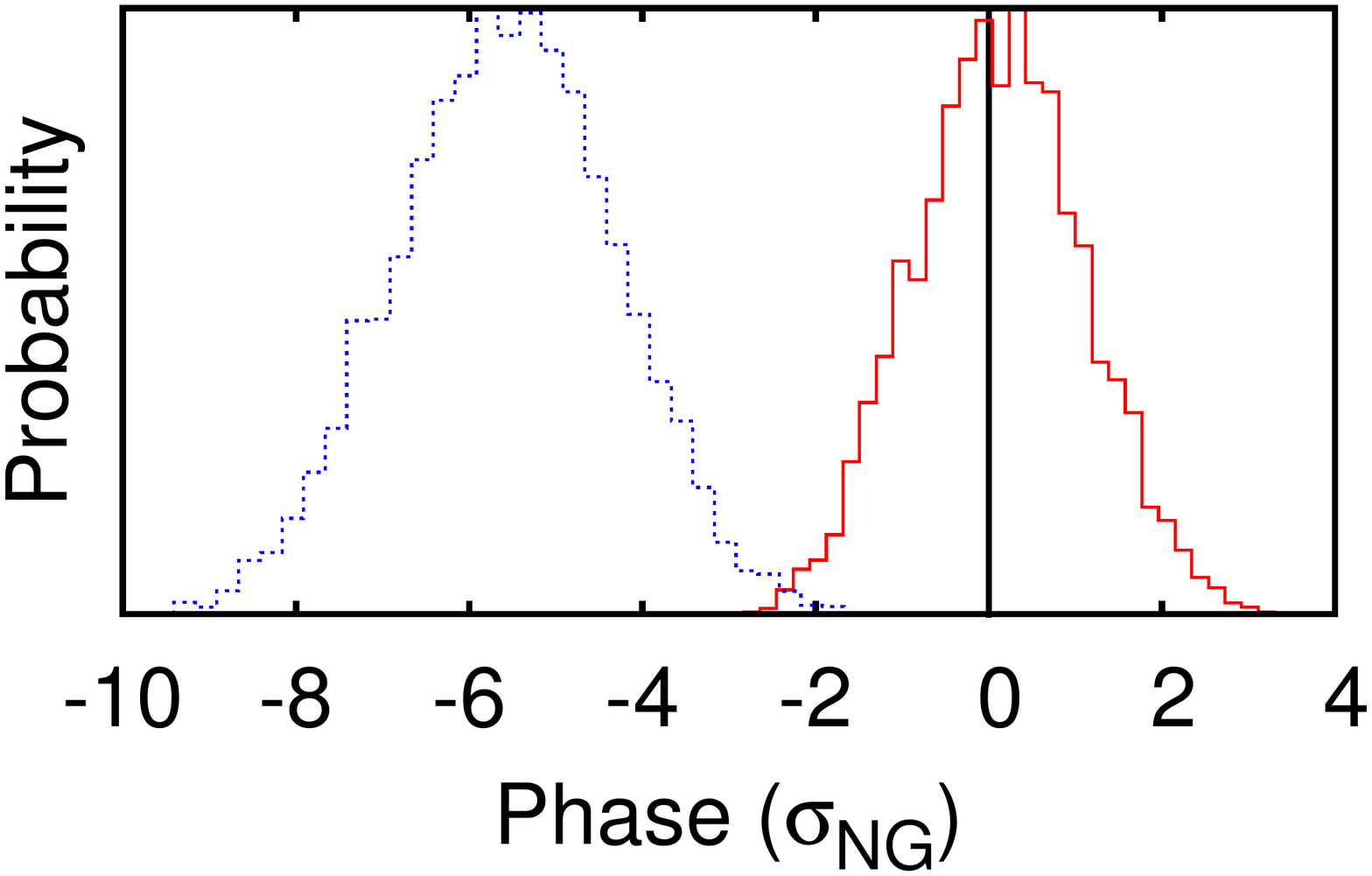} &
\includegraphics[width=60mm]{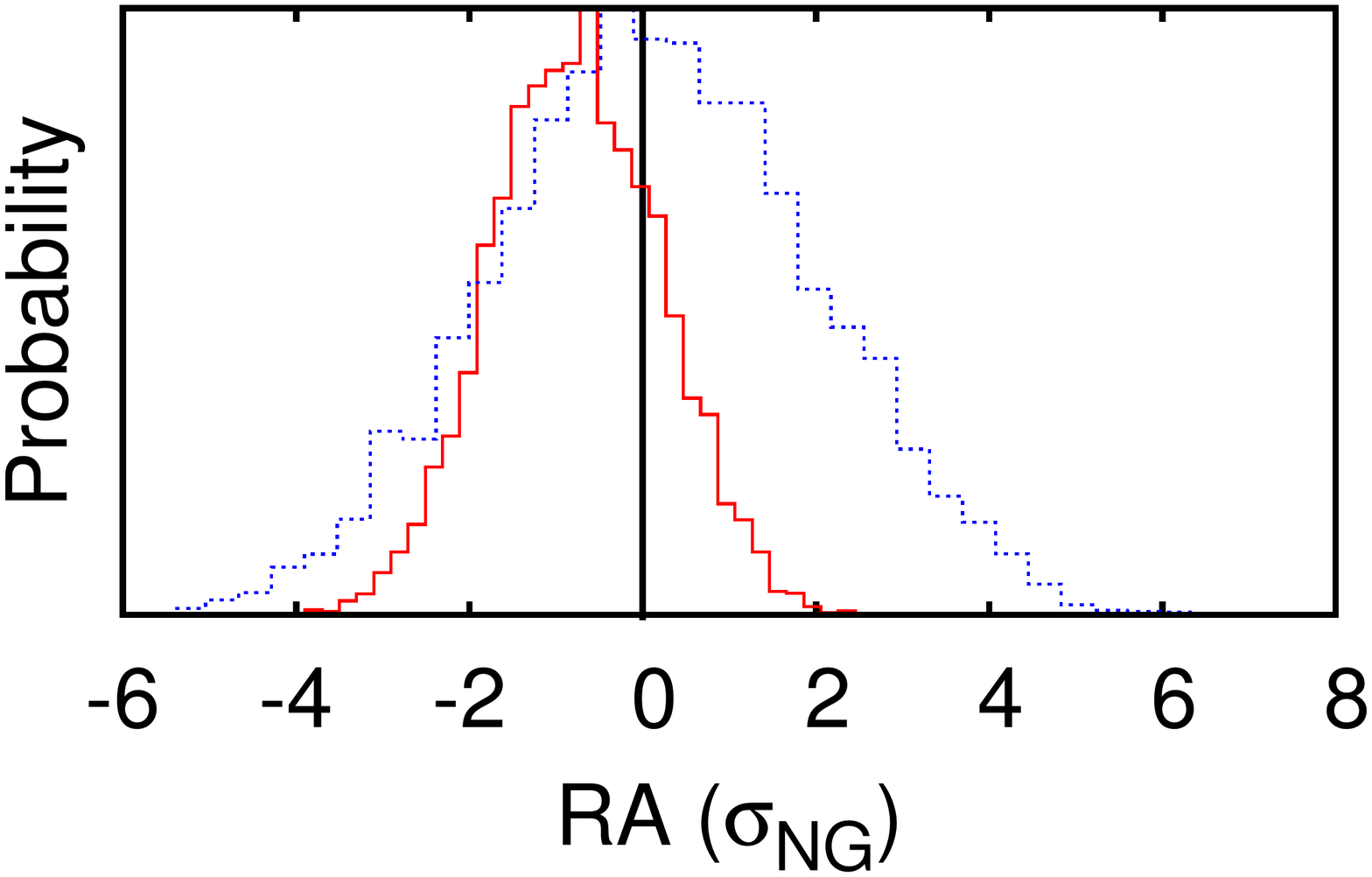} &
\includegraphics[width=60mm]{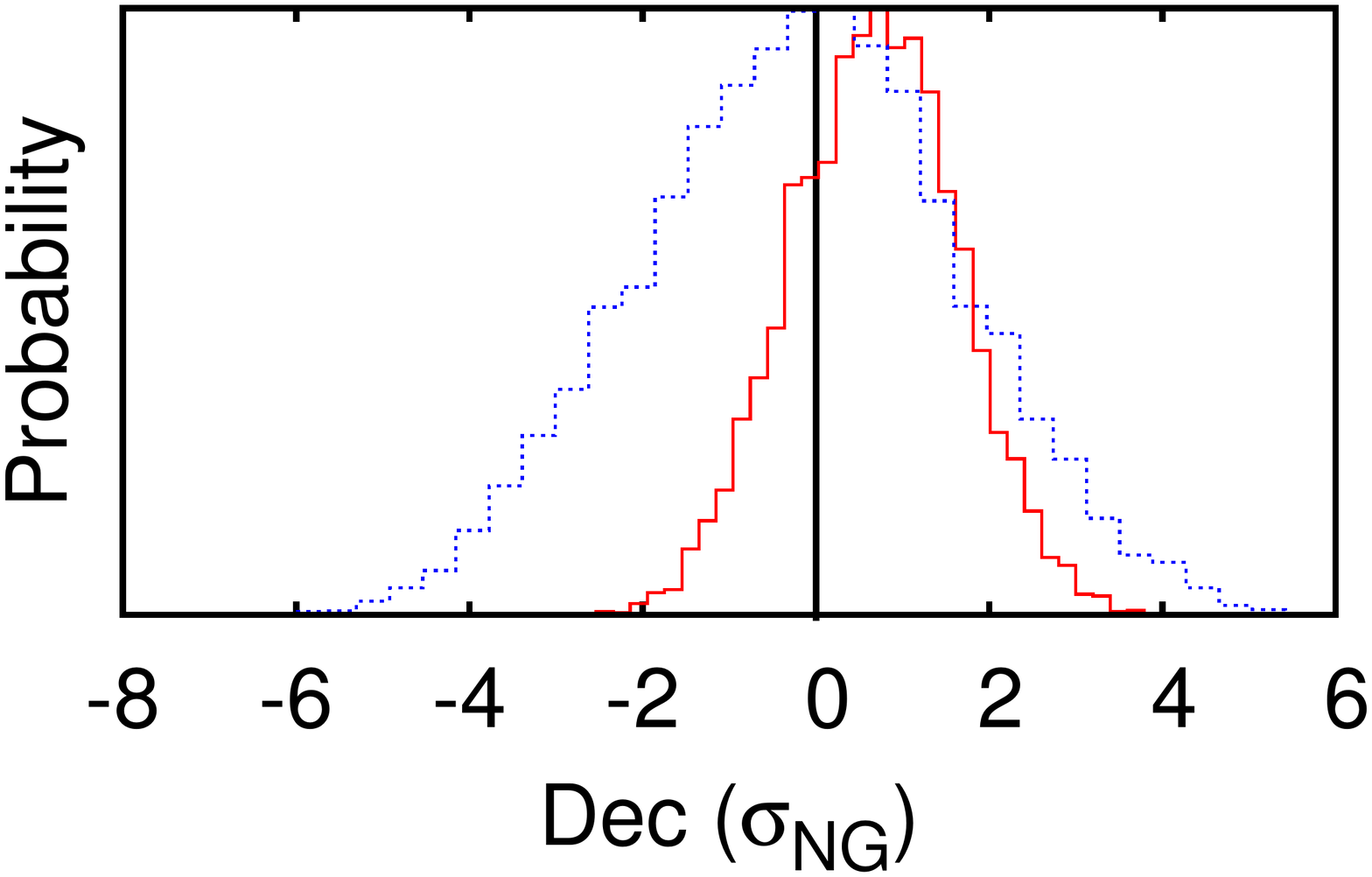} \\
\hspace{-0.5cm}
\includegraphics[width=60mm]{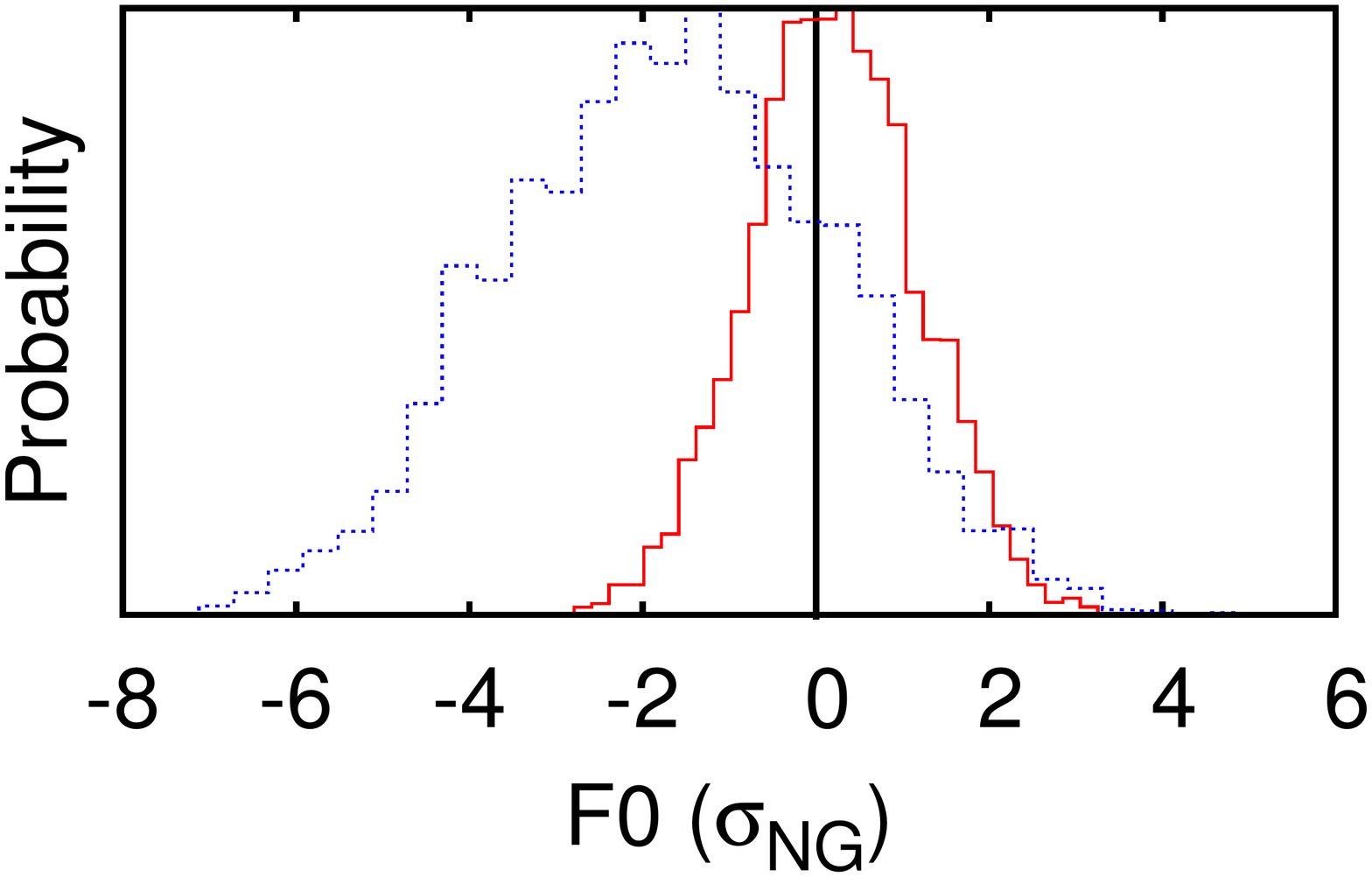} &
\includegraphics[width=60mm]{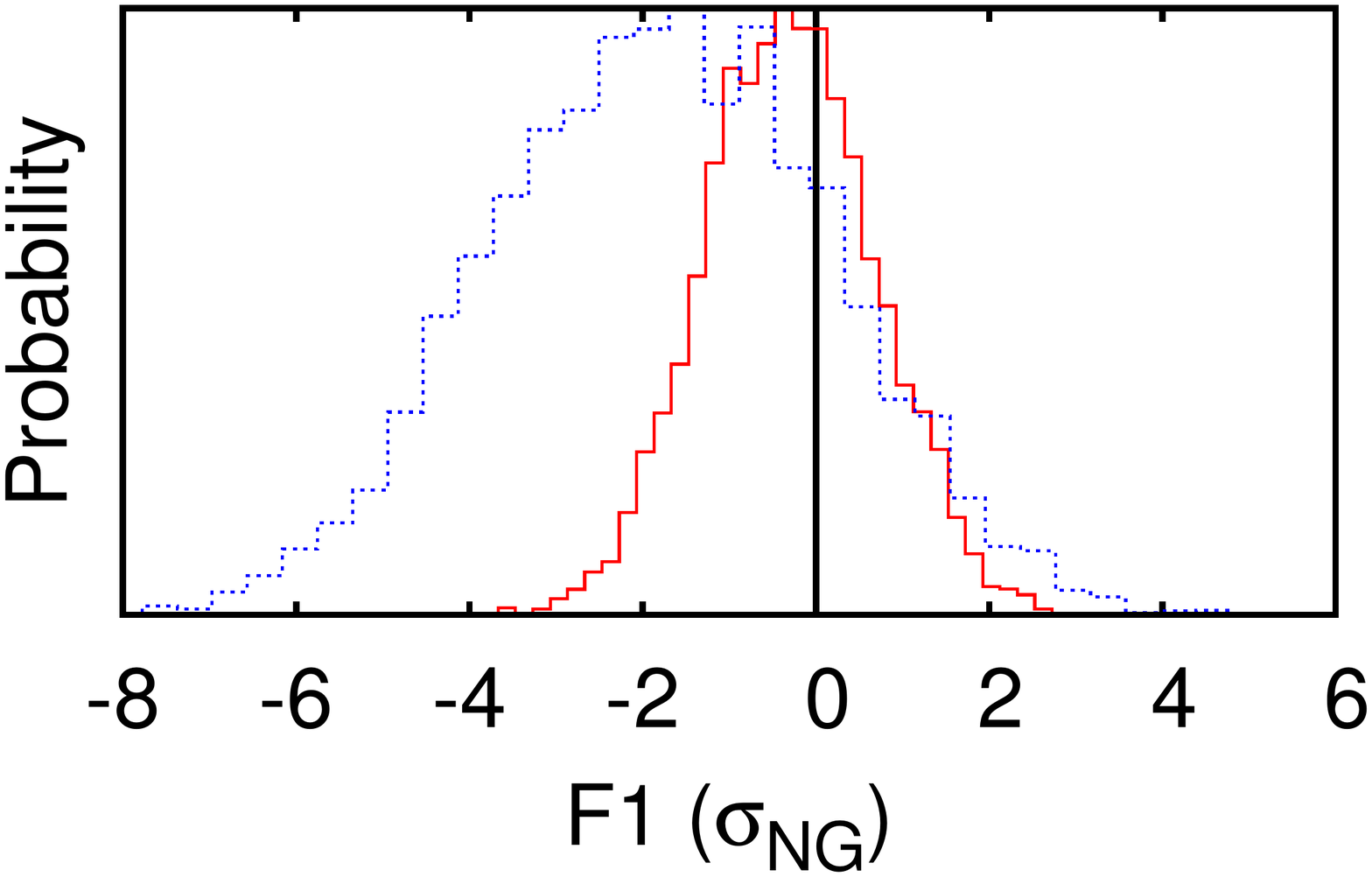} &
\includegraphics[width=60mm]{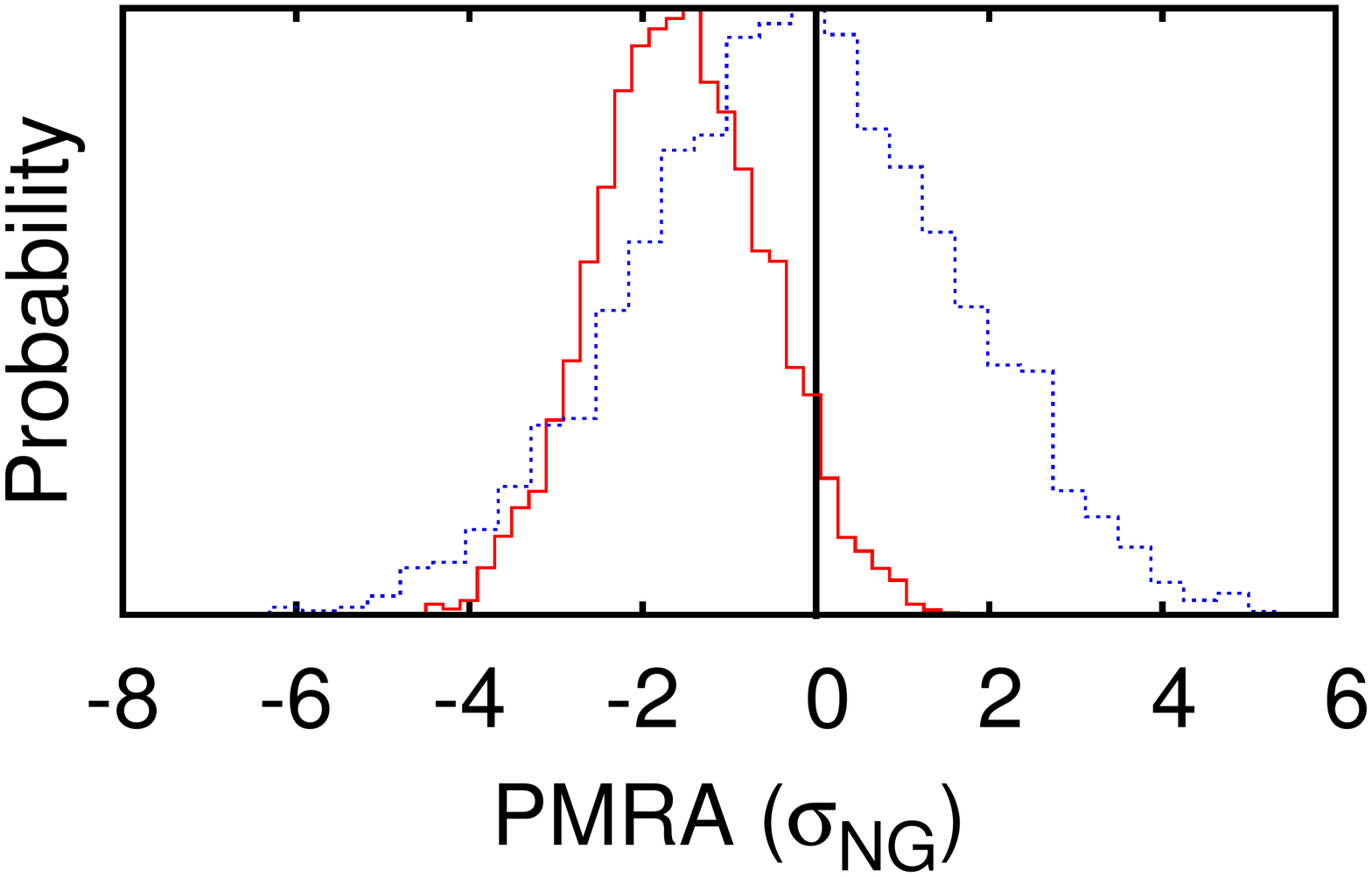}\\
\hspace{-0.5cm}
\includegraphics[width=60mm]{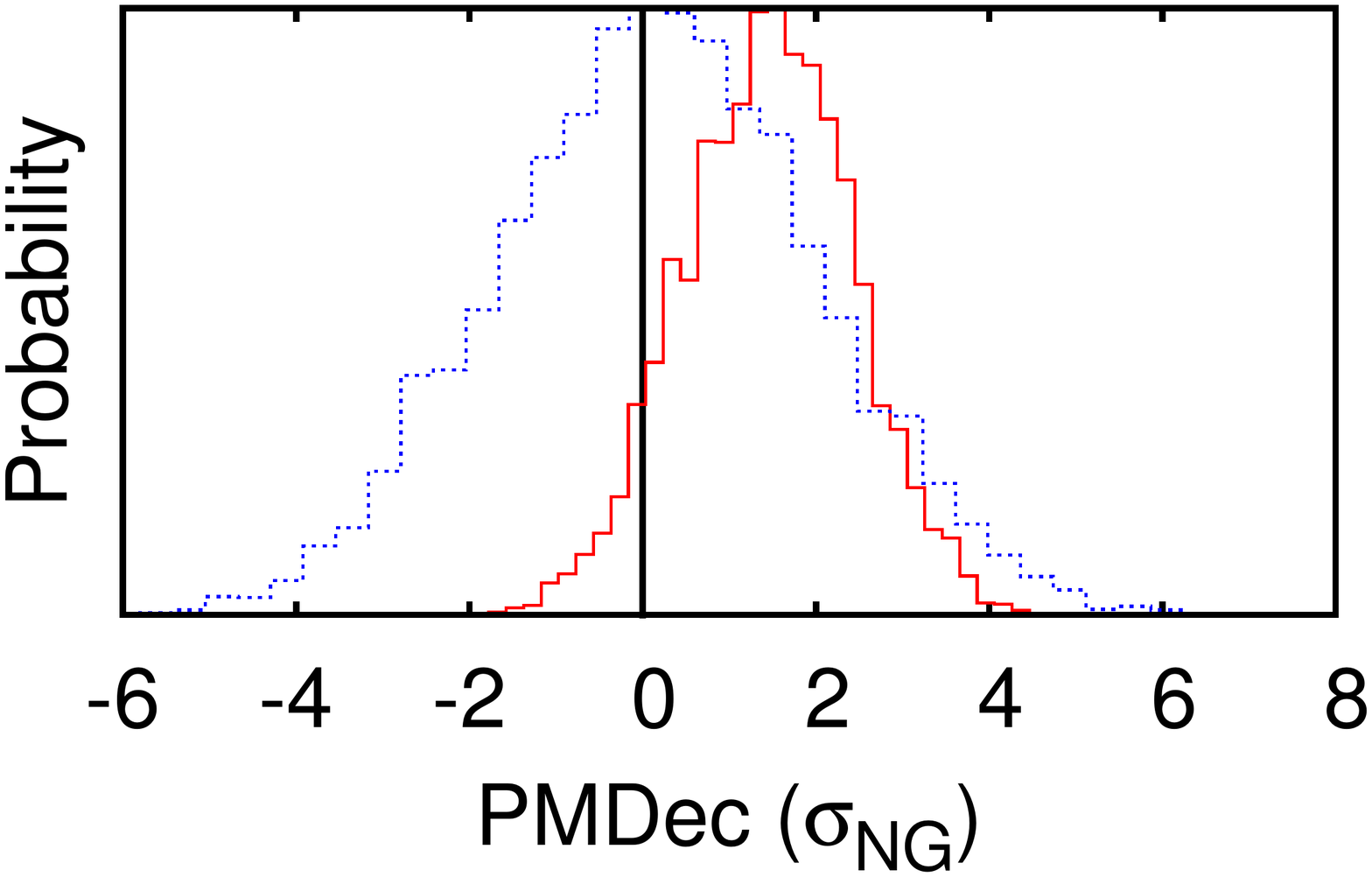} &
\includegraphics[width=60mm]{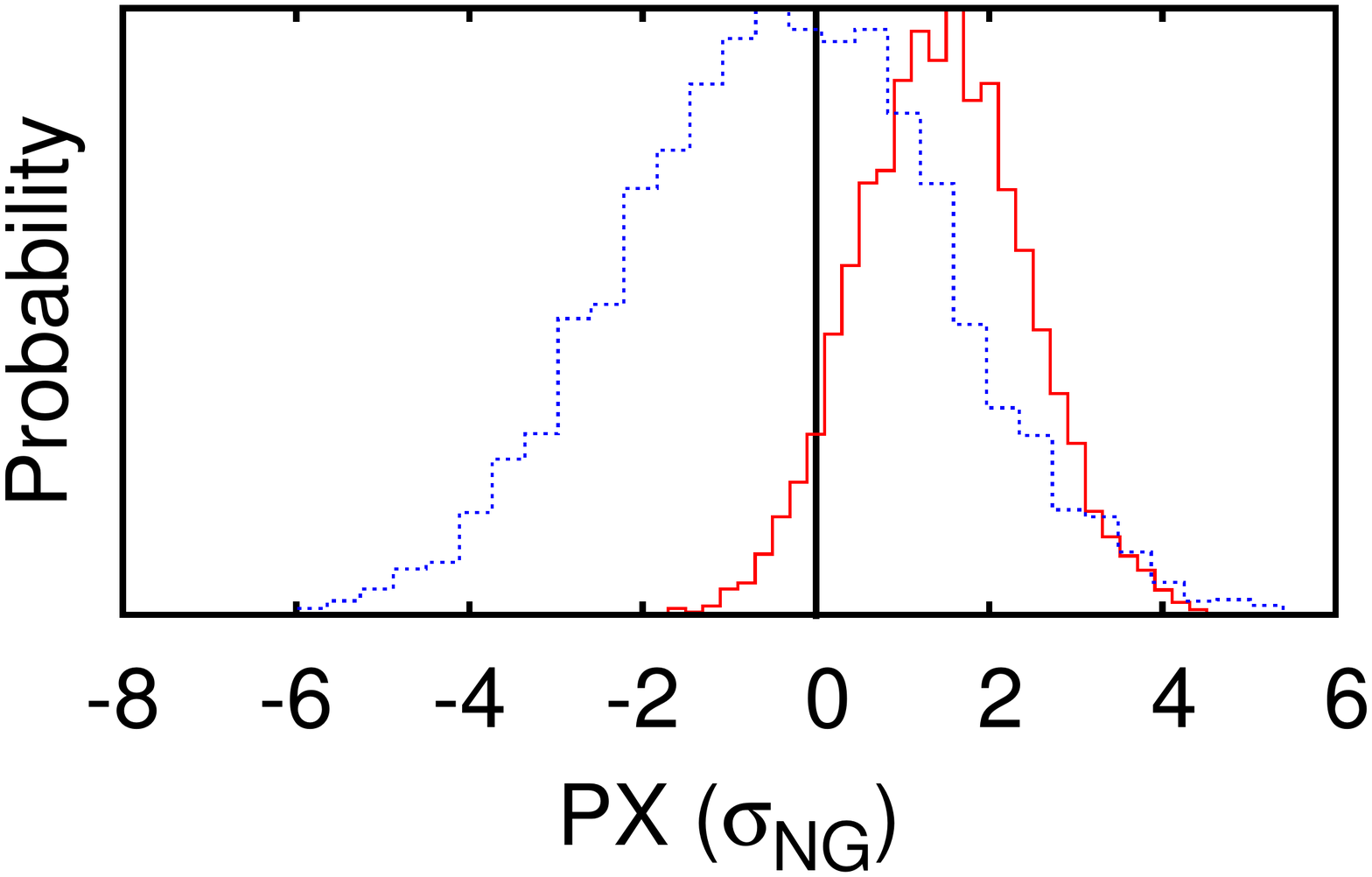} &
\includegraphics[width=60mm]{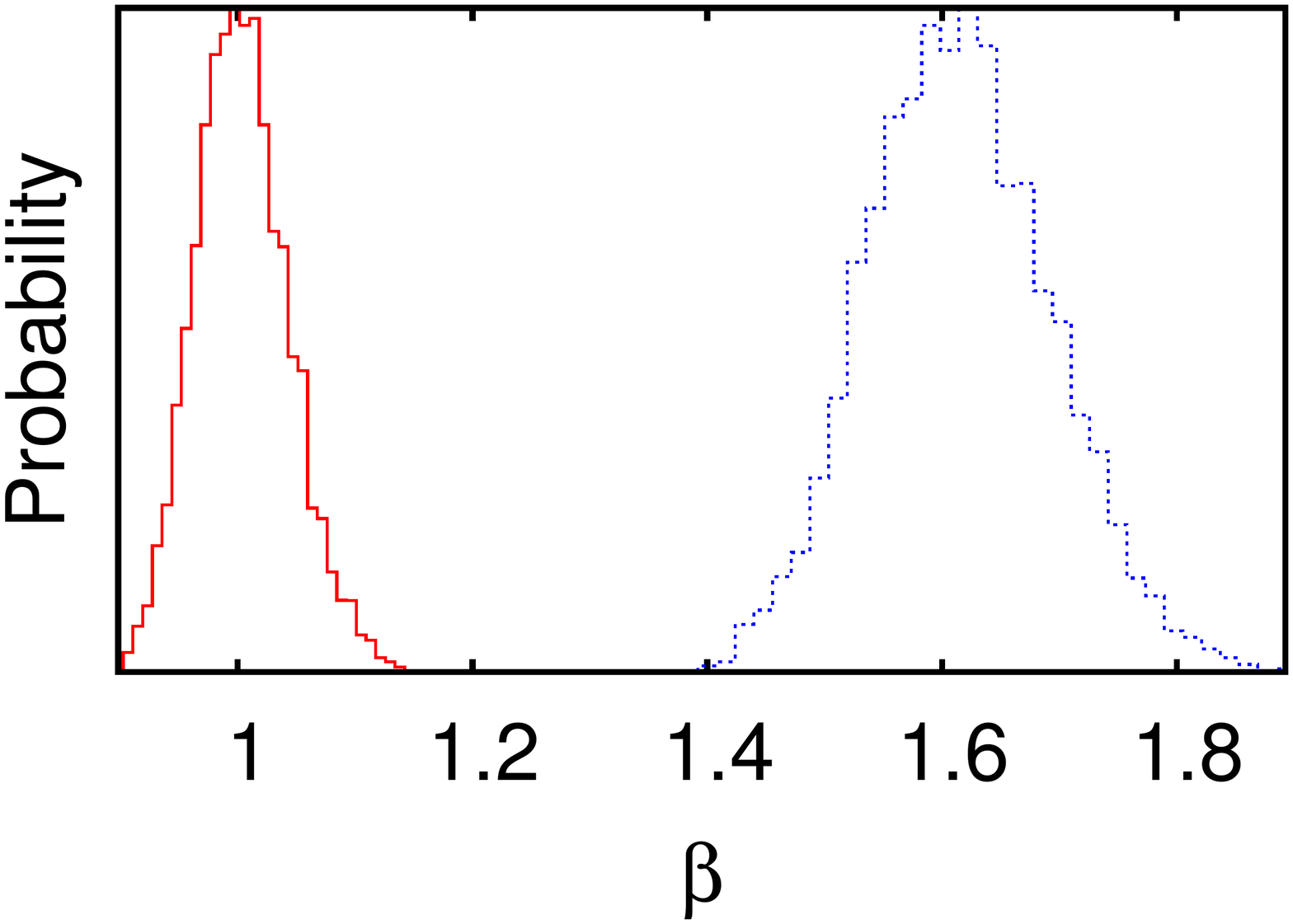} \\
\end{array}$
\end{center}
\vspace{-0.8cm}
\caption{1-dimensional marginalised posteriors for the timing model and $\beta$ parameters in simulated dataset 1 for the isolated pulsar PSR J0030+0451 for model 1 (red solid line) and model 2 (blue dotted line).  Values on the $x$-axes for the timing model parameters are given in terms of the standard deviation in that parameter returned by the analysis when including the additional terms, with the injected parameter value at 0 in all cases. The non-Gaussian nature of the noise results in a significant increase in the scaling parameter $\beta$ in model 2, leading to a decrease in the precision with which the timing model parameters are detected by a factor $\sim$ 2 relative to model 1. \label{figure:Sim1Comp}}
\end{figure*}

Table \ref{Table:Sim1} lists the mean parameter estimates and standard deviations for the two models applied to simulation 1.  In addition, Fig. \ref{figure:Sim1Comp} shows the one-dimensional marginalised posteriors for the timing model parameters and the white noise scaling parameter $\beta$ for model 1 (red solid lines) and model 2 (blue dotted lines) respectively.  In all cases the constraints on the timing model parameters are improved by a factor $\sim$ 2 when we include the additional parameters and thus correctly account for the non-Gaussian nature of the noise.  The degree to which the timing model parameters are affected will, however, naturally depend on the severity of the non-Gaussianity in the residuals.  In addition from Fig. \ref{figure:SimRes} we see that the effect of the non-Gaussian terms is to i) increase the number of outliers, and ii) to shift the mean of the residuals as the distribution is no longer symmetric.  The result of this is that when assuming a Gaussian likelihood the constant phase term is significantly offset from zero, and the EFAC term $\beta$ is increased in order to accommodate the outliers.  While the shift in the phase term is irrelevant, the increase in the EFAC term leads directly to the decrease in sensitivity to the timing model parameters observed previously.
Comparing the evidence between the two models we find that the additional parameters in model 1 are strongly favoured with $\Delta \log E = 18$, suggesting a definitive detection of non-Gaussianity in the residuals.

\subsection{Simulation 2}
\label{Section:sim2}

\begin{figure*}
\begin{center}$
\begin{array}{ccc}
\hspace{-0.5cm}
\includegraphics[width=60mm]{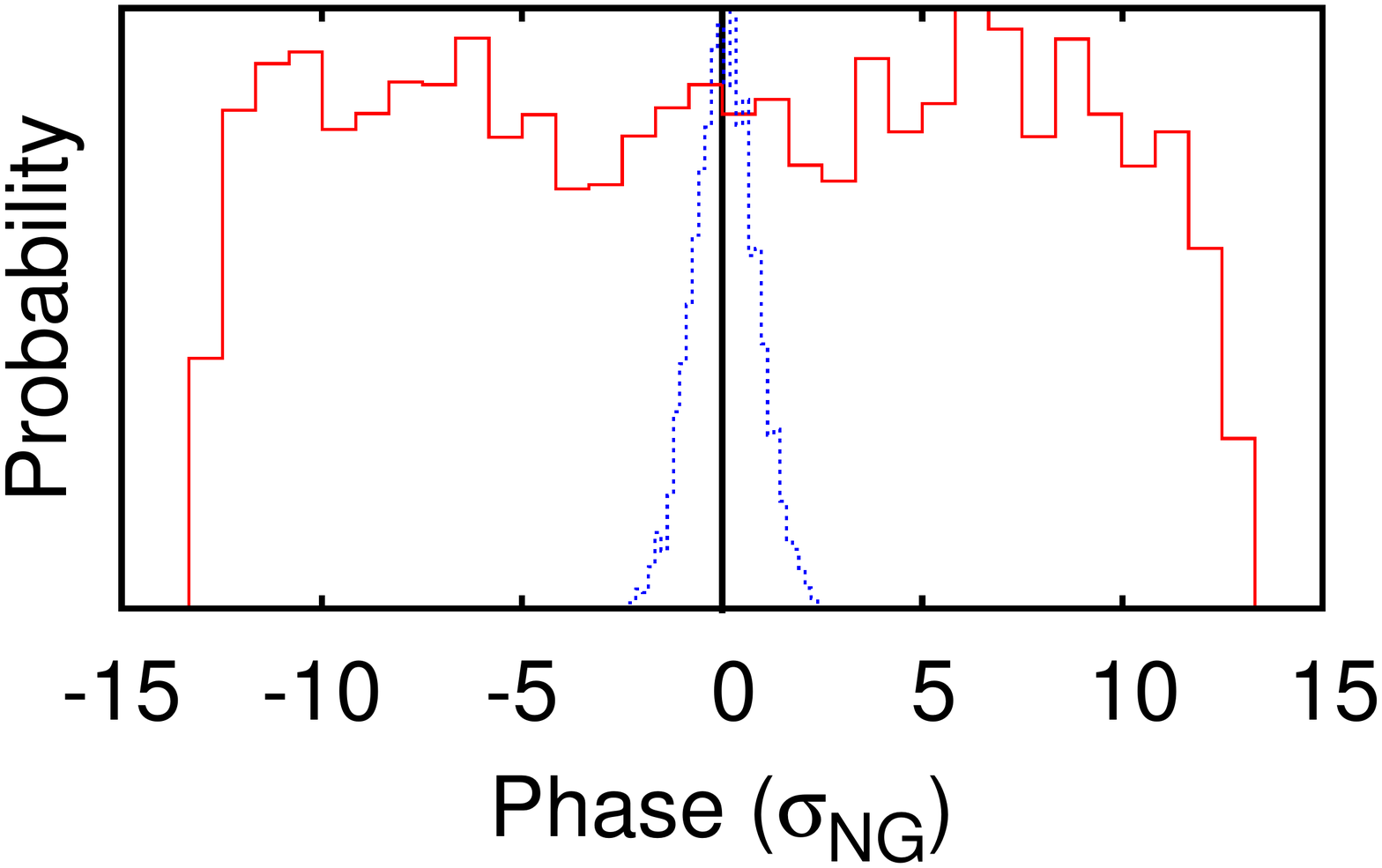} &
\includegraphics[width=60mm]{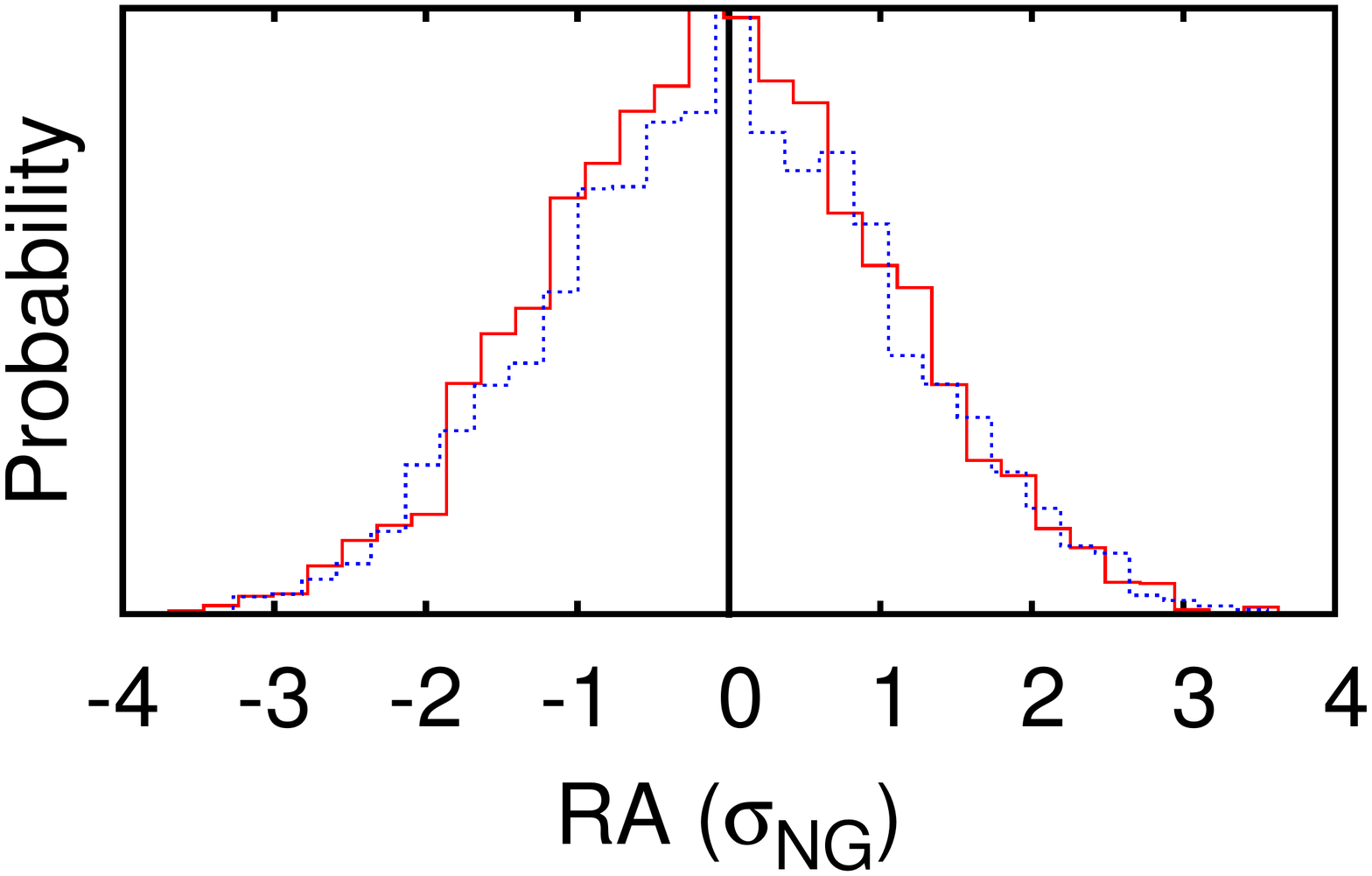} &
\includegraphics[width=60mm]{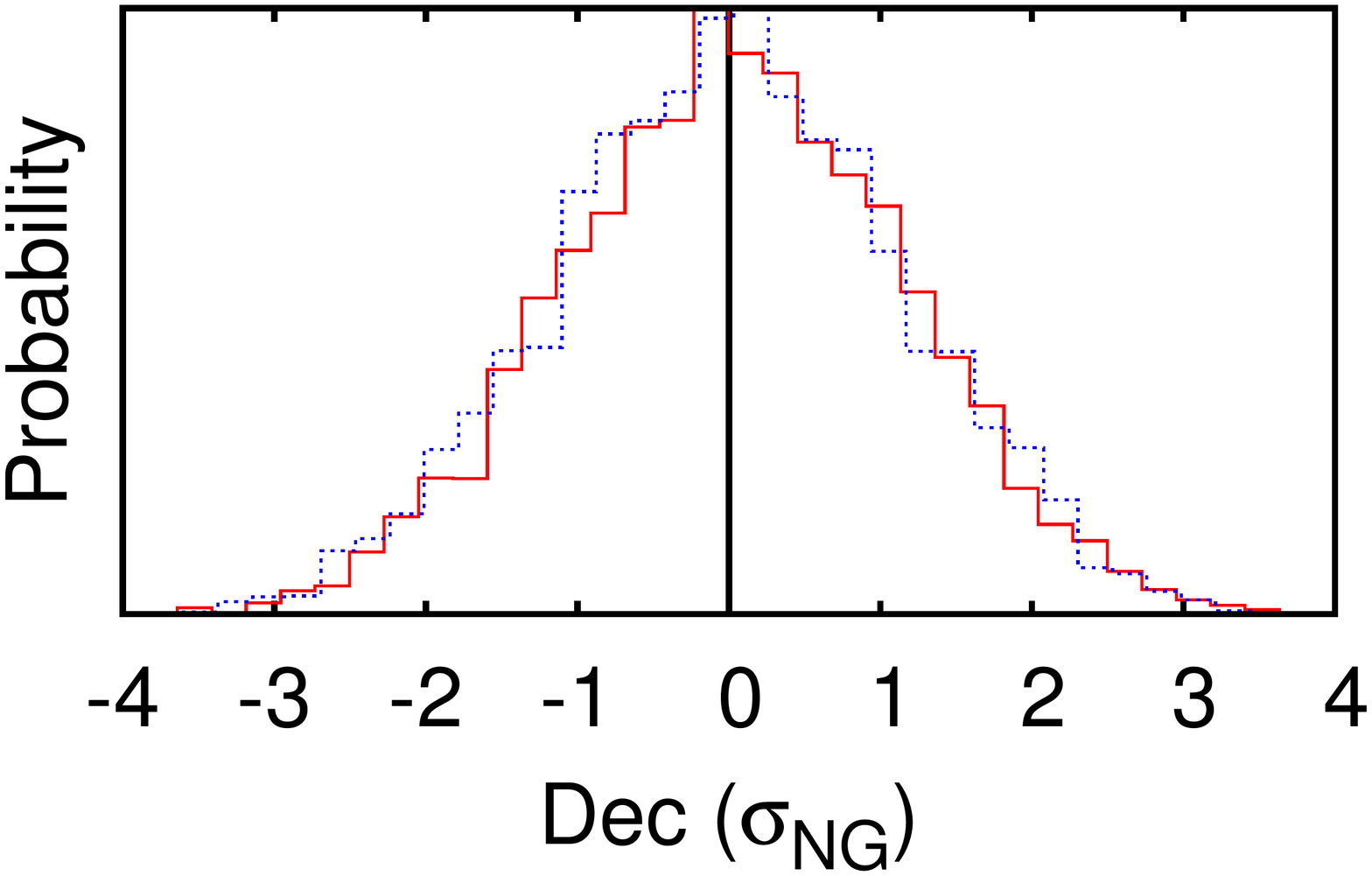} \\
\hspace{-0.5cm}
\includegraphics[width=60mm]{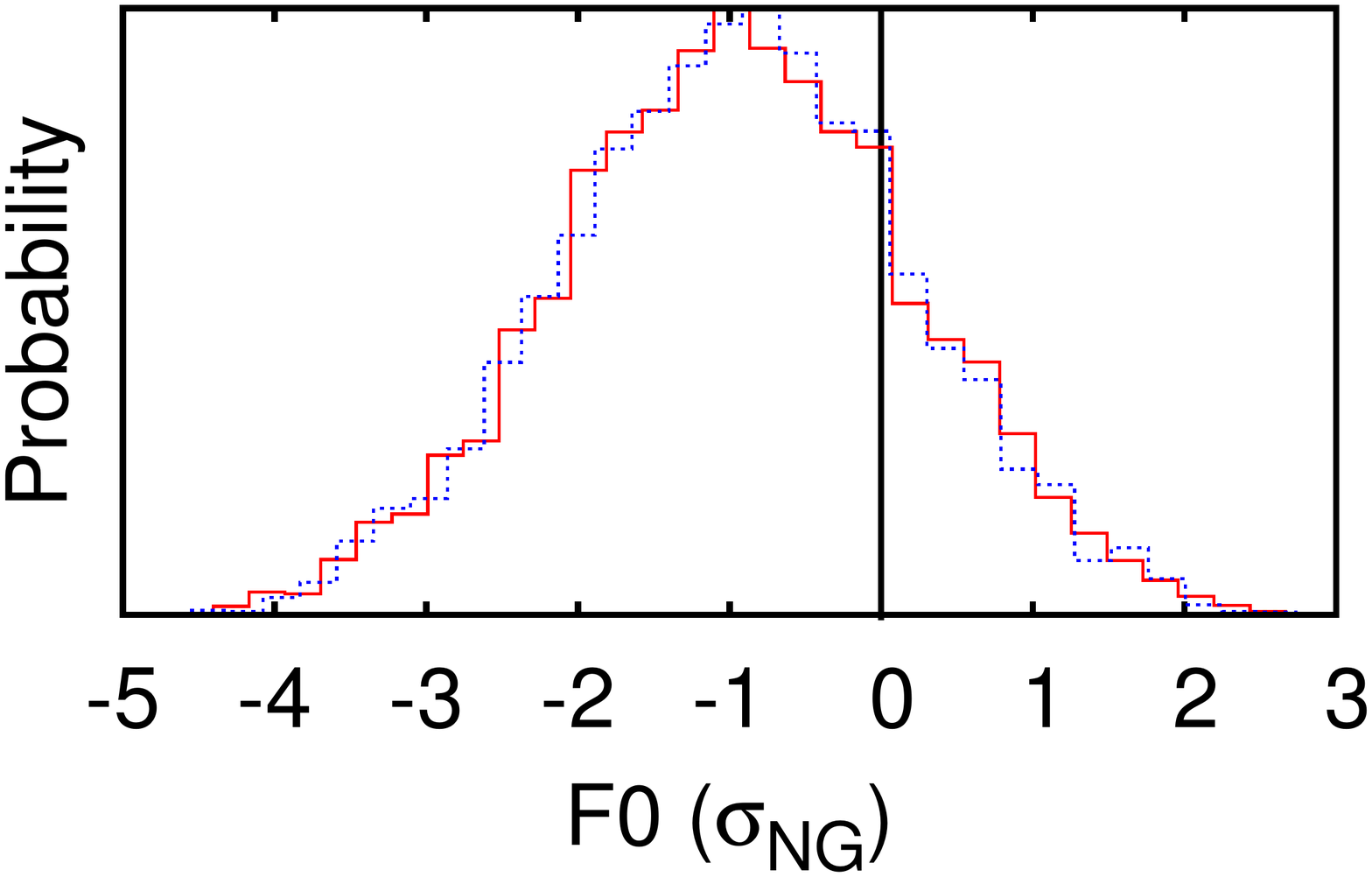} &
\includegraphics[width=60mm]{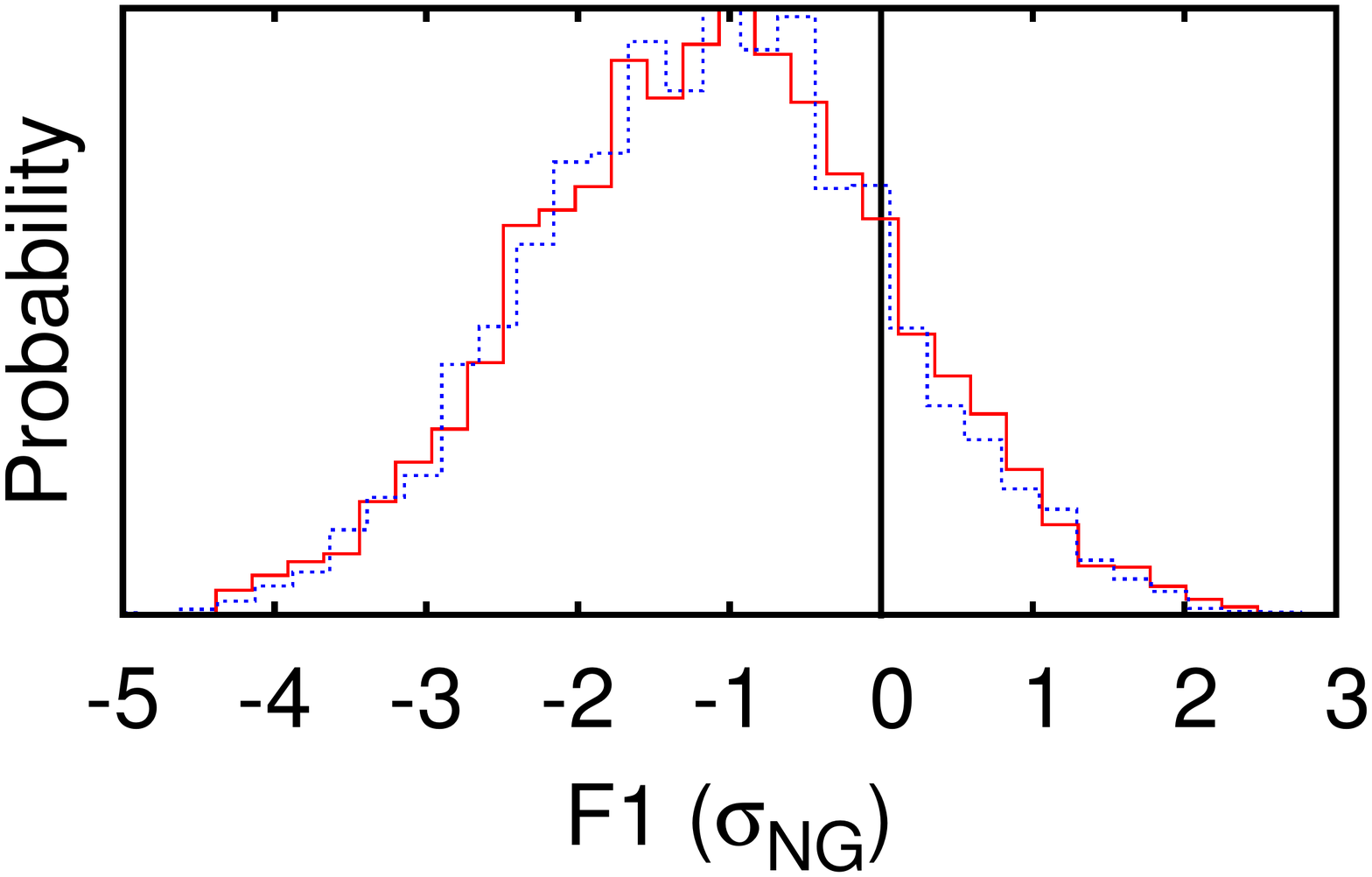} &
\includegraphics[width=60mm]{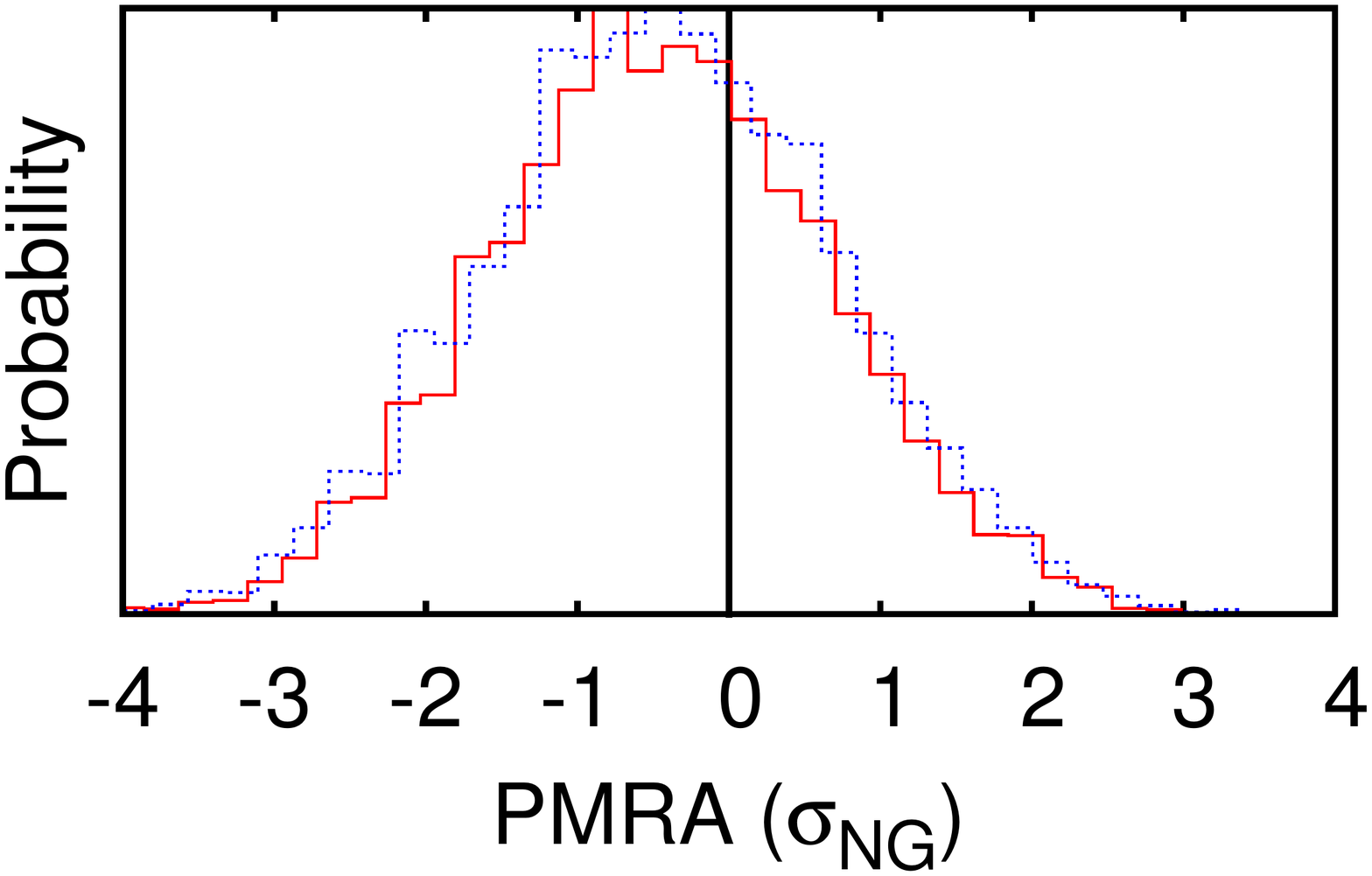}\\
\hspace{-0.5cm}
\includegraphics[width=60mm]{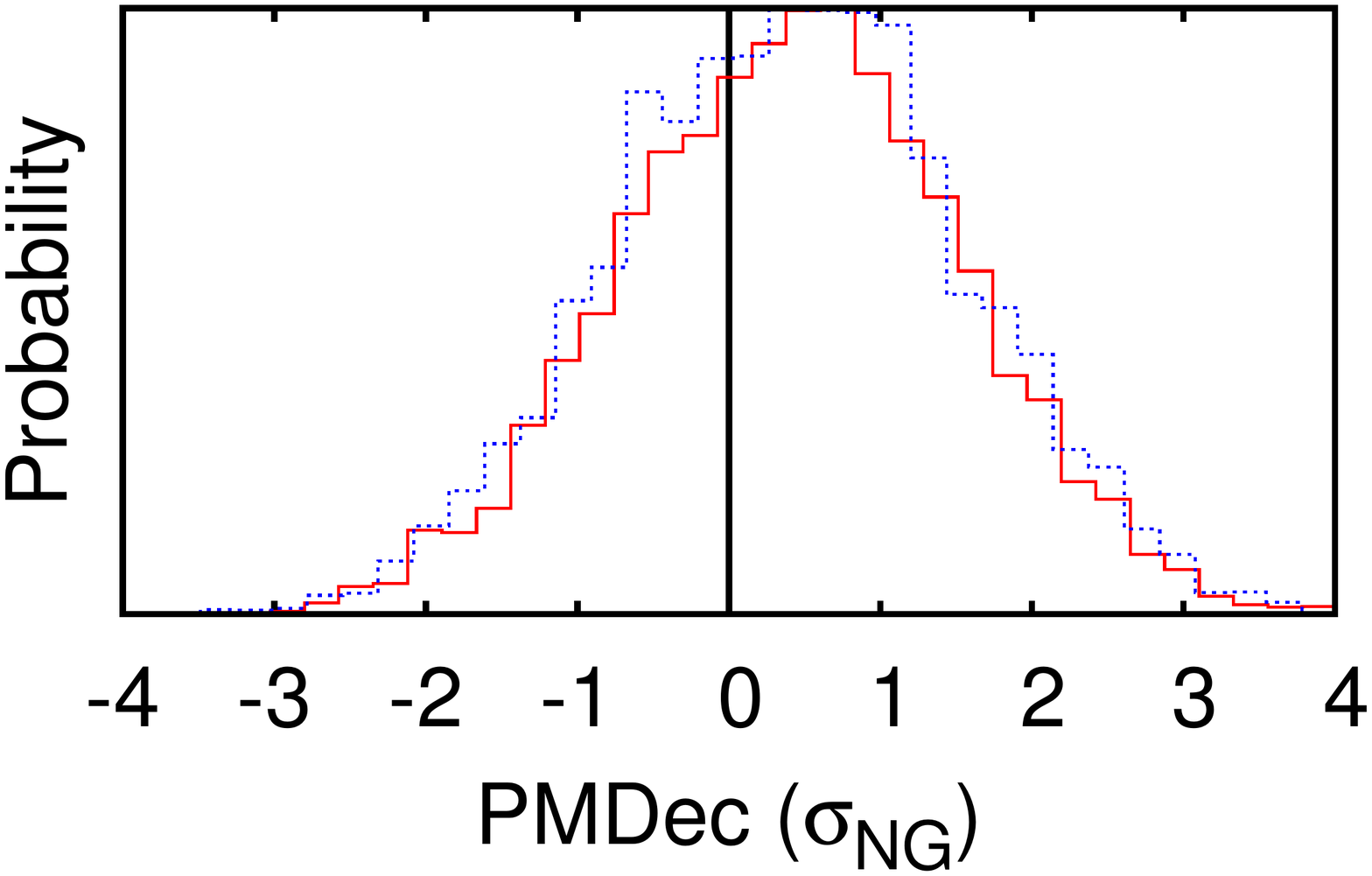} &
\includegraphics[width=60mm]{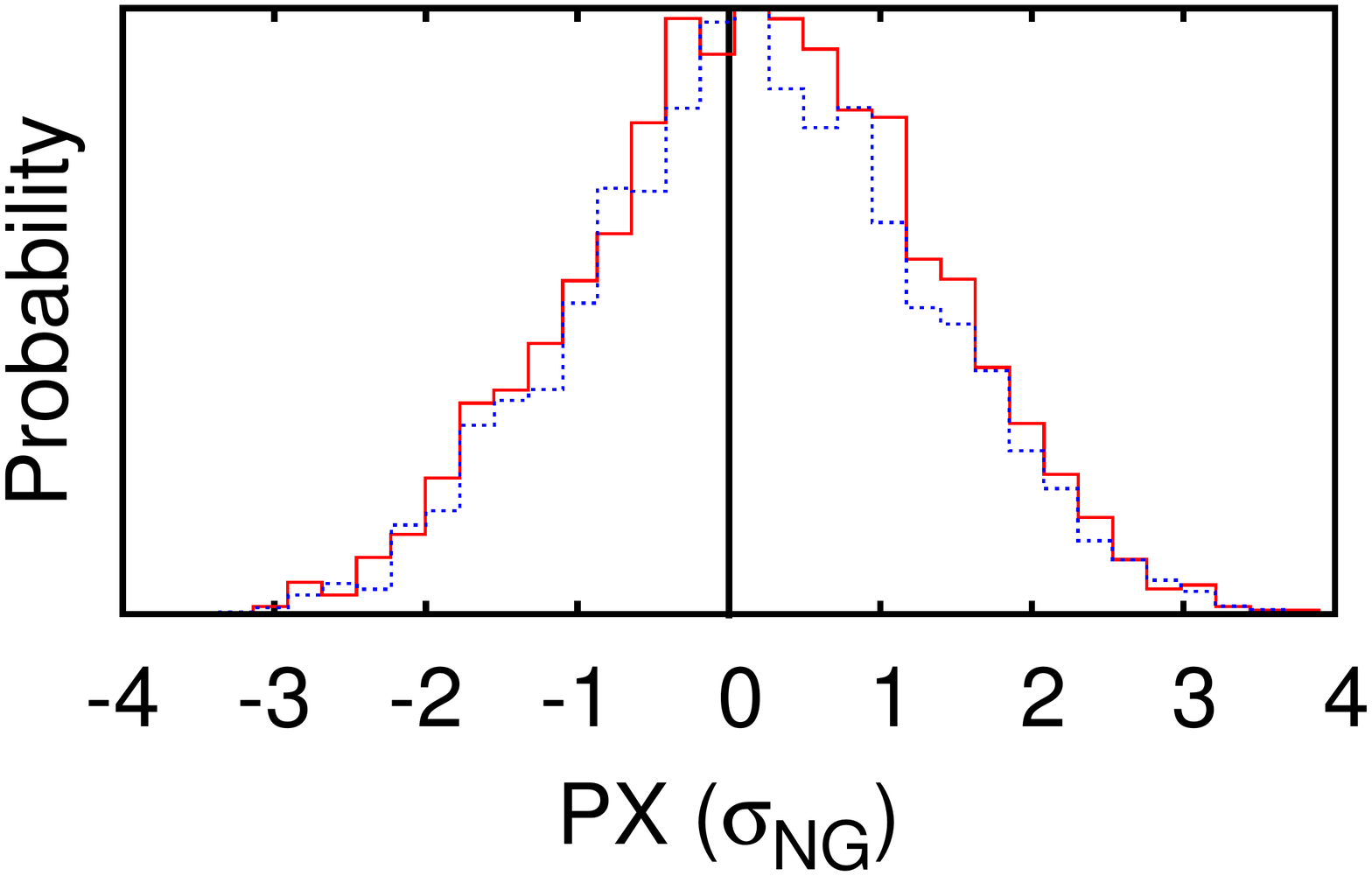} &
\includegraphics[width=60mm]{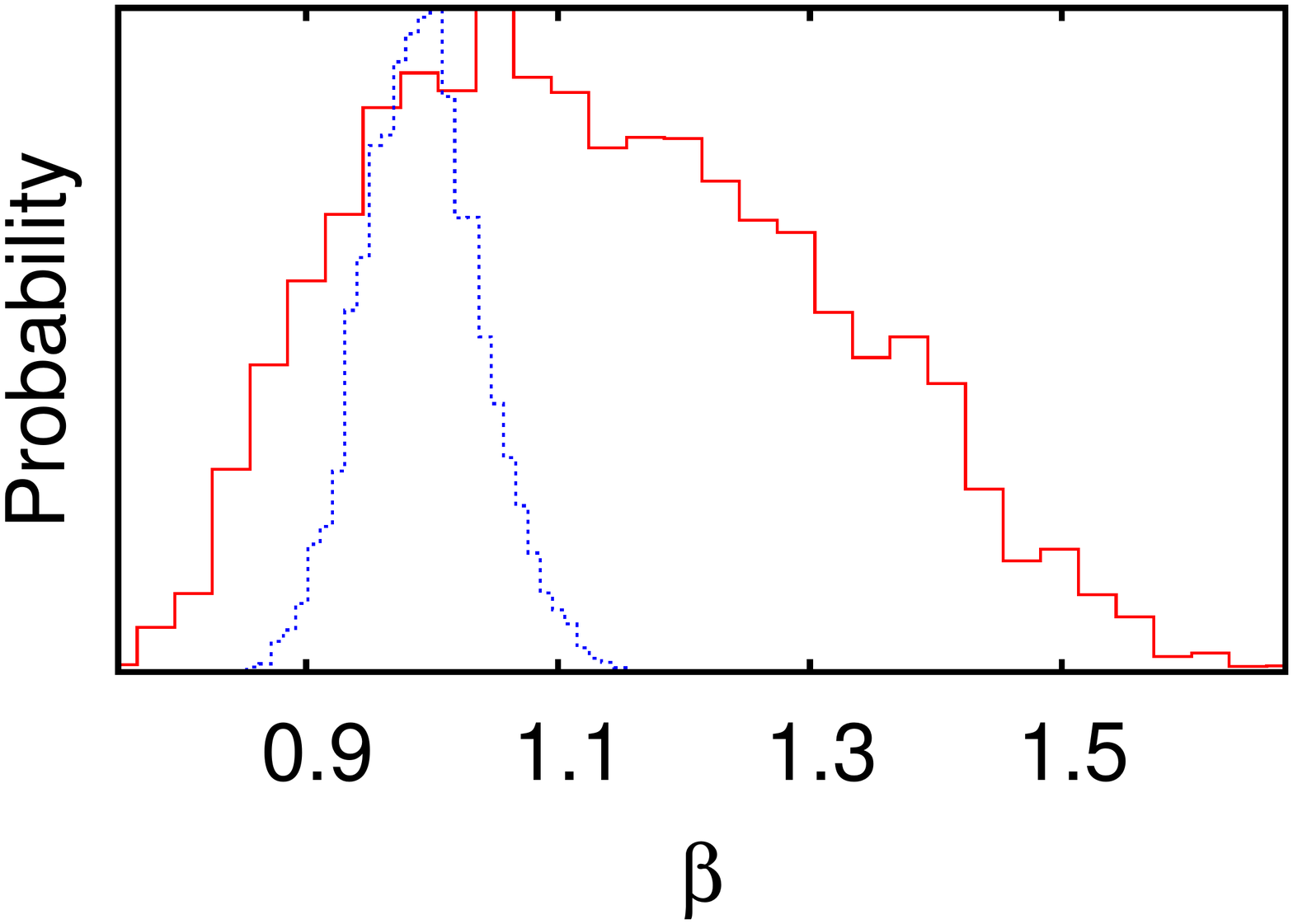} \\

\end{array}$
\end{center}
\vspace{-0.8cm}
\caption{1-dimensional marginalised posteriors for the timing model and $\beta$ parameters in simulated dataset 2 for the isolated pulsar PSR J0030+0451 for model 1 (red solid line) and model 2 (blue dotted line).  Values on the $x$-axes for the timing model parameters are given in terms of the standard deviation in that parameter returned by the analysis when including the additional terms, with the injected parameter value at 0 in all cases. Here the simulated noise was Gaussian in nature, and so with the exception of the phase offset, the parameter estimates for the timing model parameters using the two models are identical.  The disparity in the phase offset and $\beta$ parameters is discussed in Section \ref{Section:sim2}. \label{figure:Sim2Comp}}
\end{figure*}

As for simulation 1 we list the mean parameter estimates and standard deviations using models 1 and 2 for simulation 2 in Table \ref{Table:Sim1}.  Similarly Fig. \ref{figure:Sim2Comp} shows the one-dimensional marginalised posteriors for the timing model parameters and the white noise scaling parameter $\beta$ for model 1 (red solid lines) and model 2 (blue dotted lines) respectively.  In this simulation, with the exception of the phase offset, the posteriors for the timing model parameters are identical for the two models, as could be expected as the noise is now Gaussian in nature.  The phase offset however is now totally unconstrained across the prior in model 1, and the uncertainties in the scaling parameter $\beta$ are a factor $\sim$ 4 greater than in model 2.

\begin{figure*}
\begin{center}$
\begin{array}{c}
%
\includegraphics[width=150mm]{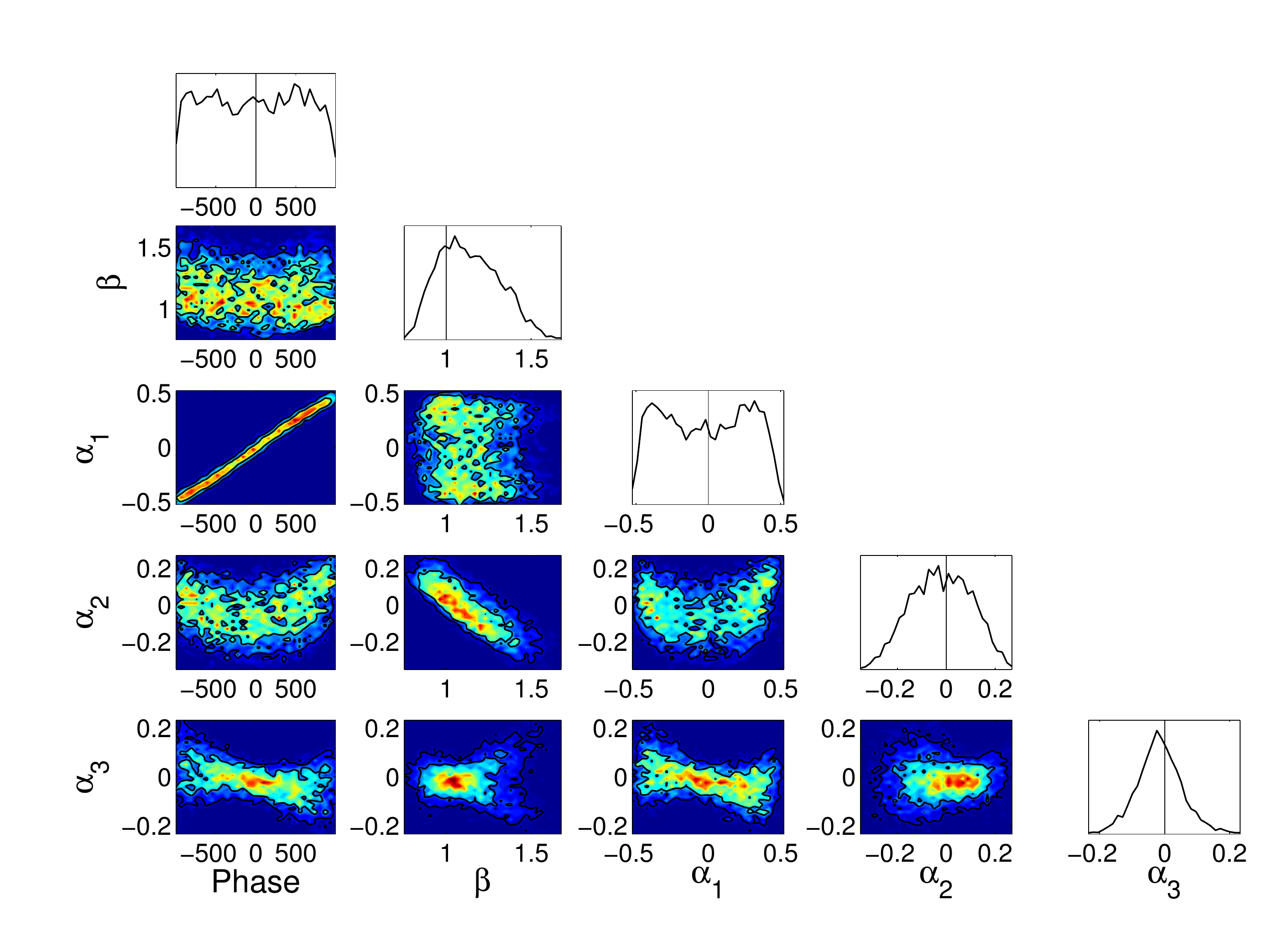} 
\end{array}$
\end{center}
\vspace{-0.8cm}
\caption{One and two-dimensional marginalised posterior distributions for the phase offset, $\beta$, and $\bmath{\alpha}$ parameters in simulation 2.  The phase offset and $\alpha_1$ parameters are in this instance completely correlated.  Qualitatively this simply represents that there is no difference between a dataset with a phase offset, and a dataset whose noise probability density is both symmetric and offset from zero.  Similarly $\beta$ and $\alpha_2$ show strong correlations, leading to the increased uncertainty in the parameter seen in Fig. \ref{figure:Sim2Comp}. \label{figure:Sim22D}}
\end{figure*}

We can understand this disparity by looking at the two-dimensional marginalised posteriors for the phase offset, $\beta$, and $\bmath{\alpha}$ parameters in Fig. \ref{figure:Sim22D}.  Here it becomes clear that the phase offset and $\alpha_1$ parameter are in this instance completely correlated.  Qualitatively this simply represents that there is no difference between a dataset with a phase offset, and a dataset whose noise probability density is both symmetric and offset from zero.  We also see that there is a strong correlation between the $\beta$ scaling parameter and $\alpha_2$ leading to the increased uncertainties in the former relative to model 2.

Comparing the evidence between the two models we find that, as expected, model 2 is favoured with $\Delta \log E = 4$ providing strong support for the simpler model.

\begin{table*}
\caption{Parameter estimates for the two simulations of PSR J0030+0451. Figures in parentheses represent one standard deviation in the least-significant digits quoted.}
\begin{tabular}{lccc}
\hline\hline
\multicolumn{4}{c}{Simulation 1} \\ 
\hline
Model Parameter & Simulation & Model 1 & Model 2 \\
\hline
$\log$ Evidence\dotfill		& - 		& 3052.9 & 3035.0 \\
Right ascension, $\alpha$\dotfill &  00:30:27.4299630 & 00:30:27.429956(7) & 00:30:27.429964(14)\\ 
Declination, $\delta$\dotfill & +04:51:39.75230 & +04:51:39.7525(3) & +04:51:39.7522(5)\\ 
Pulse frequency, $\nu$ (s$^{-1}$)\dotfill  & 205.53069608827310 &  205.53069608827315(6) & 205.53069608827300(12) \\ 
First derivative of pulse frequency, $\dot{\nu}$ (s$^{-2}$)\dotfill & $-$1.3061388e-16 & $-$1.306139(5)$\times 10^{-16}$ & $-$1.306149e-16(9)$\times 10^{-16}$\\ 
Proper motion in right ascension, $\mu_{\alpha}$ (mas\,yr$^{-1}$)\dotfill & $-$4.054 & $-$4.09(2) & $-$4.05(4)\\ 
Proper motion in declination, $\mu_{\delta}$ (mas\,yr$^{-1}$)\dotfill & $-$5.03 & $-$4.94(5) & $-$5.0(1)\\ 
Parallax, $\pi$ (mas)\dotfill & 4.023 & 4.037(14) & 4.02(3)\\ 
$\beta$\dotfill & 1 & 1.00(4) & 1.62(8)\\
$\alpha_1$\dotfill & 0.1 & 0.13(5) & -\\
$\alpha_2$\dotfill & 0.1 & 0.15(5) & -\\
$\alpha_3$\dotfill & 0.4 & 0.41(4) & -\\
\hline
\hline
\multicolumn{4}{c}{Simulation 2} \\ 
\hline
Model Parameter & Simulation & Model 1 & Model 2 \\
\hline
$\log$ Evidence\dotfill		& - 		& 3131.5 & 3136.4 \\
Right ascension, $\alpha$\dotfill &  00:30:27.4299630 & 00:30:27.429964(9) & 00:30:27.429964(9)\\ 
Declination, $\delta$\dotfill & +04:51:39.75230 & +04:51:39.7523(3) & +04:51:39.7523(3)\\ 
Pulse frequency, $\nu$ (s$^{-1}$)\dotfill  & 205.53069608827310 &  205.53069608827304(7) & 205.53069608827304(7) \\ 
First derivative of pulse frequency, $\dot{\nu}$ (s$^{-2}$)\dotfill & $-$1.3061388e-16 & $-$1.306146e-16(6)$\times 10^{-16}$ & $-$1.306146e-16(6)$\times 10^{-16}$\\ 
Proper motion in right ascension, $\mu_{\alpha}$ (mas\,yr$^{-1}$)\dotfill & $-$4.054 & $-$4.06(3) & $-$4.06(3)\\ 
Proper motion in declination, $\mu_{\delta}$ (mas\,yr$^{-1}$)\dotfill & $-$5.03 & $-$5.01(6) & $-$5.01(6)\\ 
Parallax, $\pi$ (mas)\dotfill & 4.023 & 4.037(14) & 4.025(16)\\ 
$\beta$\dotfill & 1 & 1.1(2) & 0.99(5)\\
$\alpha_1$\dotfill & 0 & $-$0.0(3) & -\\
$\alpha_2$\dotfill & 0 & $-$0.02(12) & -\\
$\alpha_3$\dotfill & 0 & $-$0.01(7) & -\\
\hline
\hline
\end{tabular}
\label{Table:Sim1}
\end{table*}

\section{Application to real data}
\label{Section:RealData}

We now apply the likelihood developed in section \ref{Section:PulsarNonGaussian} to the publicly available Parkes Pulsar Timing Array (PPTA) data release 1 dataset for the binary pulsar J0437$-$4715 \citep{2013PASA...30...17M}.  In brief, this dataset spans 15 years of observations, with approximately the first nine consisting of only single frequency, 20cm observations, and the remainder including additional 10cm and 50cm observations.  The timing model we fit includes the 21 parameters listed in Table \ref{Table:0437} in addition to a set of 12 "jumps" (offsets between different observing systems).  We then simultaneously fit for a power law red noise process and power law dispersion measure variations, including periods from the length of the dataset $T$, down to one month, at intervals of $1/T$ as described in sections \ref{Section:Red} and \ref{Section:DM} respectively.  Finally we also include an EFAC and EQUAD parameter for each observing system group. In order to explore the effects of potential non-Gaussianity in the dataset we then consider two cases.  Firstly, that both the radiometer noise term ($\beta\sigma$) and quadrature term ($\gamma$) describe Gaussian random processes, and secondly, that while the radiometer noise is Gaussian, the quadrature term is non-Gaussian.  In the latter case we parameterise this non-Gaussianity as in Section \ref{Section:White1} where we fit for the convolved PDF of the two distributions, rather than sampling from the $j$ parameters numerically.

\begin{figure*}
\centering
\subfloat{\includegraphics[width=100mm]{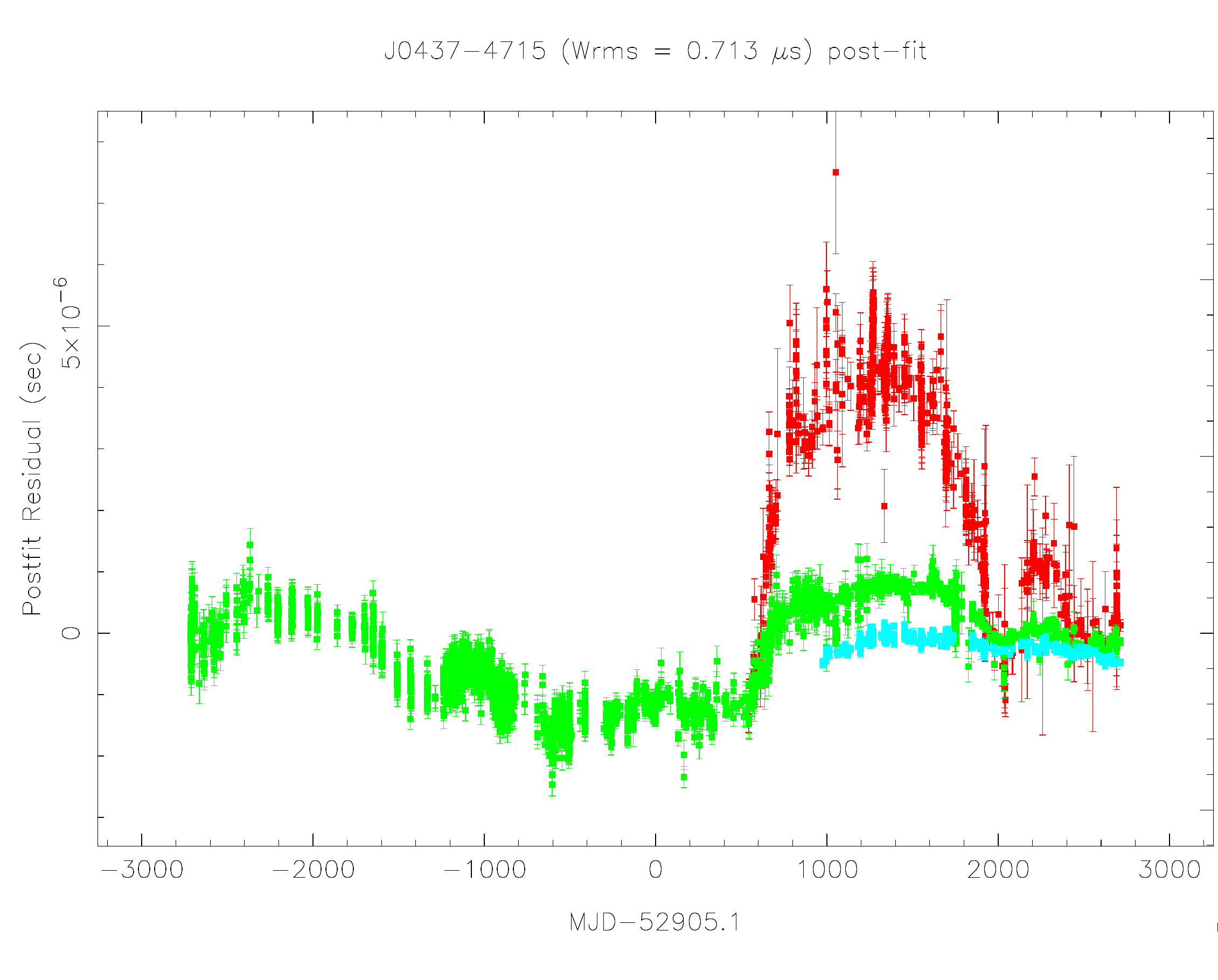}}
\\
\subfloat{\includegraphics[width=100mm]{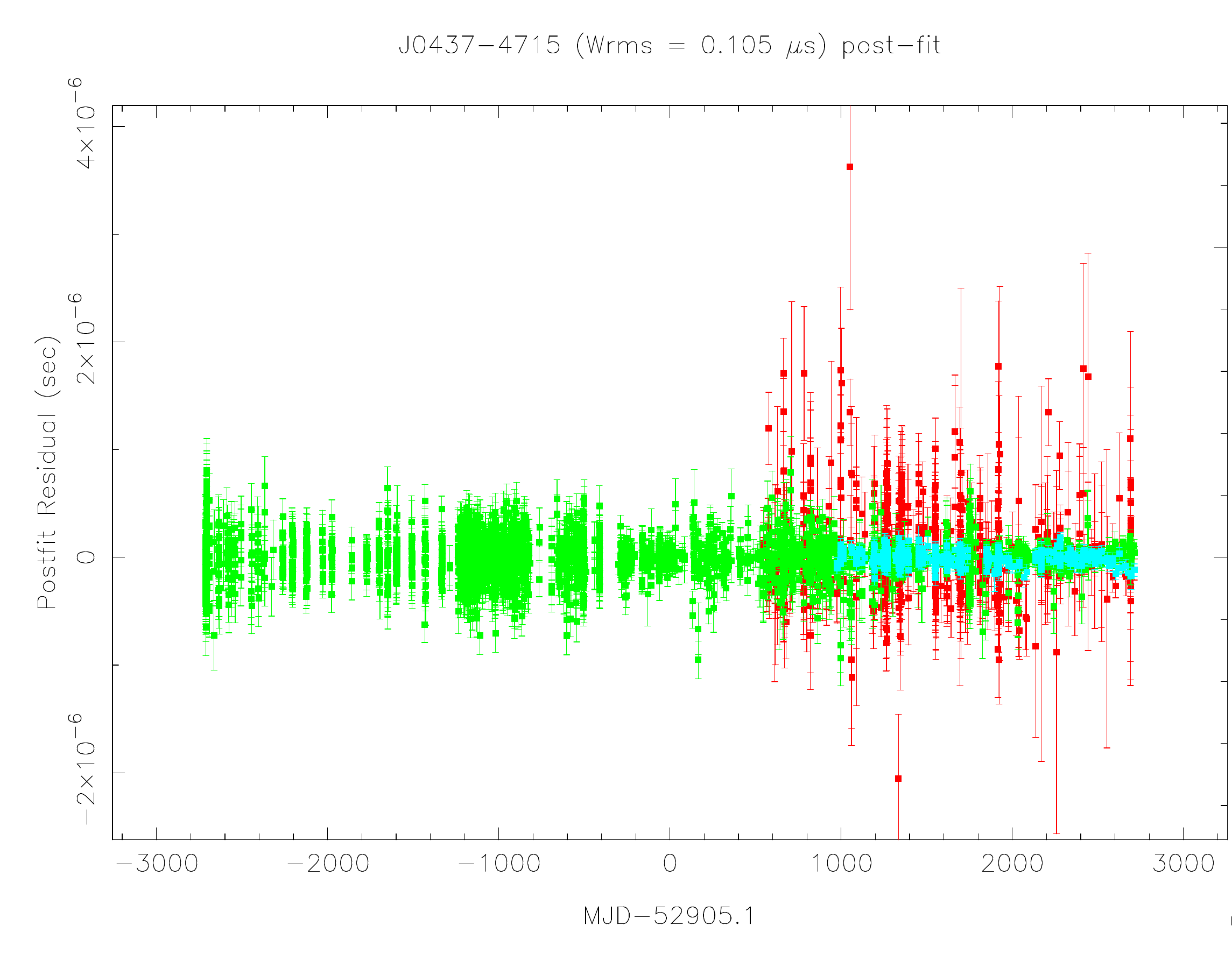}}
\\
\vspace{-0.8 cm}  
\subfloat{\includegraphics[width=80mm]{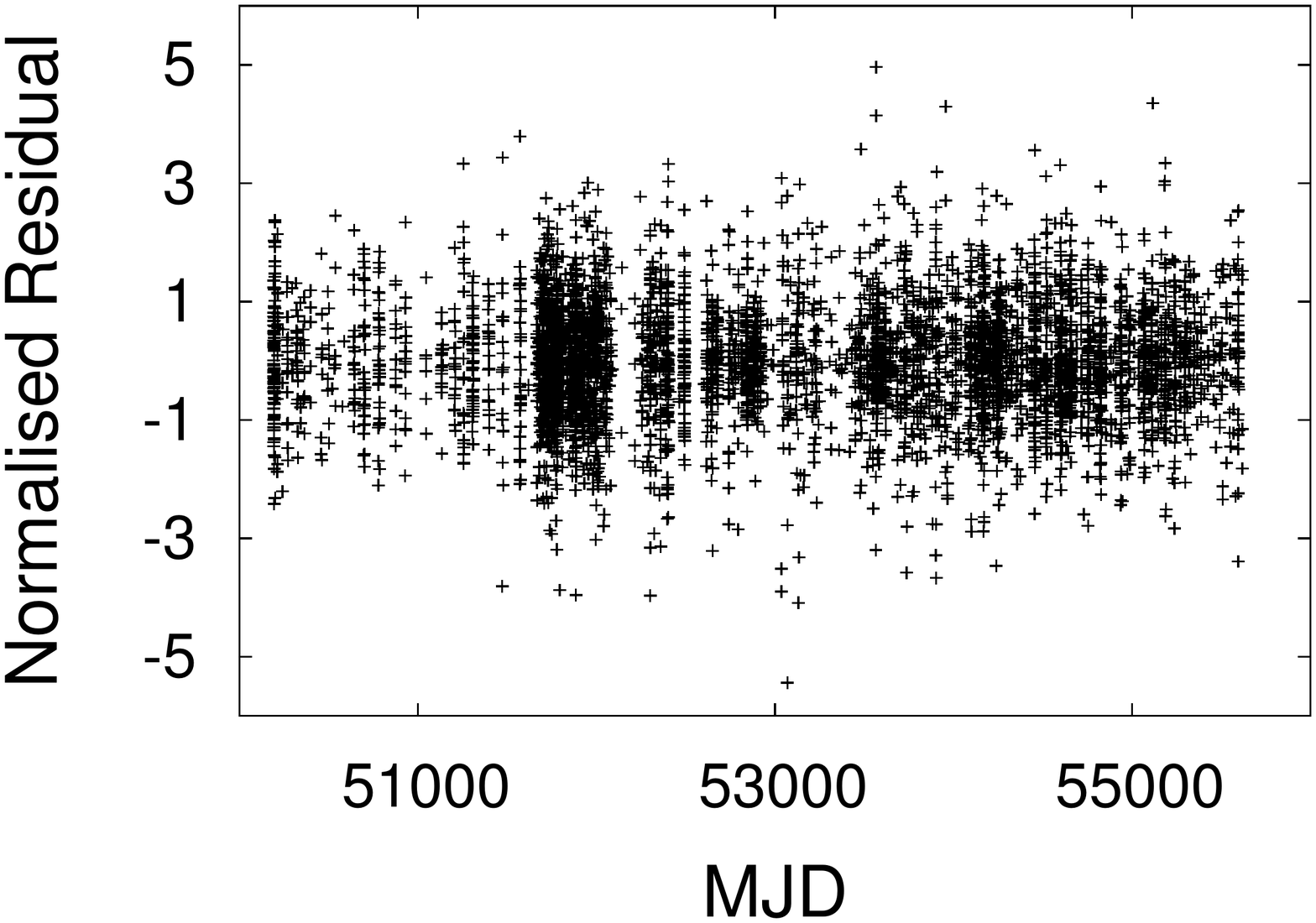}}\,
\subfloat{\includegraphics[width=80mm]{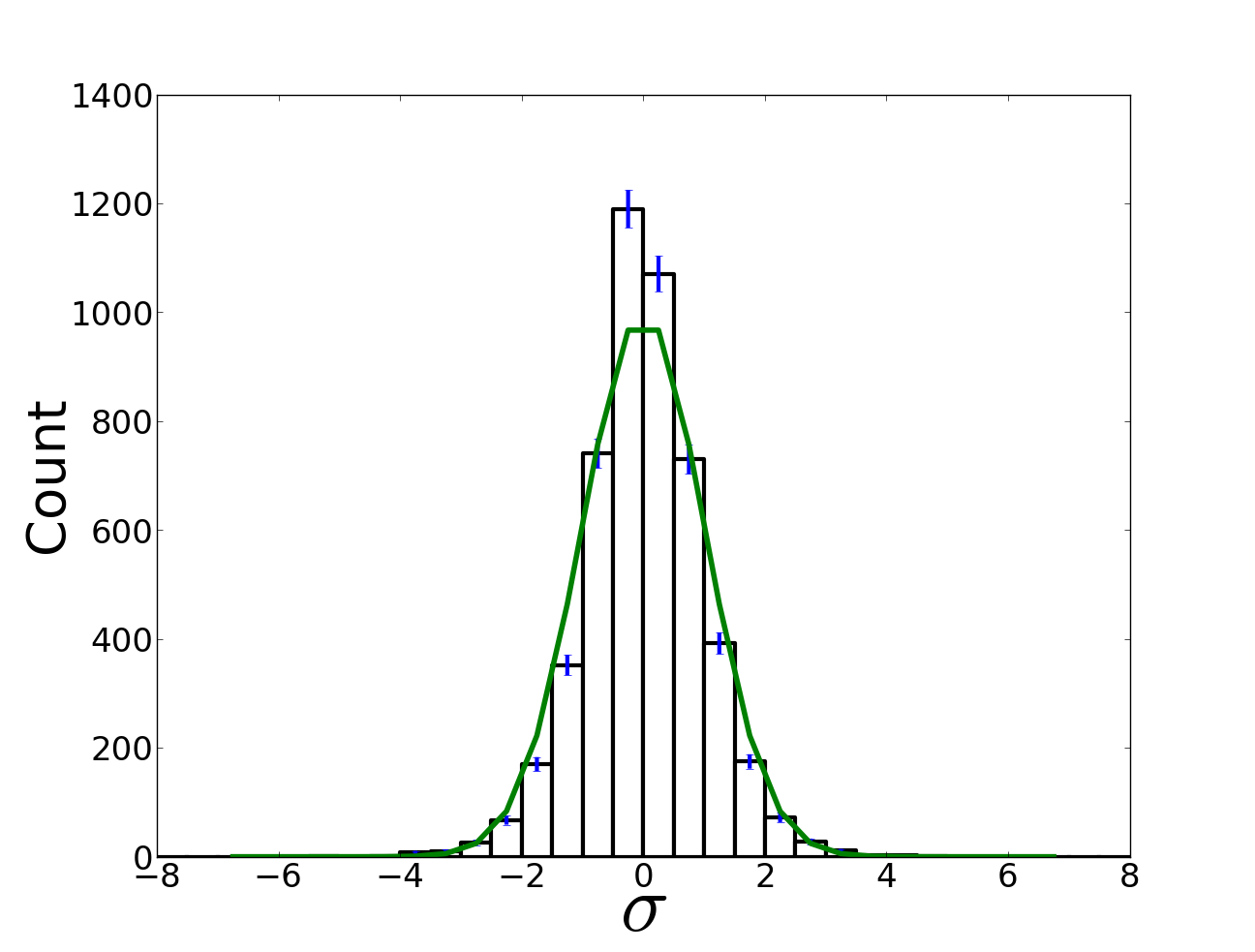}}
\vspace{-0.3 cm}  
\caption{Residuals for the publicly available PPTA data release 1 dataset for the binary pulsar J0437$-$4715 after subtracting the maximum likelihood Gaussian timing solution given in Table \ref{Table:0437} (top), and additionally after subtracting the maximum likelihood red noise and dispersion measure variations (middle).  Colours indicate 10cm (blue), 20cm (green) and 50cm (red) observing wavelengths.  In the bottom left panel we show the normalised residuals after dividing each point by its error bar, and in the bottom right panel we show a histogram of these normalised residuals, overlayed with a unit Gaussian to show the expected number counts in each bin. Error bars for each histogram bin are given by $\sqrt{N}$ with $N$ the number of points in the bin. }
\label{figure:0437Res}
\end{figure*}

Residuals for this dataset after subtracting the maximum likelihood Gaussian timing solution given in Table \ref{Table:0437} are shown in Fig. \ref{figure:0437Res} (top), and after also subtracting the maximum likelihood red noise and dispersion measure variations (middle).  In the bottom left panel we show the normalised residuals after dividing each point by its error bar, and in the bottom right panel we show a histogram of these normalised residuals, overlayed with a unit Gaussian to show the expected number counts in each bin.

From the plot of the normalised residuals it is clear that there are a significant number of outliers ($> 4-5\sigma$) compared to the number expected from a Gaussian distribution.  In addition the histogram suggests that there is also an over abundance of points at small deviations ($< 0.5\sigma$).  As seen in Section \ref{Section:Pulsarsims} this behaviour is indicative of the white noise parameters in the Gaussian fit overestimating the error bars of the points in order to best accommodate outliers.

\subsection{A post fit evidence comparison}

As in section \ref{Section:Pulsarsims} we would like to compute the difference in the Evidence for the Gaussian and non-Gaussian models, however for this dataset the total dimensionality of the problem is $\sim$ 800, which makes the calculation of the evidence both extremely expensive computationally, and also much less precise numerically than in the lower dimensional simulation.

As such we consider that even in the presence of significant non-Gaussianity, the timing model parameter estimates obtained from the Gaussian analysis did not differ significantly from the non-Gaussian analysis, only their uncertainties changed.  As such, in order to obtain an approximate value for the evidence we can use the post-fit residuals, after subtracting the maximum likelihood timing model, red noise and dispersion measure variations and then fit only for the EFAC and EQUAD parameters in those residuals and an additional offset term.  This decreases the dimensionality to $\sim 30$, allowing us to use {\sc MultiNest} to compare the evidence for the Gaussian and non-Gaussian cases as before.  When including non-Gaussian coefficients in our model, both in this test and in the subsequent full analysis, we exclude the $\alpha_1$ term in order to minimise the covariance between the offset and the non-Gaussian parameters.

\begin{table}
\centering
\caption{Log evidence values for different numbers of non-Gaussian coefficients in a post fit analysis of the J0437$-$4715 residuals.} 
\centering 
\begin{tabular}{c c c} 
\hline\hline 
non-Gaussian Coefficients Included &  $\log$ Evidence\\[0.5ex] 
\hline 
0 & 0\\
$\alpha_{2..3}$ & 38.7\\
$\alpha_{2..4}$ & 47.4\\
$\alpha_{2..5}$ & 48.3\\
$\alpha_{2..6}$ & 48.2\\
\hline
\end{tabular}
\label{Table:logevidence} 
\end{table}

Table \ref{Table:logevidence} lists the log evidence values for different sets of non-Gaussian coefficients, normalised such that the log evidence for no additional coefficients (i.e. assuming Gaussian statistics) is 0.  We see that there is a significant increase in the log evidence ($\sim$ 39) when including even just two coefficients, indicating definitive support for their inclusion in the model.  As the number increases the rise in evidence increases, reaching a maximum with 4 included coefficients.  Given the timing model, red noise and dispersion measure variation solutions that were subtracted from the data were obtained from a Gaussian analysis, we will however still include coefficients up to and including $\alpha_6$ in the full analysis.

\subsection{Joint Bayesian analysis}

Given the large dimensionality of the problem this analysis cannot be carried out using {\sc MultiNest}.  As such we make use of the 'Guided Hamiltonian Sampler' used previously in pulsar timing analysis in \citep{2013PhRvD..87j4021L}.  This sampler makes use of both gradient information in the likelihood, and also the hessian in order to efficiently sample from large parameter spaces.

\begin{figure*}
\begin{center}$
\begin{array}{cc}
\hspace{-0.5cm}
\includegraphics[width=100mm]{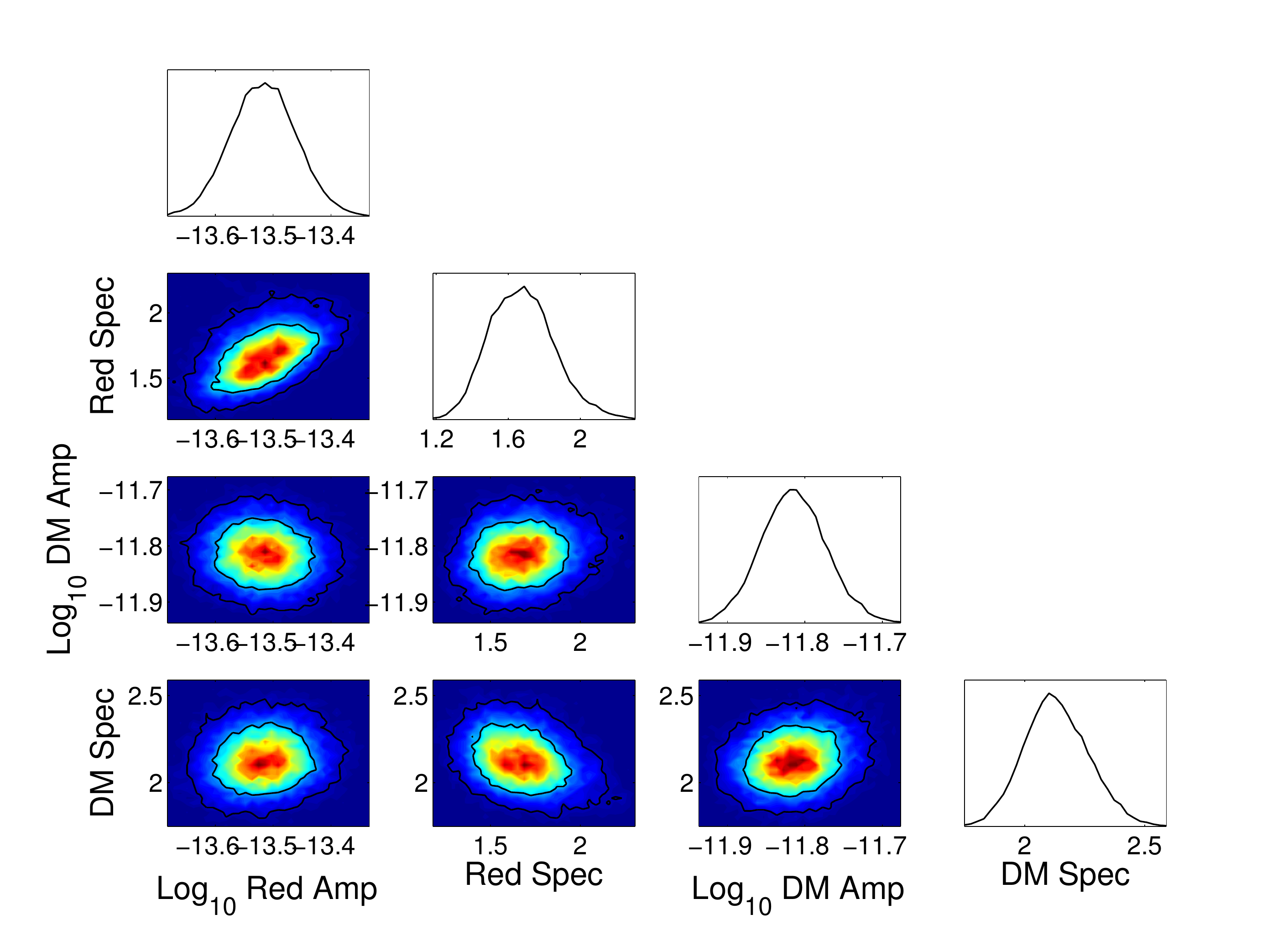} &
\includegraphics[width=100mm]{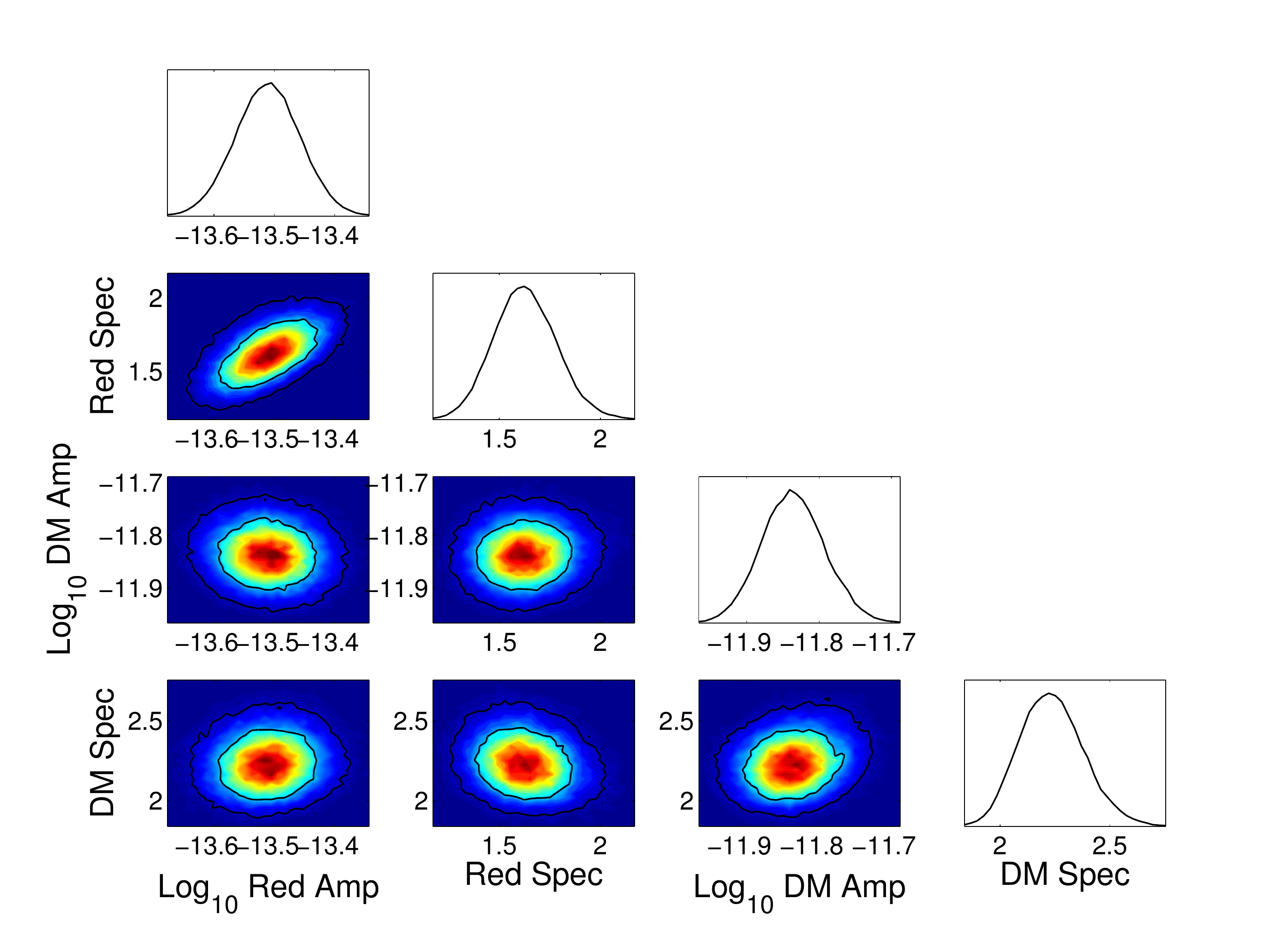}  \\
\hspace{-1.5cm}
\includegraphics[width=100mm]{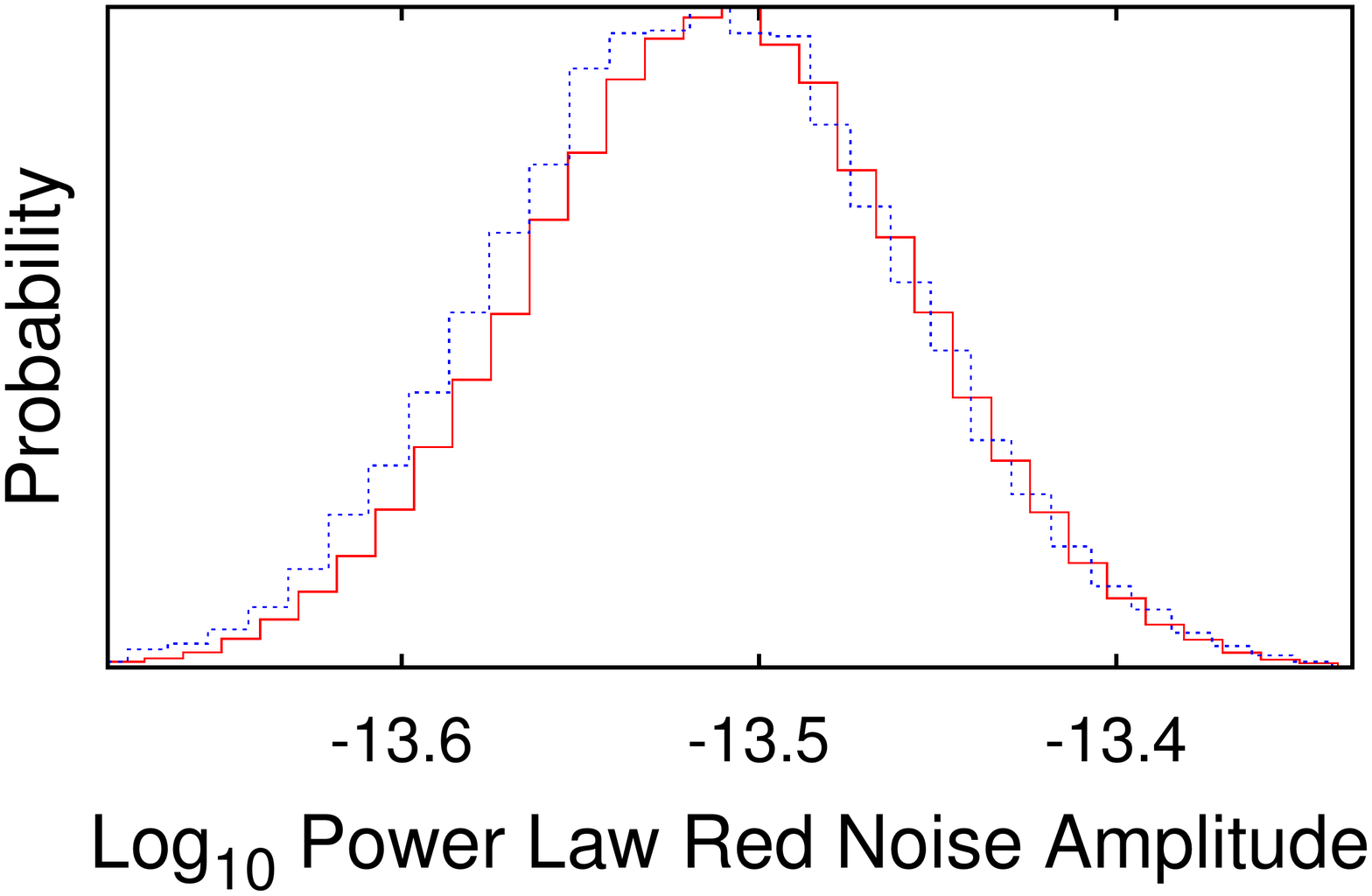} &
\hspace{-1.5cm}
\includegraphics[width=100mm]{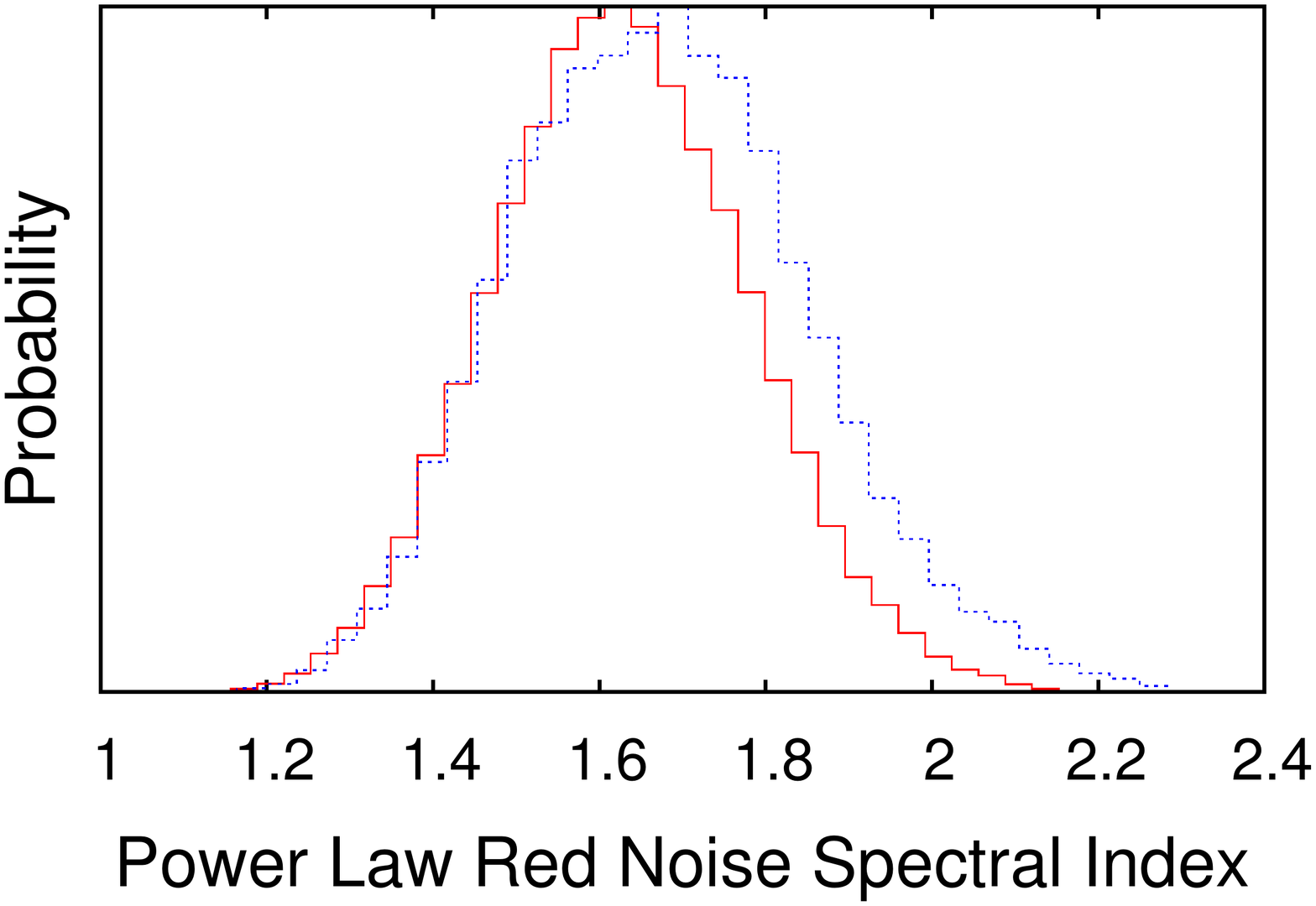}  \\
\hspace{-1.5cm}
\includegraphics[width=100mm]{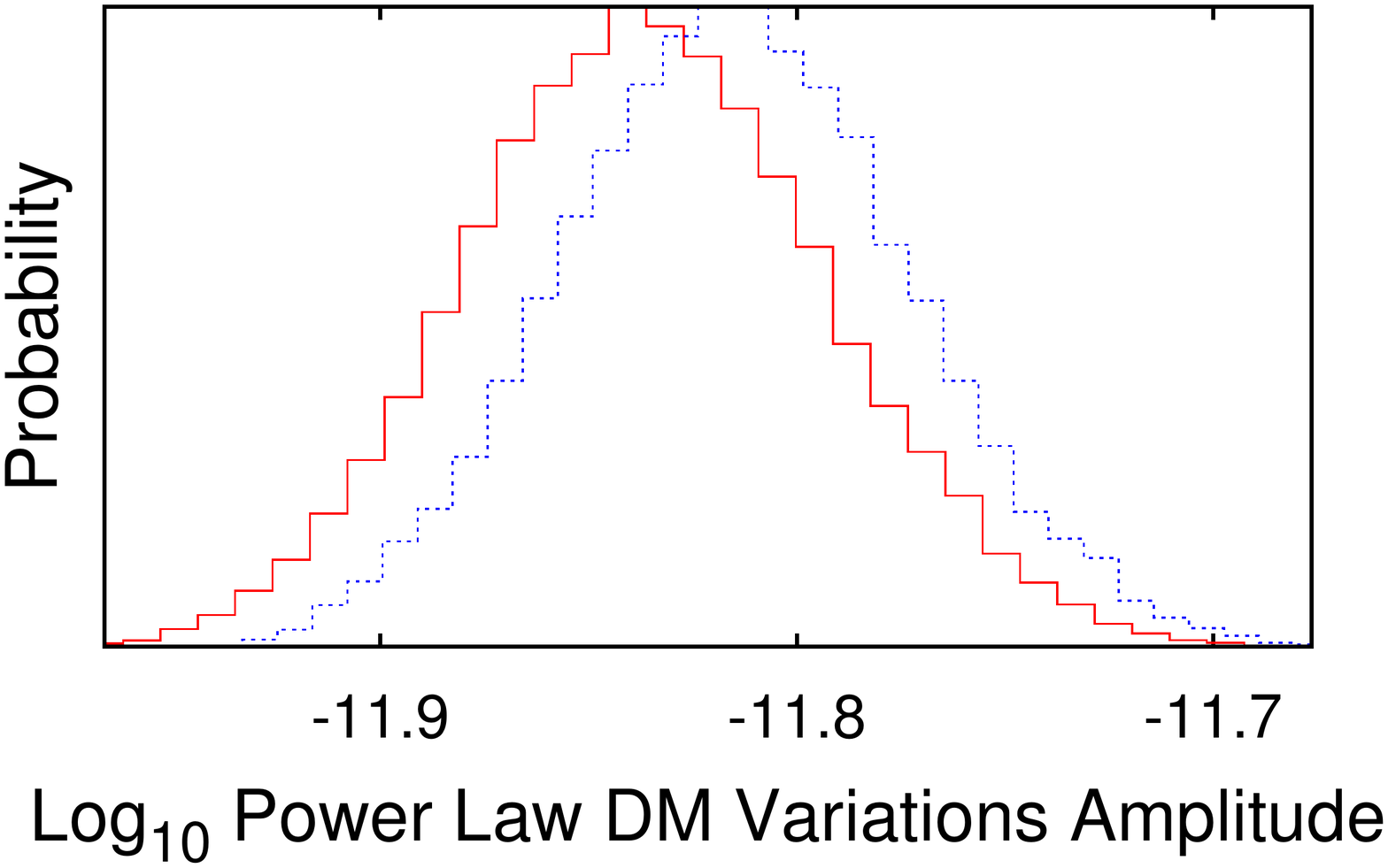} &
\hspace{-1.5cm}
\includegraphics[width=100mm]{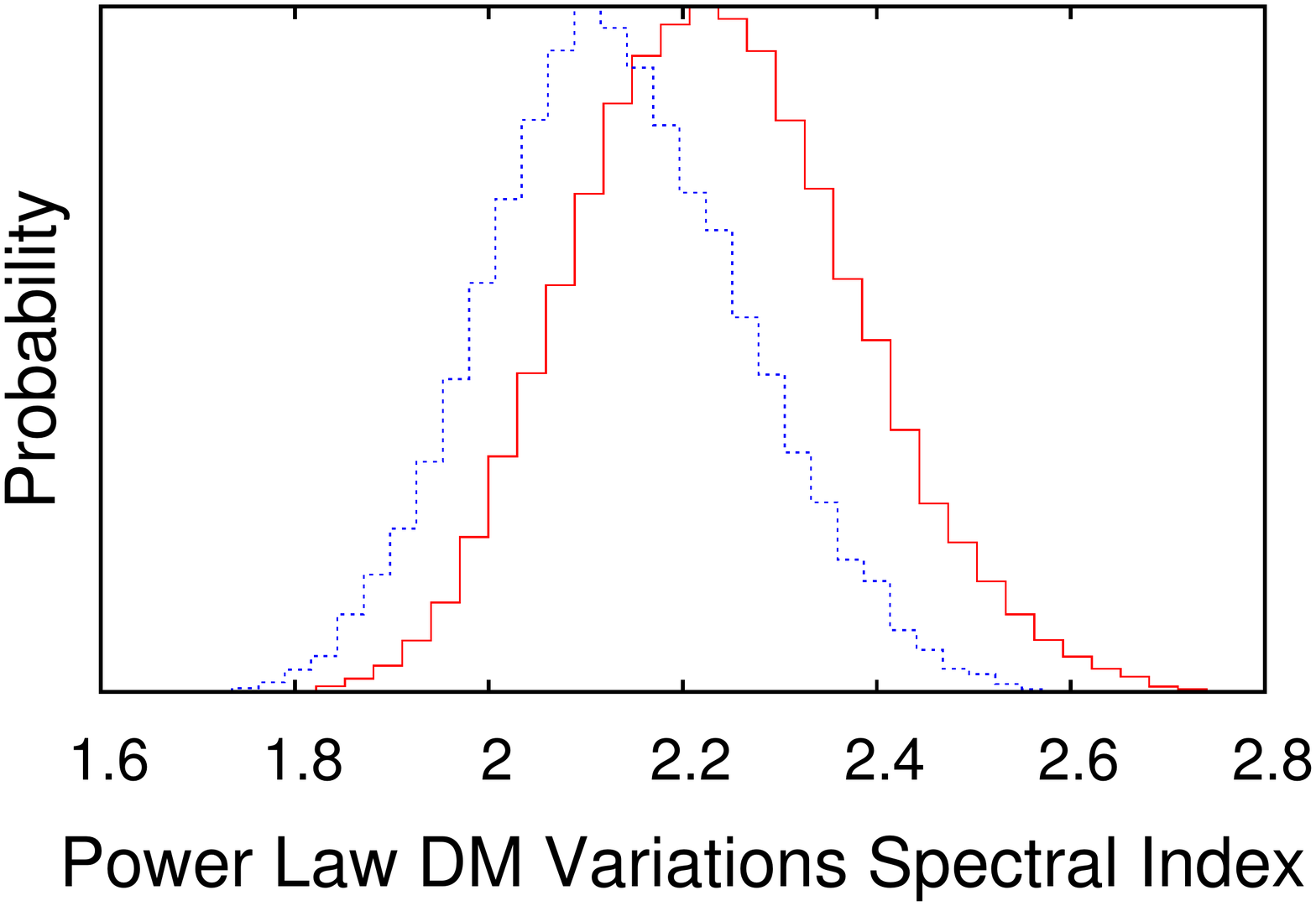}  \\
\vspace{-1.5cm}
\end{array}$
\end{center}
\caption{(Top) One and two-dimensional marginalised posterior distributions for the red noise and dispersion measure variation power law amplitudes and spectral indicies for the non-Gaussian (left) and Gaussian (right) analysis.  Both are extremely consistent with one another, however when overlaying the two sets of 1-dimensional posterior distributions for each of the 4 parameters separately (bottom 4 panels)  some differences become apparent between the non-Gaussian (blue dashed lines) and Gaussian (red solid lines) analysis.  In particular the dispersion measure variation power law parameter estimates show a slight shift towards higher amplitudes and shallower spectral indices in the non-Gaussian case.  \label{figure:0437Noise}}
\end{figure*}

Table \ref{Table:0437} lists the timing model parameter estimates and their nominal standard deviations for both the Gaussian and non-Gaussian analysis. In all cases we find the parameter estimates and their uncertainties to be consistent between both methods.  In Fig. \ref{figure:0437Noise} (top) we show the one and two-dimensional marginalised posterior distributions for the red noise and dispersion measure variation power law amplitudes and spectral indicies for the non-Gaussian (left) and Gaussian (right) analysis.  Both are also extremely consistent with one another, however when overlaying the two sets of 1-dimensional posterior distributions for each of the 4 parameters separately (bottom 4 panels)  some differences become apparent between the non-Gaussian (blue dashed lines) and Gaussian (red solid lines) analysis.  In particular the dispersion measure variation power law parameter estimates show a slight shift towards higher amplitudes and shallower spectral indices in the non-Gaussian case.

\begin{figure*}
\begin{center}$
\begin{array}{c}
%
\includegraphics[width=150mm]{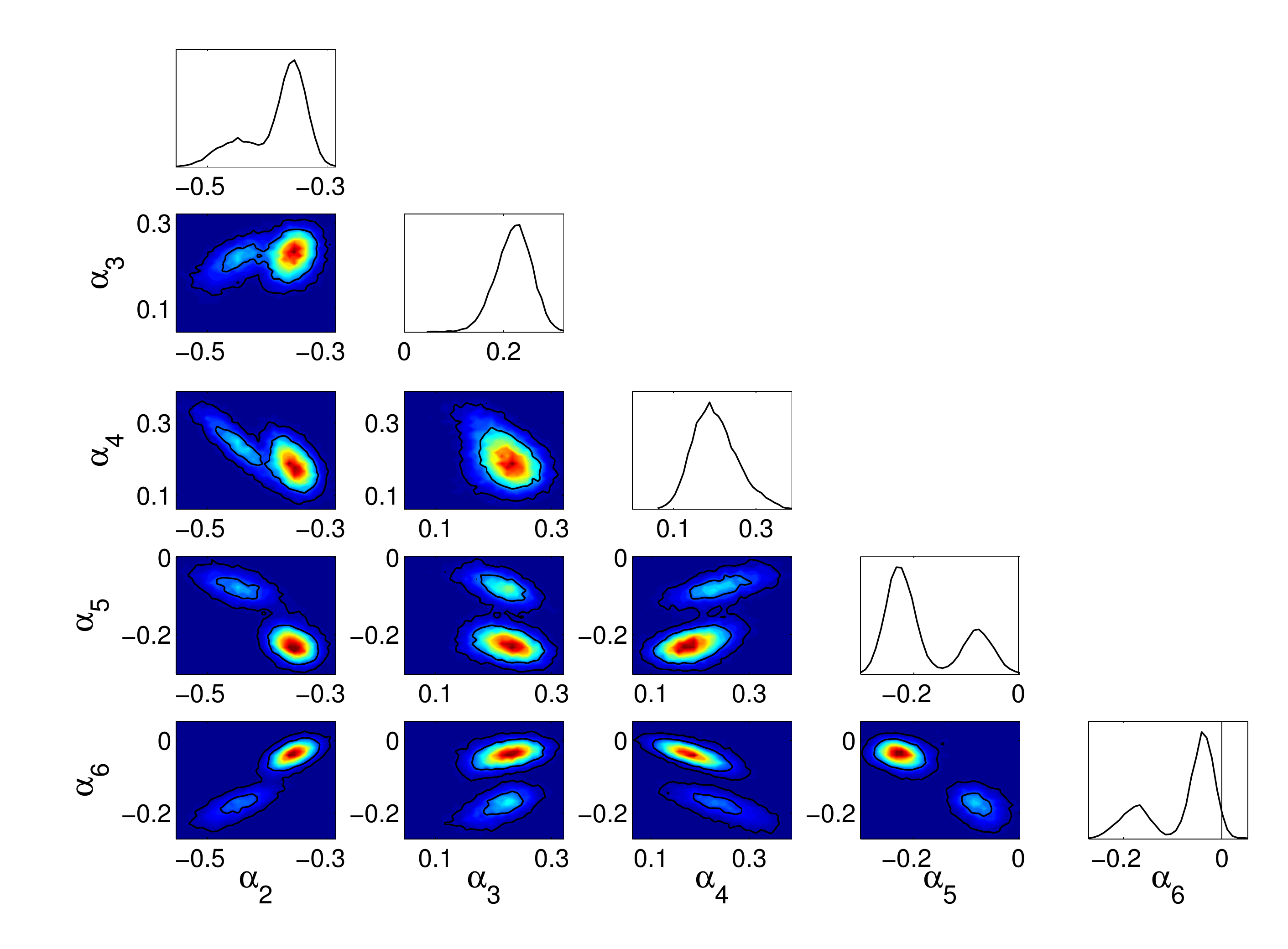} \\
\includegraphics[width=120mm]{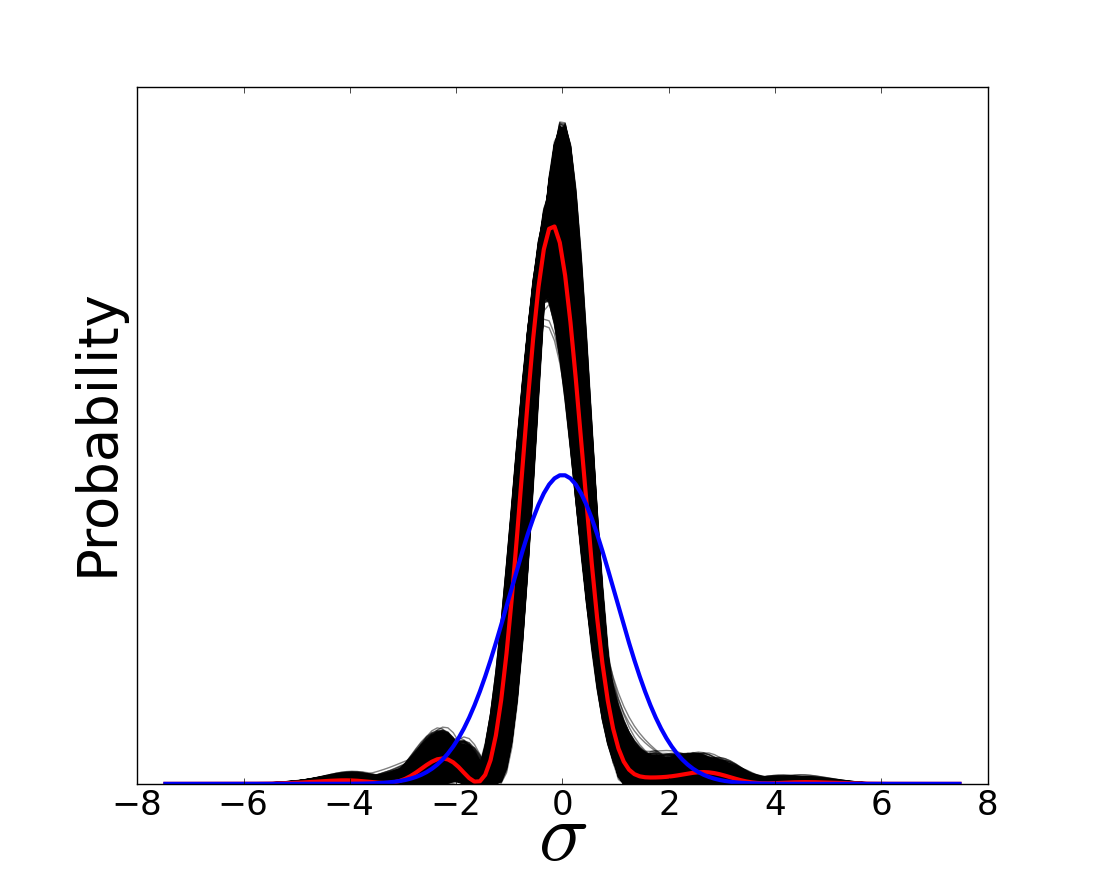}  \\
\end{array}$
\end{center}
\caption{(Top) One and two-dimensional marginalised posterior distributions for the 5 non-Gaussian coefficients fit in the analysis of J0437$-$4715.  Vertical lines are included at 0 where visible in the plot, however, except for $\alpha_5$ all the coefficients are inconsistent with this value suggesting a definitive detection of non-Gaussianity in the dataset. (Bottom) The set of equally weighted probability density functions obtained from the non-Gaussian analysis of J0437$-$4715 (black lines) with $\gamma = 1$.  In addition we show the mean of the distribution (red line) and a unit Gaussian (blue line) all normalised to have a sum of 1.  The difference between the Gaussian and fitted non-Gaussian functions is clear, with a larger probability for both small ($|\sigma| < 1$) and larger ($|\sigma| > 4$) deviations than given by the Gaussian probability function. \label{figure:0437PDF}}
\end{figure*}

\begin{figure*}
\begin{center}$
\begin{array}{cc}
\includegraphics[width=80mm]{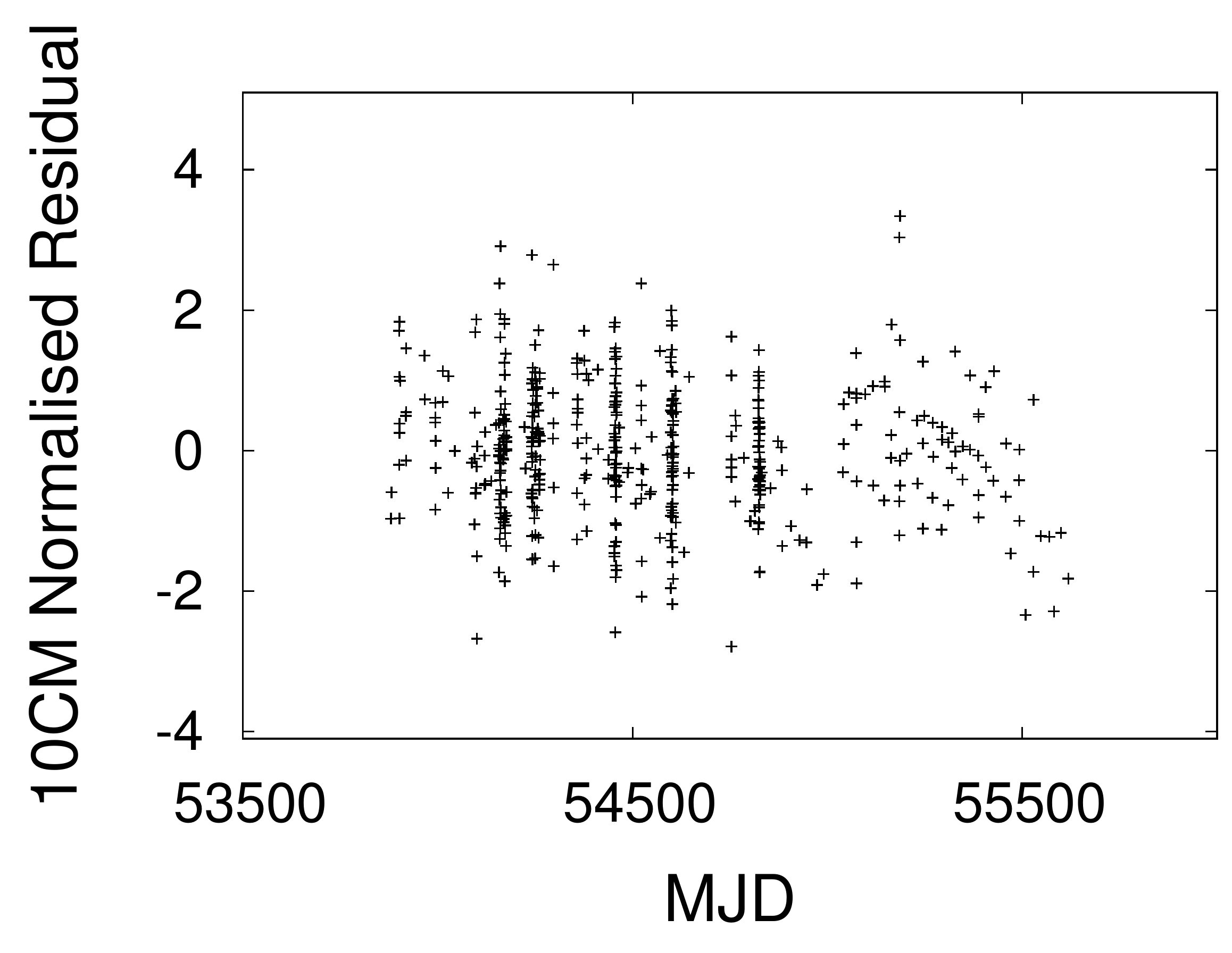} &
\includegraphics[width=85mm, height=65mm]{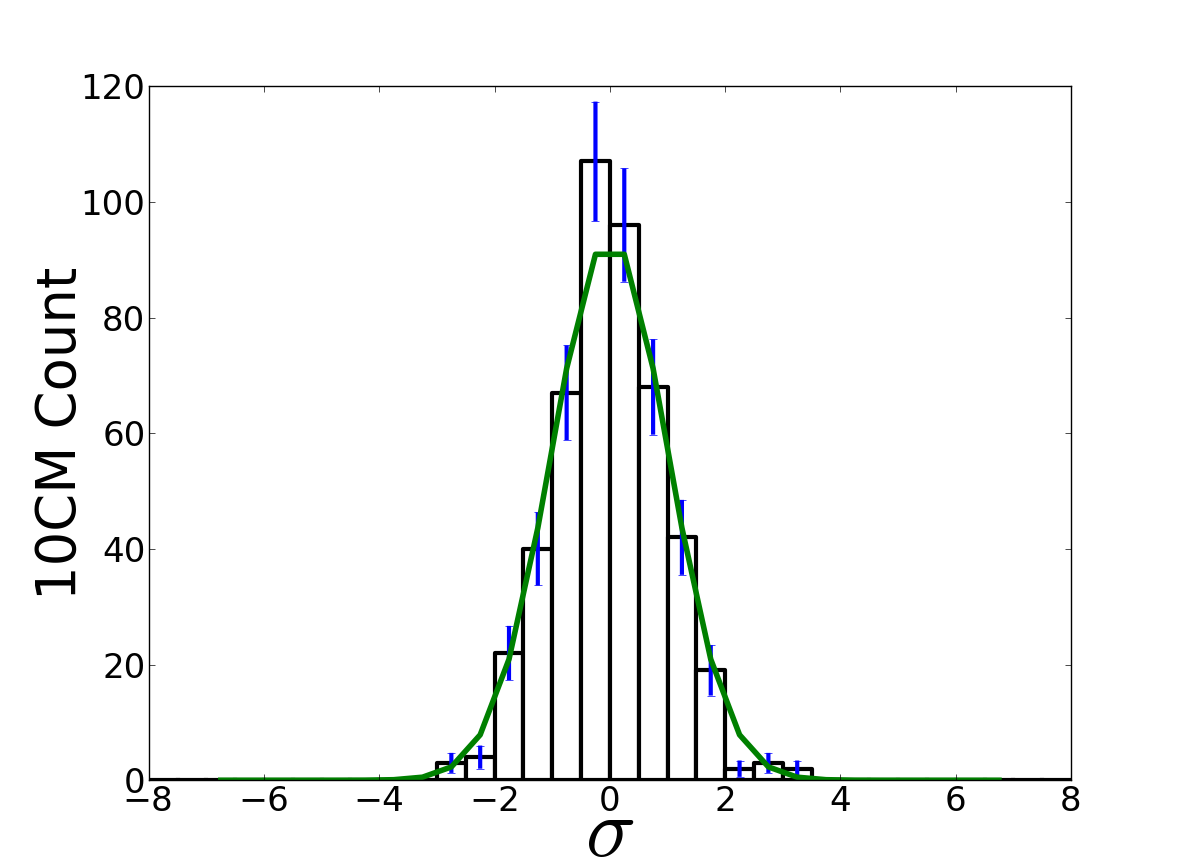} \\
\includegraphics[width=80mm]{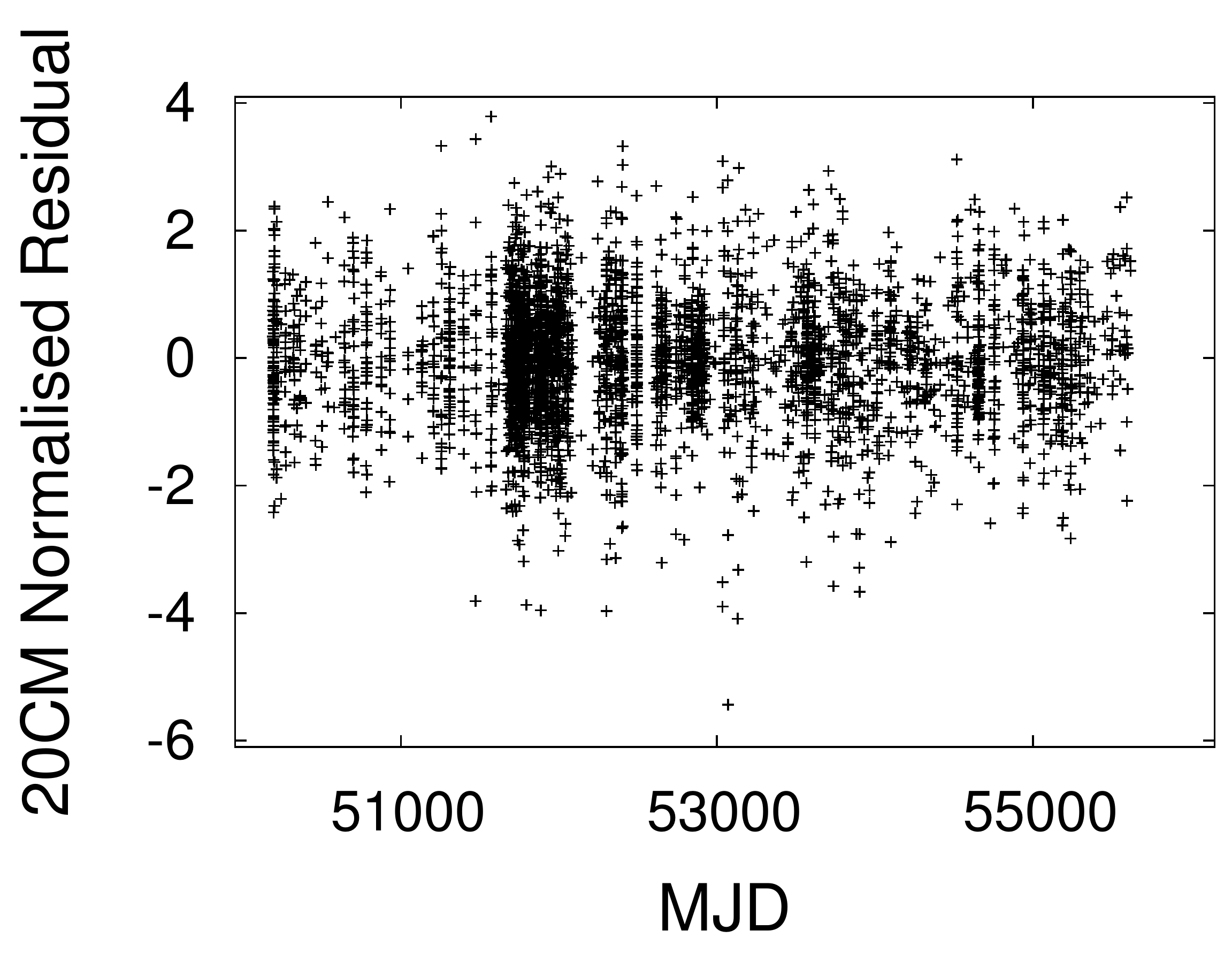}  &
\includegraphics[width=85mm, height=65mm]{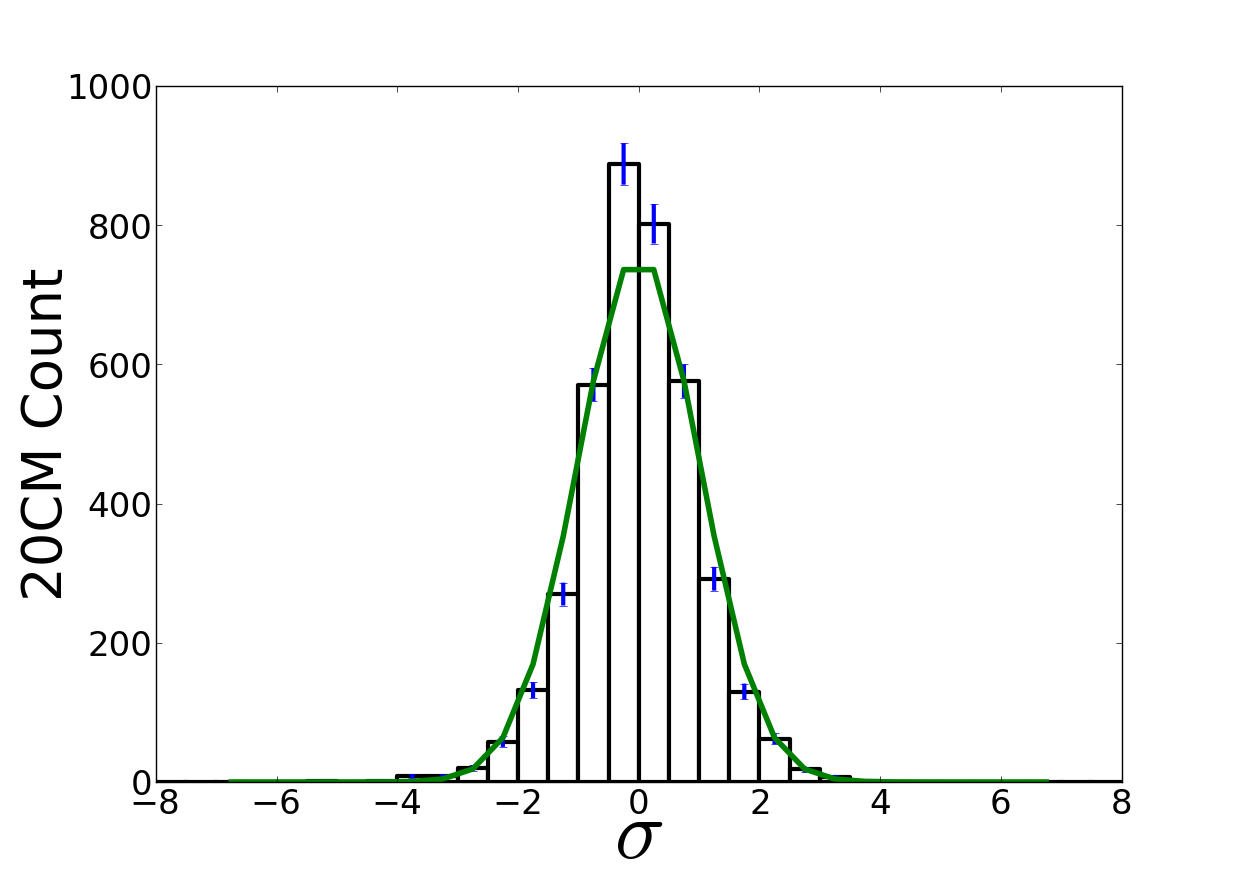}  \\
\includegraphics[width=80mm]{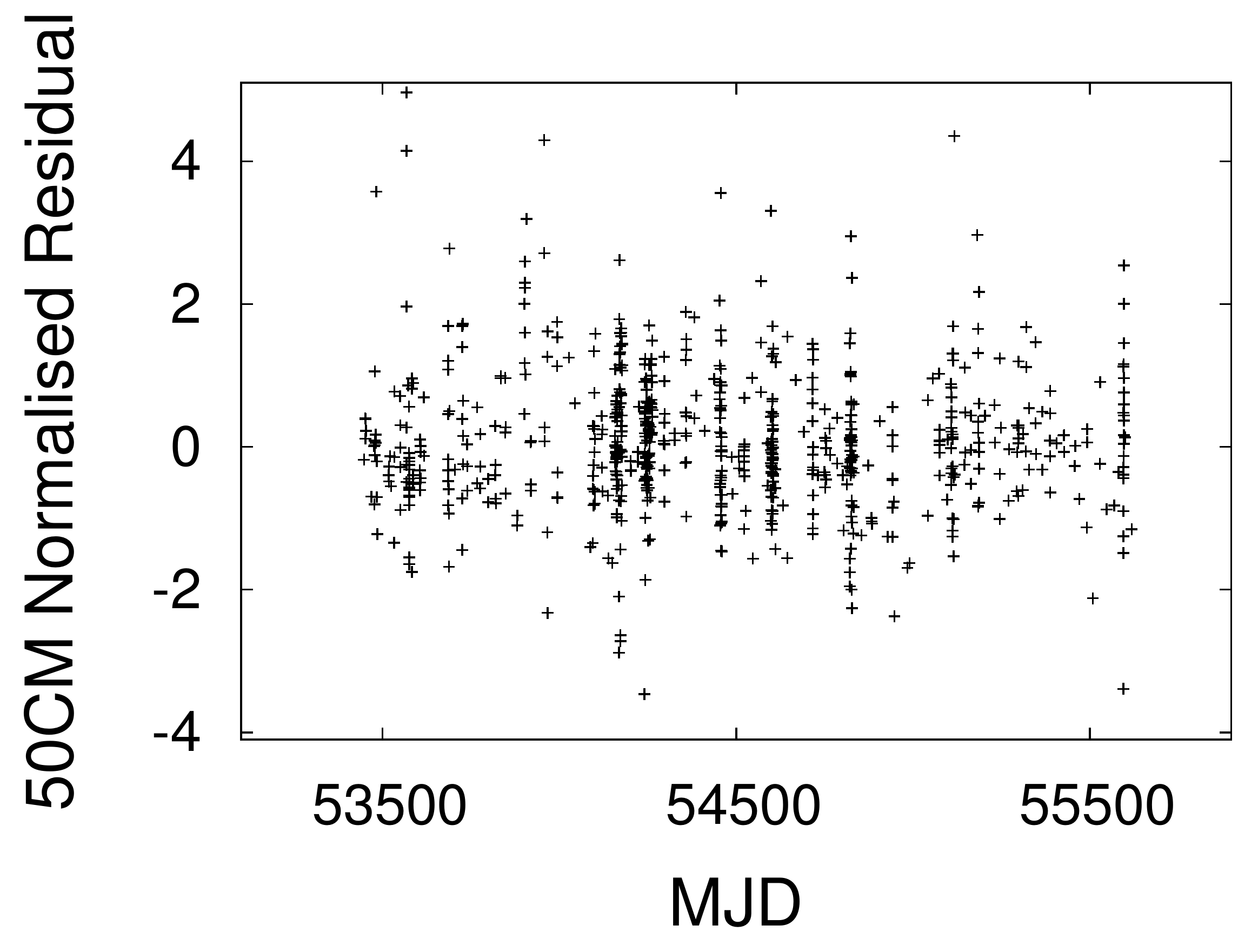}  &
\includegraphics[width=85mm, height=65mm]{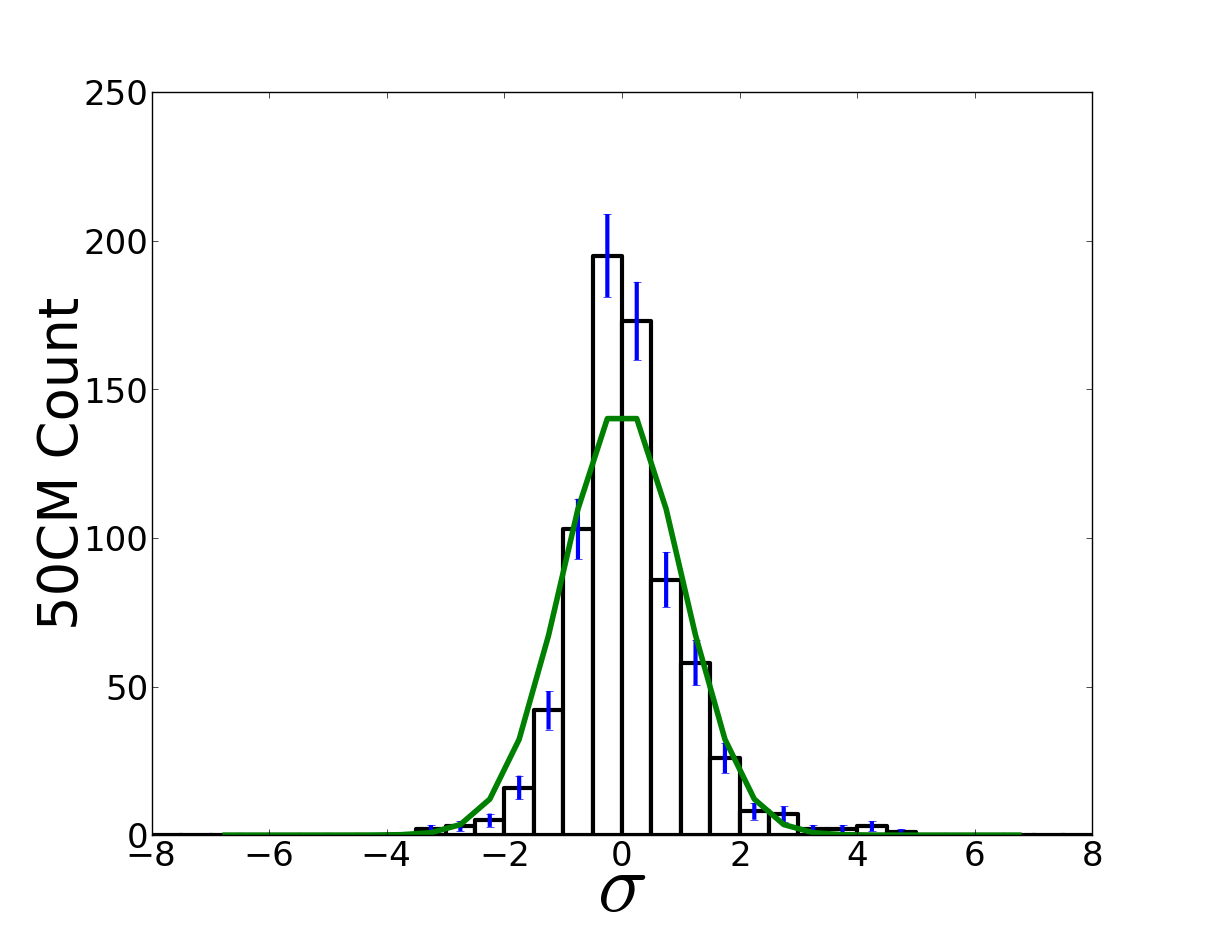}  \\
\end{array}$
\end{center}
\caption{(Left) Normalised Residuals for the publicly available PPTA data release 1 dataset for the binary pulsar J0437$-$4715 after subtracting the maximum likelihood Gaussian timing solution given in Table \ref{Table:0437} and the maximum likelihood red noise and dispersion measure variations for 10CM (top), 20CM (middle), and 50CM (bottom) datapoints. (Right)  Histogram of the normalised residuals for 10CM (top), 20CM (middle), and 50CM (bottom) datapoints, overlayed with a unit Gaussian to show the expected number counts in each bin. Error bars for each histogram bin are given by $\sqrt{N}$ with $N$ the number of points in the bin. \label{figure:0437NormRes}}
\end{figure*}

Despite these similarities in the timing and stochastic parameter estimates between the Gaussian and non-Gaussian analysis, Fig. \ref{figure:0437PDF} indicates a definitive detection of non-Gaussianity in the dataset, in agreement with the difference in the log evidence for the noise only analysis.  In the top plot we show the one and two-dimensional marginalised posterior distributions for the 5 non-Gaussian coefficients fit in the analysis of J0437$-$4715.  Vertical lines are included at 0 where visible in the plots, however, except for $\alpha_5$ all the coefficients are inconsistent with this value.  In the bottom plot we then show the set of equally weighted PDFs obtained from the non-Gaussian analysis  (black lines) setting $\gamma = 1$.  In addition we over plot the mean of the distribution (red line) and a unit Gaussian (blue line) all of which have been normalised to have a sum of 1.  The difference between the Gaussian and non-Gaussian PDFs is clear, with a larger probability for both small ($|\sigma| < 1$) and larger ($|\sigma| > 4$) deviations than given by the Gaussian PDF.  
That such a significant detection of non-Gaussianity does not lead to larger changes in the parameter estimates can potentially be attributed to a frequency dependence on the significance of the $\alpha$ parameters.  In \cite{2014MNRAS.443.1463S} the 10cm J0437$-$4715 data was found to be describable through Gaussian statistics alone.  This would suggest that the non-Gaussianity we detect exists primarily at low frequencies.  In Figure \ref{figure:0437NormRes} we show the normalised residuals from  Fig. \ref{figure:0437Noise} separated into its 10cm, 20cm and 50cm components, along with histograms for each wavelength.  Here the increase in non-Gaussian behaviour can clearly be seen as the wavelength increases.  Given the lowest frequencies have the greatest degree of non-Gaussianity it is less surprising that there is little impact on the timing or red spin noise parameters, as the low frequency data contributes the least to these parts of the model.  The low frequencies do, however, contribute greatly to the constraints on dispersion measure variations, and it is here we see the greatest difference between the Gaussian and non-Gaussian models.

\begin{table*}

\caption{Parameters for PSR J0437$-$4715. Figures in parentheses are  the nominal standard deviations in the least-significant digits quoted.}
\begin{tabular}{lcc}
\hline\hline
\multicolumn{2}{c}{Fit and data-set} \\
\hline
Pulsar name\dotfill & J0437$-$4715 \\ 
MJD range\dotfill & 50191.0---55619.2 \\ 
Data span (yr)\dotfill & 14.86 \\ 
Number of TOAs\dotfill & 5052 \\
\hline
\multicolumn{3}{c}{Measured Quantities} \\ 
\hline
Model Parameter & Non--Gaussian & Gaussian \\
\hline
Right ascension, $\alpha$ (rad)\dotfill &  1.20979650940(10) & 1.20979650943(11)\\ 
Declination, $\delta$ (rad)\dotfill & $-$0.82471224153(8) &  $-$0.82471224154(8)\\ 
Pulse frequency, $\nu$ (s$^{-1}$)\dotfill & 173.6879458121850(3) & 173.6879458121849(4)\\ 
First derivative of pulse frequency, $\dot{\nu}$ (s$^{-2}$)\dotfill & $-$1.728365(4)$\times 10^{-15}$ & $-$1.728365(4)$\times 10^{-15}$ \\ 
Dispersion measure, DM (cm$^{-3}$pc)\dotfill & 2.64462(11) & 2.64461(11)\\ 
First derivative of dispersion measure, $\dot{DM}$ (cm$^{-3}$pc\,yr$^{-1}$)\dotfill & $-$6(6)$\times 10^{-5}$ &  $-$7(7)$\times 10^{-5}$ \\ 
DM2 (cm$^{-3}$ pc yr$^{-2}$)\dotfill & $-$1(2)$\times 10^{-6}$ &  $-$1(2)$\times 10^{-6}$\\ 
Proper motion in right ascension, $\mu_{\alpha} \cos \delta$ (mas\,yr$^{-1}$)\dotfill & 121.439(3) & 121.441(3) \\ 
Proper motion in declination, $\mu_{\delta}$ (mas\,yr$^{-1}$)\dotfill & $-$71.474(3) & $-$71.474(3) \\ 
Parallax, $\pi$ (mas)\dotfill & 6.4(2) &  6.3(2) \\ 
Orbital period, $P_b$ (d)\dotfill & 5.7410462(3) & 5.7410461(3) \\ 
Epoch of periastron, $T_0$ (MJD)\dotfill & 54530.1722(3) & 54530.1721(3) \\ 
Projected semi-major axis of orbit, $x$ (lt-s)\dotfill & 3.36671463(8) & 3.36671464(8) \\ 
Longitude of periastron, $\omega_0$ (deg)\dotfill & 1.35(2) & 1.36(2)\\ 
Orbital eccentricity, $e$\dotfill & 1.91800(14)$\times 10^{-5}$ & 1.91796(15)$\times 10^{-5}$\\ 
First derivative of orbital period, $\dot{P_b}$\dotfill & 3.724(6)$\times 10^{-12}$ & 3.724(6)$\times 10^{-12}$\\ 
First derivative of $x$, $\dot{x}$ ($10^{-12}$)\dotfill & 1(2)$\times 10^{-15}$ &  1(2)$\times 10^{-15}$\\ 
Periastron advance, $\dot{\omega}$ (deg/yr)\dotfill & 0.0150(12) & 0.0150(13) \\ 
Companion mass, $M_c$ ($M_\odot$)\dotfill & 0.223(14) & 0.223(15) \\ 
Longitude of ascending node, $\Omega$ (degrees)\dotfill & 208.0(12) &  208.3(13)\\ 
Orbital inclination angle, $i$ (degrees)\dotfill & 137.1(8) &  137.3(8)\\ 
\hline
\hline
\label{Table:0437}
\end{tabular}
\end{table*}

\section{Conclusion}
\label{Section:Conclusion}

In this paper we have introduced a method of performing a robust Bayesian analysis of non-Gaussianity present in the residuals in pulsar timing analysis, simultaneously with the pulsar timing model, and additional stochastic parameters such as those describing the red noise, and dispersion measure variations present in the data.  
Deviations from Gaussianity are described using a set of parameters $\bmath{\alpha}$ that act to modify the probability density of the noise, such that $\bmath{\alpha} = 0$ describes Gaussian noise, and any non zero values provide support for non-Gaussian behaviour.  The advantage of this method is that one may use a finite set of non-zero $\alpha_m$ to model the non-Gaussianity, without mathematical inconsistency.  Any truncation of the series still yields a proper distribution, in contrast to the more commonly used Edgeworth expansion (e.g. \cite{2000ApJ...534...25C}).

We applied this method to two simulated datasets.  In simulation one the noise was drawn from a non-Gaussian distribution, and in simulation 2 it was purely Gaussian.
In simulation 1, the effect of the non-Gaussianity was to introduce a higher proportion of outliers relative to a Gaussian distribution. This resulted in an overestimation of the TOA uncertainties when assuming a Gaussian likelihood, and decreased the precision with which the timing model parameters could be extracted compared to an analysis that correctly incorporated the non-Gaussian behaviour on the noise.  

In the second case we showed that the parameter estimates of the timing model parameters of interest were consistent when including, or not, the $\bmath{\alpha}$ parameters, as is to be expected when the noise is Gaussian.

We then applied this method to the publicly available Parkes Pulsar Timing Array (PPTA) data release 1 dataset for the binary pulsar J0437$-$4715.  We detect a significant non-Gaussian component in the non-thermal component of the uncorrelated noise, however as the non-Gaussianity is most dominant in the lowest frequency data the impact on the timing precision in the pulsar is minimal, with only the parameter estimates of the power law dispersion measure variations being visible changed between the Gaussian and non-Gaussian analysis.


\begin{thebibliography}{}
\setlength{\labelwidth}{0pt} 



\bibitem[\protect\citeauthoryear{Coles et al.}{2011}]{2011MNRAS.418..561C} 
Coles W., Hobbs G., Champion D.~J., Manchester R.~N., Verbiest J.~P.~W., 
2011, MNRAS, 418, 561 



\bibitem[\protect\citeauthoryear{Contaldi et 
al.}{2000}]{2000ApJ...534...25C} Contaldi C.~R., Ferreira P.~G., Magueijo 
J., G{\'o}rski K.~M., 2000, ApJ, 534, 25 


\bibitem[\protect\citeauthoryear{Demorest et 
al.}{2013}]{2013ApJ...762...94D} Demorest P.~B., et al., 2013, ApJ, 762, 94 


\bibitem[\protect\citeauthoryear{Edwards, Hobbs, 
\& Manchester}{2006}]{2006MNRAS.372.1549E} Edwards R.~T., Hobbs G.~B., Manchester R.~N., 2006, MNRAS, 372, 1549 


\bibitem[\protect\citeauthoryear{Feroz 
\& Hobson}{2008}]{2008MNRAS.384..449F} Feroz F., Hobson M.~P., 2008, MNRAS, 384, 449 


\bibitem[\protect\citeauthoryear{Feroz, Hobson, 
\& Bridges}{2009}]{2009MNRAS.398.1601F} Feroz F., Hobson M.~P., Bridges M., 2009, MNRAS, 398, 1601 

\bibitem[\protect\citeauthoryear{Hall}{1989}]{1989AnnStat.17} Hall P., 1989, Ann. Statist. Volume 17, no. 2, 589--605 


\bibitem[\protect\citeauthoryear{Hellings 
\& Downs}{1983}]{1983ApJ...265L..39H} Hellings R.~W., Downs G.~S., 1983, ApJ, 265, L39 


\bibitem[\protect\citeauthoryear{Hobbs et al.}{2009}]{2009MNRAS.394.1945H} 
Hobbs G., et al., 2009, MNRAS, 394, 1945 


\bibitem[\protect\citeauthoryear{Hobbs, Edwards, 
\& Manchester}{2006}]{2006MNRAS.369..655H} Hobbs G.~B., Edwards R.~T., Manchester R.~N., 2006, MNRAS, 369, 655 


\bibitem[\protect\citeauthoryear{Jaffe 
\& Backer}{2003}]{2003ApJ...583..616J} Jaffe A.~H., Backer D.~C., 2003, ApJ, 583, 616 


\bibitem[\protect\citeauthoryear{Janssen et 
al.}{2008}]{2008AIPC..983..633J} Janssen G. H., Stappers B. W., Kramer M., Purver M., Jessner A., Cognard
I., 2008, in Bassa C.,Wang Z., Cumming A., KaspiV.M., eds, AIP Conf.
Proc. Vol. 983, 40 Years of Pulsars: Millisecond Pulsars, Magnetars and
More. Am. Inst. Phys., New York, p. 633


\bibitem[\protect\citeauthoryear{Kaspi, Taylor, 
\& Ryba}{1994}]{1994ApJ...428..713K} Kaspi V.~M., Taylor J.~H., Ryba M.~F., 1994, ApJ, 428, 713 


\bibitem[\protect\citeauthoryear{Kramer et al.}{2006}]{2006Sci...314...97K} 
Kramer M., et al., 2006, Science, 314, 97 


\bibitem[\protect\citeauthoryear{Lee et al.}{2012}]{2012MNRAS.423.2642L} 
Lee K.~J., Bassa C.~G., Janssen G.~H., Karuppusamy R., Kramer M., Smits R., 
Stappers B.~W., 2012, MNRAS, 423, 2642 


\bibitem[\protect\citeauthoryear{Lentati et 
al.}{2014}]{2014MNRAS.437.3004L} Lentati L., Alexander P., Hobson M.~P., 
Feroz F., van Haasteren R., Lee K.~J., Shannon R.~M., 2014, MNRAS, 437, 
3004 


\bibitem[\protect\citeauthoryear{Lentati et 
al.}{2013}]{2013PhRvD..87j4021L} Lentati L., Alexander P., Hobson M.~P., 
Taylor S., Gair J., Balan S.~T., van Haasteren R., 2013, Phys. Rev. D, 87, 104021 


\bibitem[\protect\citeauthoryear{Lorimer et 
al.}{2004}]{2004hpa..book.....L} Lorimer D. R., Kramer M., 2004, Ellis R., Huchra J., Kahn S., Rieke G.,
Stetson P. B., eds, Handbook of Pulsar Astronomy. Cambridge Univ. Press, Cambridge

\bibitem[\protect\citeauthoryear{Manchester et 
al.}{2013}]{2013PASA...30...17M} Manchester R.~N., et al., 2013, PASA, 30, 
17 

\bibitem[\protect\citeauthoryear{Matsakis, Taylor, 
\& Eubanks}{1997}]{1997A&A...326..924M} Matsakis D.~N., Taylor J.~H., Eubanks T.~M., 1997, A\&A, 326, 924 

\bibitem[\protect\citeauthoryear{O'Ruanaidh \& Fitzgerald}{1996}]{Thermo} O'RuanaidhJ. J. K., Fitzgerald W. J., 1996, Numerical Bayesian Methods
Applied to Signal Processing. Springer-Verlag, New York


\bibitem[\protect\citeauthoryear{Phinney}{2001}]{2001astro.ph..8028P} 
Phinney E. S., 2001, preprint (astro-ph/0108028)

\bibitem[\protect\citeauthoryear{Rocha et al.}{2001}]{2001PhRvD..64f3512R} 
Rocha G., Magueijo J., Hobson M., Lasenby A., 2001, Phys. Rev. D, 64, 063512 


\bibitem[\protect\citeauthoryear{Shannon 
\& Cordes}{2010}]{2010ApJ...725.1607S} Shannon R.~M., Cordes J.~M., 2010, ApJ, 725, 1607 

\bibitem[\protect\citeauthoryear{Shannon et 
al.}{2014}]{2014MNRAS.443.1463S} Shannon R.~M., et al., 2014, MNRAS, 443, 
1463 


\bibitem[\protect\citeauthoryear{Skilling}{2004}]{2004AIPC..735..395S} 
Skilling J., 2004, in Fischer R., Preuss R., von Toussaint U., eds, AIP Conf.
Proc. Vol. 735, Bayesian Inference and Maximum Entropy Methods in
Science and Engineering. Am. Inst. Phys., New York, p. 395


\bibitem[\protect\citeauthoryear{Taylor 
\& Weisberg}{1989}]{1989ApJ...345..434T} Taylor J.~H., Weisberg J.~M., 1989, ApJ, 345, 434 


\bibitem[\protect\citeauthoryear{van Haasteren et 
al.}{2011}]{2011MNRAS.414.3117V} van Haasteren R., et al., 2011, MNRAS, 
414, 3117 





\end{thebibliography}
\end{document}